\documentclass[11pt]{article}
\usepackage{fullpage}

\usepackage{braket,amsfonts,url}

\makeatletter
\let\NAT@parse\undefined
\makeatother

\usepackage[square, sort&compress, numbers]{natbib}
\usepackage{amsmath, amsthm}
\usepackage{array}

\usepackage[caption=false]{subfig}
\usepackage{caption}
\usepackage{algorithmic}

\usepackage{graphicx,epstopdf}

\usepackage{amsopn}

\usepackage{scalerel}
\usepackage{amssymb}
\usepackage{mathtools}
\usepackage{xspace}
\usepackage{bold-extra}
\usepackage[most]{tcolorbox}

\colorlet{texcscolor}{blue!50!black}
\colorlet{texemcolor}{red!70!black}
\colorlet{texpreamble}{red!70!black}
\colorlet{codebackground}{black!25!white!25}

\newcommand{\until}[1]{\{1,\dots, #1\}}

\newcommand{\subscr}[2]{#1_{\textup{#2}}}

\newcommand\oprocendsymbol{\hbox{$\square$}}
\newcommand\oprocend{\relax\ifmmode\else\unskip\hfill\fi\oprocendsymbol}

\def\real{\mathbb{R}}

\newtheorem{theorem}{Theorem}

\newtheorem{proposition}{Proposition}
\newtheorem{lemma}{Lemma}

\date{}
\title{\Large An Incremental Approach to Online Dynamic Mode Decomposition 
for Time-Varying Systems with Applications to EEG Data Modeling
\thanks{
This work has been supported by NSF Award IIS-1734272.
}}

\author{Mustaffa Alfatlawi \hspace{1in} Vaibhav Srivastava \thanks{Electrical and Computer Engineering Department, College of Engineering, Michigan State University, East Lansing, MI ({alfatlaw@msu.edu}, {vaibhav@egr.msu.edu }).}}

\begin{document}
\maketitle

\begin{abstract}
Dynamic Mode Decomposition (DMD) is a data-driven technique to identify a low dimensional linear
time invariant dynamics underlying high-dimensional data. For systems in which such underlying
low-dimensional dynamics is time-varying, a time-invariant approximation of such dynamics computed through standard DMD techniques may not be appropriate. We focus on DMD techniques
for such time-varying systems and develop incremental algorithms for systems without and with
exogenous control inputs. {We build upon the work in~\cite{zhang2019online} to scenarios in which high dimensional data are governed by low dimensional time-varying dynamics}.  
We consider two classes of algorithms that rely on (i) a discount factor on
previous observations, and (ii) a sliding window of observations. Our algorithms leverage existing
techniques for incremental singular value decomposition and allow us to determine an appropriately
reduced model at each time and are applicable even if data matrix is singular. We apply the developed
algorithms for autonomous systems to Electroencephalographic (EEG) data and demonstrate their
effectiveness in {terms of reconstruction} and prediction. Our algorithms for non-autonomous systems
are illustrated using randomly generated linear time-varying systems.
\end{abstract}


\section{Introduction}\label{S:1}
{Emergence of low cost sensors and their widespread deployment has led to an unprecedented amount of data. Extraction of actionable information from such plethora of data remains a challenge.  It is often the case that a low dimensional dynamical system governs the high-dimensional spatiotemporal sensory data. Dynamic Mode Decomposition (DMD) has emerged as a popular data-driven technique to efficiently compute such low dimensional dynamics \cite{schmid2010dynamic,kutz2016dynamic,tu2014dynamic}. One of the attractive features of the DMD approach is that it requires minimal assumptions on the data. Furthermore, its strong connections with the Koopman operator~\cite{rowley2009spectral,mezic2005spectral,mezic2013analysis,mezic2004comparison,koopman1931hamiltonian} makes it further appealing and theoretically grounded.  
The efficacy and simplicity of the DMD has inspired its application in a wide range of areas from fluid dynamics~\cite{schmid2010dynamic,kutz2016dynamic} to video processing~\cite{grosek2014dynamic, surana2015koopman} to epidemiology~\cite{proctor2015discovering} to neuroscience~\cite{brunton2016extracting}.

For high-dimensional data that is generated by an underlying low dimensional time-varying  dynamics, the standard DMD approach may be applicable only locally in time and appropriate time-varying DMD operators can be computed by discounting old data or considering a sliding window of observations~\cite{zhang2019online}. Even in scenarios when underlying dynamics is time-invariant and nonlinear, such time-varying linear approximations have been shown to be very effective~\cite{costa2019adaptive}. In this paper, we develop incremental methods for the computation of the DMD for time-varying systems and demonstrate the utility of the developed algorithm using Electroencephalographic (EEG) data. 

Incremental approaches to DMD refer to methods that allow for efficiently computing the DMD when data is provided sequentially instead of being provided as a batch. The sequential arrival might be {due to the inherent} nature of the application or it may be used for computational efficiency even if all the data is accessible. One such technique is streaming DMD~\cite{hemati2014dynamic,hemati2016improving} that first computes the projection of new data on the current orthogonal basis for the range of the data matrix and compares the norm of the difference between the projection and the new data with a threshold. If the difference is larger than the threshold, then it appends the basis with an additional element. Incremental total DMD introduced in~\cite{matsumoto2017fly} applies incremental Singular Value Decomposition (SVD)~\cite{brand2002incremental,brand2006fast,oxberry2017limited} for incremental computation of dynamic modes and subsequent identification of dominant modes.
% to update the SVD of the data as a prior step to the computation of the total DMD. 
{Unlike the streaming DMD and the total DMD, our focus in this paper is on systems with time-varying dynamics.}

{Within the context of time-varying systems, an incremental approach is the Online DMD ~\cite{zhang2019online}, which adapts the sequential least squares method~\cite{kay1993fundamentals, isermann2010identification} to incrementally update the DMD computation}. The online DMD incrementally computes the DMD operator using {the Sherman-Morrison} identity~\cite{sherman1950adjustment}. Authors of~\cite{zhang2019online} consider two strategies in their DMD computation to account {for time-varying} systems: (i) discounting the old data, and (ii) using a sliding window of observations. They also consider DMD for systems with control input~\cite{proctor2016dynamic,korda2018linear}. In particular, they extend the DMD with control approach proposed in~\cite{proctor2016dynamic} to develop an online DMD with control algorithm for time-varying system. 

In this paper, we build upon the work in~\cite{zhang2019online}, and develop incremental SVD based techniques for the computation of the online DMD for time-varying systems. 
{The proposed approach provides access to singular values of the data matrix at each time and hence allows it to determine a reduced order model. For the scenarios in which the high-dimensional data is governed by a low-dimensional dynamics, such reduced order models provide better future prediction accuracy.} Furthermore, for scenarios where {the Sherman-Morrison identity} cannot be applied, e.g., if the data matrix remains singular even after sufficiently long time, the access to singular values allows for efficient computation of appropriate pseudo-inverse.  The major contributions of this work are fourfold
% \textcolor{blue}{Unlike the streaming DMD which updates the spatial structure of large datasets, the proposed approach 
% implements weighted and windowed incremental SVD to update the spatial and temporal structure of the dataset to handle the time-varying dynamics. Also, the proposed approach is different from the Incremental TDMD by its capability to eliminate the representation of the old data either gradually or abruptly, and the the update to the DMD computation is performed recursively.}
\begin{enumerate}
    \item We leverage incremental SVD techniques to develop two algorithms for the computation of the online DMD for time-varying systems that rely on discounting old data and a sliding window of data, respectively;
    \item We extend both these algorithms to develop incremental SVD based algorithms for the computation of the online DMD with control input for time-varying systems; 
    \item We apply our algorithms for the case without control input on an EEG dataset and show their efficacy in reconstructing and predicting error related potential, a slow cortical potential seen in the EEG signal that is elicited by an unexpected outcome; 
    \item We apply our algorithms for the case with control input  on randomly generated linear time-varying dynamical systems, and demonstrate their efficacy.  
\end{enumerate}

% The proposed algorithms can be considered a generalization for the incremental TDMD algorithm for which it can be applied to both total and conventional DMD problems.

The remainder of the paper is organized as follows. In Section~\ref{DMD theory}, we {present} some background on the DMD for time-invariant and time-varying systems. In Section~\ref{Incremental SVD}, we present some background for the incremental SVD algorithms. In Section \ref{Incremental DMD}, we leverage the incremental SVD algorithms to develop novel incremental DMD algorithms.  We apply these algorithms to EEG data in Section~\ref{ErrPs}. We present incremental SVD-based algorithms for online DMD with control and illustrate its utility using numerical examples in Section~\ref{Incremental DMD with input}. Finally, we conclude in Section~\ref{sec:conclusions}.

\medskip
% {\color{red} Introduce general notation: bold upper case letters for matrices, bold lowercase for vectors, $\real^n$, $\mathbb C^n$, $(\cdot)^{\text{T}}$, $(\cdot)^*$, etc}

\noindent
\emph{Notation:}
We denote the set of real and complex matrices of size $n \times m$ by $\mathbb{R}^{n \times m}$ and $\mathbb{C}^{n \times m}$, respectively.  The set of real and complex $n$-dimensional vectors are denoted by $\mathbb{R}^{n}$ and $\mathbb{C}^{n}$, respectively. We denote matrices with bold upper case letters, vectors with bold lower case letters, and scalar with normal lower case letters. We denote the matrix transpose and the matrix  conjugate transpose by $(\cdot)^{\text{T}}$ and $(\cdot)^*$, respectively.  The kernel  and range of a matrix $\mathbf{X}$ are denoted by $\text{Ker} \mathbf{X}$, and $\text{Ran} \mathbf{X}$, respectively. 
}

%%%%%%%%%%%%%%%%%%
%%%%%%%%%%%%%%%%%%
%%%%%%%%%%%%%%%%
\section{Dynamic Mode Decomposition for Time-Varying Systems}\label{DMD theory}
{In this section, we recall the DMD setup and describe the time-varying DMD problem that we study in this paper. }

\subsection{Dynamic Mode Decomposition} \label{DMD}
{Consider the following discrete-time system:}
\begin{eqnarray}\label{NTI}
\mathbf{x}_{k+1}&=&\mathbf{f}(\mathbf{x}_k),   
\end{eqnarray}
where $\mathbf{x}_k\in \mathbb{R}^n$ is a high-dimensional state vector ($n\gg1$) 
% composed of $n$ spatially distributed measurements 
sampled at $t_{k}= k \Delta t$,  $k \in \until{(m+1)}$, and $\mathbf{f}$ is an unknown map which describes the evolution of the state vector between two subsequent sampling times. Suppose that the evolution of the high-dimensional state $\mathbf x$ is governed by some underlying low-dimensional dynamics. Then, the DMD computes a data-driven linear approximation to the system~\eqref{NTI} as follows~\cite{kutz2016dynamic, tu2014dynamic}.

Consider a collection of $(m+1)$ sequential measurements arranged in the following two datasets:
\begin{eqnarray}
\mathbf{X}=
\begin{bmatrix}
\mathbf{x}_1 & \mathbf{x}_2 & \cdots &\mathbf{x}_{m}
\end{bmatrix},
\mathbf{Y}=
\begin{bmatrix}
\mathbf{x}_2 & \mathbf{x}_3 & \cdots &\mathbf{x}_{m+1} 
\end{bmatrix}.\nonumber
\end{eqnarray}
Let the projection of $\mathbf{X}$ onto its leading $r$ singular vectors be $\bar{\mathbf{X}}=\bar{\mathbf{U}}_x \bar{\mathbf{\Sigma}}_x \bar{\mathbf{V}}_x^*$ 
%  Assume that the dataset $\mathbf{X}$ has the SVD of $\mathbf{X}=\mathbf{U}_r \mathbf{\Sigma}_r \mathbf{V}^*_r$
, where $r$ is the targeted dimension of the underlying low-dimensional dynamical system,  $\bar{\mathbf{U}}_x \in\mathbb{C}^{n\times r}$, $\bar{\mathbf{V}}_x \in \mathbb{C}^{r\times m}$, and $\bar{\mathbf{\Sigma}}_x = \text{diag} \{\sigma_1,\dots,\sigma_r \}  \in \mathbb{C}^{r \times r}$. Then, the DMD
framework approximates the dynamics~\eqref{NTI} by 
\[
\mathbf{x}_{k+1} = \bar{\mathbf{A}}\mathbf{x}_k, 
\]
where $\bar{\mathbf{A}} \in \real^{n \times n}$ is called the DMD operator and is given by 
\begin{equation}\label{eq:dmd-operator}
\bar{\mathbf{A}}= \mathbf{Y} \bar{\mathbf{V}}_x \bar{{\mathbf{\Sigma}}}_x^{-1} \bar{{\mathbf{U}}}_x^*.
\end{equation}

Let the projection of the state $\mathbf x$ and the DMD operator $\bar{\mathbf{A}}$ onto the space spanned by leading $r$ singular vectors of $\mathbf X$ be $\tilde{\mathbf x}_k = \bar{\mathbf{U}}^\ast \mathbf x_k$ and 
\begin{equation}
 \tilde{\mathbf{A}}= \bar{\mathbf{U}}_x^* \bar{\mathbf{A}} \bar{\mathbf{U}}_x
                  =\bar{\mathbf{U}}_x^* \mathbf{Y} \bar{\mathbf{V}}_x \bar{{\mathbf{\Sigma}}}_x^{-1}, \nonumber
\end{equation}
respectively. Then, the approximation to the underlying $r$-dimensional dynamics is
\begin{equation}\label{eq:dmd-projected}
\tilde{\mathbf x}_{k+1} =  \tilde{\mathbf{A}}\tilde{\mathbf x}_{k}. 
\end{equation}

Let $\mathbf{\Lambda}$ be the diagonal matrix of the eigenvalues of $\tilde{\mathbf{A}}$ and $\mathbf{W}$ be the matrix of the associated eigenvectors such that
\begin{equation} \label{dmdeigendecomposition}
\tilde{\mathbf{A}} \mathbf{W} = \mathbf{W} \mathbf{\Lambda}.
\end{equation}
Then, two types of DMD modes can be identified using the eigen-decomposition in \eqref{dmdeigendecomposition}: the projected DMD modes $\hat{\mathbf{\Phi}}= \bar{\mathbf{U}}_x \mathbf{W}$ and the exact DMD modes $\mathbf{\Phi}= \mathbf{Y} \bar{\mathbf{V}}_x \bar{{\mathbf{\Sigma}}}_x^{-1} \mathbf{W}$ \cite{tu2014dynamic}. Either of these DMD modes and the associated eigenvalues describe the evolution of the high-dimensional system using low-dimensional dynamics~\eqref{eq:dmd-projected}. 

% For both types of DMD modes, the associated DMD eigenvalues are the entries of $\mathbf{\Lambda}$. 

% can projected to obtain the lower dimensional state , and lower-dimensional DMD operator $\tilde{\mathbf{A}}$ given by:
% % Given that the spatial space of $\mathbf{X}$ has an $r$-dimensional basis consists of the leading $r$ left singular vectors, then  lower-dimensional DMD operator $\tilde{\mathbf{A}}$ in this basis is given by:
% \begin{equation}
%  \tilde{\mathbf{A}}= {\mathbf{U}}^* {\mathbf{A}} \mathbf{U}
%                   ={{\mathbf{U}}}^* \mathbf{Y} {\mathbf{V}} {{\mathbf{\Sigma}}}^{-1}. \nonumber
% \end{equation}

% The leading left singular vectors can be specified in a way to eliminate the modes generated by noisy measurements \cite{dawson2016characterizing,hemati2017biasing}. 

\subsection{Dynamic Mode Decomposition with Control}
Consider the following non-autonomous discrete-time system:
\begin{eqnarray} \label{NTI with control}
\mathbf{x}_{k+1}&=&\mathbf{f}^c(\mathbf{x}_k,\boldsymbol{\gamma}_k), 
\end{eqnarray}
where $\mathbf{x}_k\in \mathbb{R}^n$ is a high-dimensional state vector ($n\gg1$) sampled at $t_{k}= k \Delta t$,  $k \in \until{(m+1)}$, and $f^c$ is an unknown time-varying map which describes the evolution of the state vector between two subsequent sampling times, and $\boldsymbol{\gamma}_k \in \real^l$ is the exogenous input. Then,  the  Dynamic Mode Decomposition with Control (DMDc)  computes  a  data-driven  linear  approximation  to  the system \eqref{NTI with control} as follows \cite{proctor2016dynamic}.

Consider a collection of $m$ sequential measurements $\mathbf{x}_i \in \real^n$  and the associated exogeneous inputs $\boldsymbol{\gamma}_i \in \real^l$,   $i \in \until{m}$. DMDc
approximates the non-autonomous dynamics underlying these measurements by
\[
\mathbf{x}_{k+1} = \mathbf{A}\mathbf{x}_k + \mathbf{B} \boldsymbol{\gamma}_k = \mathbf{G} \begin{bmatrix}
\mathbf{x}_k \\\boldsymbol{\gamma}_k
\end{bmatrix}, 
\]
where $\mathbf{A} \in \real^{n \times n}$ is called the DMD operator, $\mathbf{B} \in \real^{n \times l}$ is called the input matrix, and $\mathbf G=\begin{bmatrix}
\mathbf{A} & \mathbf{B}
\end{bmatrix} \in \real^{n \times (n+l)}$. In particular, the DMDc algorithm constructs matrices $\mathbf{X}=\begin{bmatrix} \mathbf{x}_1 & \cdots & \mathbf{x}_{m-1} \end{bmatrix}$, $\mathbf{Y}=\begin{bmatrix} \mathbf{x}_2 & \cdots & \mathbf{x}_{m} \end{bmatrix}$, and $\mathbf{\Gamma}=\begin{bmatrix} \boldsymbol{\gamma}_1 & \cdots & \boldsymbol{\gamma}_{m-1} \end{bmatrix}$.
% , where $\mathbf{X},\mathbf{Y} \in \mathbb{R}^{n \times (m-1)}$ and $\mathbf{\Gamma} \in \real^{l \times (m-1)}$. 
Let $\mathbf{X}^{c} =\begin{bmatrix}\mathbf{X} \\ \mathbf{\Gamma} \end{bmatrix} \in \mathbb{R}^{(n+l) \times (m-1)}$, and $\mathbf{X}^{c}= \mathbf{U}^{c} \mathbf{\Sigma}^{c} \mathbf{V}^{c*}$ be its singular value decomposition, where $\mathbf{U}^{c} \in \mathbb{C}^{(n+l) \times (n+l)}$, $\mathbf{\Sigma}^{c} \in \mathbb{C}^{(n+l) \times (n+l)}$, and $\mathbf{V}^{c} \in \mathbb{C}^{(m-1)\times (n+l)}$. 

The DMDc algorithm estimates $\mathbf{G}$ by 
% Since $\mathbf{Y}= \mathbf{G} \mathbf{X}^{c}$ by construction, an estimate of $\mathbf{G}$ is:
 \begin{equation}
\mathbf{G}= \mathbf{Y} \mathbf{V}^{c} (\mathbf{\Sigma}^{c})^{-1} \mathbf{U}^{c*}. \nonumber
\end{equation}
Finally, $\mathbf{U}^{c*}$ can be written in a partitioned form of $\mathbf{U}^{c*}=\begin{bmatrix}
\mathbf{U}^{c_a*}  & \mathbf{U}^{c_b*} \end{bmatrix}$, where $\mathbf{U}^{c_a}\in \mathbb{C}^{n\times (n+l)}$ and $\mathbf{U}^{c_b}  \in \mathbb{C}^{l\times (n+l)}$. Then, the DMDc algorithm computes the matrices $\mathbf{A}^c$ and $\mathbf{B}^c$ given by
\begin{equation}
\mathbf{A}^c= \mathbf{Y} \mathbf{V}^{c} (\mathbf{\Sigma}^{c})^{-1} \mathbf{U}^{c_a*} \quad \text{and} \quad \mathbf{B}^c= \mathbf{Y} \mathbf{V}^{c} (\mathbf{\Sigma}^{c})^{-1} \mathbf{U}^{c_b*}.\nonumber
\end{equation}

Similar to the case of the DMD, the DMDc algorithm also enables identification of the low-dimensional non-autonomous system underlying the high-dimensional measurements. Assume that the matrix $\mathbf{X}^{c}$ can be approximated by its projection onto the leading $p$ singular vectors by $\bar{\mathbf{X}}^{c}=\bar{\mathbf{U}}^{c} \bar{\mathbf{\Sigma}}^{c} \bar{\mathbf{V}}^{c*}$, where $\bar{\mathbf{U}}^{c} \in \mathbb{C}^{(n+l) \times p}$, $\bar{\mathbf{\Sigma}}^{c} \in \mathbb{C}^{p \times p}$, and $\bar{\mathbf{V}}^{c} \in \mathbb{C}^{(m-1)\times p}$, and the matrix $\mathbf{X}$ can be approximated by its projection onto the leading $r \le p$ singular vectors by $\bar{\mathbf{X}}=\bar{\mathbf{U}} \bar{\mathbf{\Sigma}} \bar{\mathbf{V}}^*$, where $\bar{\mathbf{U}} \in \mathbb{C}^{n \times r}$, $\bar{\mathbf{\Sigma}} \in \mathbb{C}^{r \times r}$, and $\bar{\mathbf{V}} \in \mathbb{C}^{(m-1) \times r}$. Then, a reduced order model can be represented as follows:
\begin{eqnarray}
\tilde{\mathbf{x}}_{k+1}= \tilde{\mathbf{A}}^c\tilde{\mathbf{x}}_{k}+\tilde{\mathbf{B}}^c {\boldsymbol{\gamma}_k}, \nonumber
\end{eqnarray}
where the lower-dimensional matrices $ \tilde{\mathbf{A}}^c \in \mathbf{R}^{r \times r}$ and $\tilde{\mathbf{B}}^c \in \mathbf{R}^{r \times l}$ can be calculated by:
\begin{equation}
\tilde{\mathbf{A}}^c= \bar{\mathbf{U}}^{*} \mathbf{Y} \bar{\mathbf{V}}^{c} (\bar{\mathbf{\Sigma}}^{c})^{-1} \bar{\mathbf{U}}^{c_a*}\bar{\mathbf{U}}\quad \text{and} \quad \tilde{\mathbf{B}}^c= \bar{\mathbf{U}}^*\mathbf{Y} \mathbf{V}^{c} (\mathbf{\Sigma}^{c})^{-1} \mathbf{U}^{c_b*}.\nonumber
\end{equation}

{
\subsection{Problem Formulation}\label{S2.2}
In this paper, we study incremental algorithms for  the time-varying DMD and DMDc problems defined below. 
% For the dynamical system in \eqref{NTI}, the DMD modes and eigenvalues provide an approximation with in the vicinity of operating point defined by the sampling time $t_{k}$, which limits the accuracy of approximation as the operating point moves to $\left\lbrace t_{k}+l \Delta t \mid l\geq 1 , l \in \mathbb{Z}_{+} \right\rbrace$. 
\subsubsection{Time-varying DMD}
Consider the time-varying discrete time system of the form
% The time-varying aspect can be casted by assuming the following alternative discrete-time representation:    
\begin{eqnarray} \label{eq:prob-setup}
\mathbf{x}_{k+1}&=&\mathbf{f}(t_k, \mathbf{x}_k), 
\end{eqnarray}
where $\mathbf{x}_k\in \mathbb{R}^n$ is a high-dimensional state vector ($n\gg1$) 
% composed of $n$ spatially distributed measurements 
sampled at $t_{k}= k \Delta t$,  $k \in \until{(m+1)}$, and $\mathbf{f}$ is an unknown time-varying map which describes the evolution of the state vector between two subsequent sampling times. Similar to the standard DMD setup, we assume that the high-dimensional dynamics~\eqref{eq:prob-setup} is generated by a low-dimensional time-varying dynamics. 

Suppose that, at sampling time $t_{k+1}$, we have access to a collection of $(w+1)$ sequential measurements $\begin{bmatrix} \mathbf{x}_{k-w+1} & \mathbf{x}_{k-w+2} & \mathbf{x}_{k-w+3} & \cdots &\mathbf{x}_{k+1} \end{bmatrix}$. Our objective is to design efficient computational  techniques to identify the time-varying DMD operator $\mathbf{A}_k$ that {approximates the system}~\eqref{eq:prob-setup} by the following linear time-varying system
\begin{align}\label{A(k)_sol}
\mathbf{x}_{k+1}=\mathbf{A}_k \mathbf{x}_k, 
\end{align}
such that the cost function
\begin{align}\label{cost function}
\mathbf{J}(\mathbf{A}_k)= \sum_{j=k-w+1}^{k} \left\|\rho^{(k-j)}(\mathbf{y}_{j} - \mathbf{A}_{k} \mathbf{x}_{j})\right\|^2, 
\end{align}
is minimized, where $\rho\in (0,1]$ {is a discounting factor}.
% such that 
% approximates~\eqref{eq:prob-setup}.
% for $j\in \{k-w+1,\cdots,k\}$, and 
% such that $\mathbf{A}_k $ minimizes 

\subsubsection{Time-varying DMDc}
Consider the following variant of the system in \eqref{eq:prob-setup}
\begin{eqnarray} \label{eq:control prob-setup}
\mathbf{x}_{k+1}&=&\mathbf{f}^c(t_k, \mathbf{x}_k,\boldsymbol{\gamma}_k), 
\end{eqnarray}
where $f^c$ is an unknown time-varying map which describes the evolution of the state vector between two subsequent sampling times, and $\boldsymbol{\gamma}_k \in \real^l$ is the exogenous input.  
% for which the definitions in \eqref{eq:prob-setup} for $\mathbf{x}_k$, $\mathbf{t}_k$, and $\mathbf{f}$ hold. However, in this system, the effects of exogenous factors are introduced as $\boldsymbol{\gamma}_k$. 

Assume that, at sampling time $\mathbf{t}_{k+1}$, two collections of sequential measurements are available, one for the system states $\begin{bmatrix} \mathbf{x}_{k-w+1} & \mathbf{x}_{k-w+2} & \mathbf{x}_{k-w+3} & \cdots &\mathbf{x}_{k+1} \end{bmatrix}$, and the other one for the exogenous inputs $\begin{bmatrix} \boldsymbol{\gamma}_{k-w+1} & \boldsymbol{\gamma}_{k-w+2} & \boldsymbol{\gamma}_{k-w+3} & \cdots &\boldsymbol{\gamma}_{k+1} \end{bmatrix}$. Our objective is to design efficient computational techniques to identify the time-varying DMDc operator $\mathbf{A}^c_k$ and input matrix $\mathbf{B}^c_k$ that {approximates the system}  \eqref{eq:control prob-setup} by the following linear time-varying system

\begin{align}\label{AB(k)_sol}
\mathbf{x}_{k+1}=\mathbf{A}^c_k \mathbf{x}_k+\mathbf{B}^c_k \boldsymbol{\gamma}_k, 
\end{align}
such that the cost function
\begin{align}\label{cost function for control}
\mathbf{J}^c(\mathbf{A}^c_k,\mathbf{B}^c_k)= \sum_{j=k-w+1}^{k} \left\|\rho^{(k-j)}\left(\mathbf{y}_{j} - \mathbf{A}^c_{k} \mathbf{x}_{j}-\mathbf{B}^c_k \boldsymbol{\gamma}_j\right)\right\|^2, 
\end{align}
is minimized, where $\rho\in (0,1]$ is a discounting factor.

Let $\mathbf{G}^*_k= \begin{bmatrix} \mathbf{A}_k^{c*} & \mathbf{B}_k^{c*}\end{bmatrix}$, and $\mathbf{x}^{c*}_j= \begin{bmatrix} \mathbf{x}_j^* & \boldsymbol{\gamma}_j^*\end{bmatrix}$. Then, the linear model \eqref{AB(k)_sol} can be written as
\begin{align}\label{G(k)_sol}
\mathbf{x}^c_{k+1}=\mathbf{G}_k \mathbf{x}^c_k, 
\end{align}
% for $j\in \{k-w+1,\cdots,k\}$, 
and the cost function \eqref{cost function for control} can be equivalently written as
\begin{align}\label{G cost function for control}
\mathbf{J}^c(\mathbf{G}_k)= \sum_{j=k-w+1}^{k} \left\| \rho^{(k-j)}\left(\mathbf{y}_{j} - \mathbf{G}_{k} \mathbf{x}^c_{j}\right) \right\|^2. 
\end{align}

{Online DMD approach~\cite{zhang2019online} also minimizes the cost functions \eqref{cost function} and \eqref{cost function for control}, and computes the time-varying DMD and DMDc operators.} In addition to the computation of these operators, the computational techniques developed in this paper also enable efficient computation of (i) the associated DMD eigenvalues and modes, and (ii) the linear time-varying systems that approximate the low-dimensional dynamics underlying~\eqref{eq:prob-setup} and~\eqref{eq:control prob-setup}, respectively.

% also includes the efficient computation of , as well as the efficient computation of , as well as .

% Along with the time-varying DMD operator $\mathbf{A}_k$, these techniques should also enable efficient computation of the time-varying low-dimensional DMD operator $\tilde{\mathbf{A}}_k$ that describes the underlying low-dimensional dynamics, as well as the associated DMD eigenvalues and modes. 

% on the data measurements by limiting the number of represented measurements to be $w<k$, and neutralizing the discounting factor by assigning the value of $\rho =1$.  
\subsection{Block Computation of Time-varying DMD and DMDc} 
We refer to a computation that requires all the data until the sampling time $t_k$ to compute the desired solution at time $t_k$ as a block computation. In contrast, an incremental computation uses only the data at time $t_k$ and the solution at time $t_{k-1}$ to compute the desired solution. 

{In this section, we describe block computation for time-varying DMD and DMDc in Lemma~\ref{block DMD lemma} and~\ref{block DMDc lemma}, respectively. These lemmas are immediate consequence of least square solutions of a linear system of equations~\cite[Section 5.13]{CDM:01} and we omit their proofs.} We will use these block computations to derive the incremental computations later in the paper. 

% The incremental techniques developed later in the paper should be consistent with these block computations. 

% The block algorithm can be described as an intuitive algorithm for identifying the time-varying DMD models because it requires to use all accessible measurements at each sampling time. The following lemma provide the necessary framework for the block algorithm for time varying DMD.

\begin{lemma}\label{block DMD lemma}
Consider a sequence of $(w+1)$ measurements $\{ \mathbf{x}_{k-w+1}, \ldots, \mathbf{x}_{k+1} \}$  at sampling time $t_{k+1}$ that is arranged in the following two matrices 
\begin{eqnarray}
\label{XY(k)}
\mathbf{X}_{k}&=
\begin{bmatrix}
\rho^{k}\mathbf{x}_{k-w+1} & \rho^{k-1}\mathbf{x}_{k-w+2} &\cdots& \mathbf{x}_{k} 
\end{bmatrix}, \\
\mathbf{Y}_{k}&=
\begin{bmatrix}
\rho^{k}\mathbf{y}_{k-w+1} & \rho^{k-1}\mathbf{y}_{k-w+2} &\cdots& \mathbf{y}_{k}
\end{bmatrix},\nonumber
\end{eqnarray}  
where $\mathbf{x}_j \in \mathbb{R}^n$, for each $j\in \{k-w+1, \cdots, k+1\}$, and $\mathbf{y}_j=\mathbf{x}_{j+1}$. Then, the following statements hold for the cost function~\eqref{cost function}
\begin{enumerate}
{\item if $\text{Ker}\mathbf{X}_k^*$ is trivial, then the unique minimizer of \eqref{cost function} is 
\begin{align} \label{unique lse Ak}
  \mathbf{A}_{k}=\mathbf{Y}_{k}\mathbf{X}_{k}^* \left(\mathbf{X}_{k}\mathbf{X}_{k}^*\right)^{-1}
  =\mathbf{Y}_{k}\mathbf{V}_{x_{k}} \mathbf{\Sigma}_{x_{k}}^{-1} \mathbf{U}_{x_{k}}^*,
\end{align}
where $\mathbf{U}_{x_{k}}\in\mathbb{C}^{n\times n}$ is a unitary matrix, $\mathbf{\Sigma}_{x_{k}} =\text{diag}\{\sigma_1,\dots,\sigma_n\}\in\mathbb{C}^{n\times n}$, $\mathbf{V}_{x_{k}}\in\mathbb{C}^{k\times n}$ has orthonormal columns, and are defined by the reduced SVD of $\mathbf{X}_{k} = \mathbf{U}_{x_{k}}\mathbf{\Sigma}_{x_{k}}\mathbf{V}_{x_{k}}^*$. }
 \item if $\text{Ker}\mathbf{X}_k^*$ is non-trivial, then there exists infinitely many minimizers of \eqref{cost function}, and the unique minimizer with the smallest induced two-norm is
 \begin{align}\label{min norm LS A(k)}
    \bar{\mathbf{A}}_k= \mathbf{Y}_{k}\bar{\mathbf{V}}_{x_{k}} \bar{\mathbf{\Sigma}}_{x_{k}}^{-1} \bar{\mathbf{U}}_{x_{k}}^*,
\end{align}
where $\bar{\mathbf{U}}_{x_{k}}\in\mathbb{C}^{n\times r}$, $\bar{\mathbf{\Sigma}}_{x_{k}} =\text{diag}\{\sigma_1,\dots,\sigma_r\}\in\mathbb{C}^{r\times r}$, $\bar{\mathbf{V}}_{x_{k}}\in\mathbb{C}^{k\times r}$, and $r=n-\text{dim}\left(\text{Ker}\mathbf{X}_k^*\right)$, are defined by the reduced SVD of $\mathbf{X}_{k} \approx \bar{\mathbf{U}}_{x_{k}}\bar{\mathbf{\Sigma}}_{x_{k}}\bar{\mathbf{V}}_{x_{k}}^*$. 
\end{enumerate} 
% \oprocend 
% a least squares solution for $\mathbf{A}_k$ that minimizes the cost function in \eqref{cost function} 
% can be obtained by solving the following system of linear equations
% \begin{align} \label{system of normal equations}
%     \mathbf{Y}_k \mathbf{X}^*_k = \mathbf{A}_k(\mathbf{X}_k \mathbf{X}_k^*),
% \end{align}
% A unique least square solution is given by
% \begin{align} \label{unique lse Ak}
%   \mathbf{A}_k = \mathbf{Y}_k \mathbf{X}^*_k (\mathbf{X}_k \mathbf{X}_k^*)^{-1},
% \end{align}
% if and only if $\text{Ker}\mathbf{X}_k^*$ is trivial. 
% \\Furthermore,  then there exists a unique solution $\bar{\mathbf{A}}_k$ such that $\|\bar{\mathbf{A}}_k\| \le \|\mathbf{A}_k\|$ for all $\mathbf{A}_k$ satisfies \eqref{system of normal equations}, given by
% \begin{align}\label{min norm LS A(k)}
%     \bar{\mathbf{A}}_k= \mathbf{Y}_{k}\bar{\mathbf{V}}_{x_{k}} \bar{\mathbf{\Sigma}}_{x_{k}}^{-1} \bar{\mathbf{U}}_{x_{k}}^*
% \end{align}
% where $\bar{\mathbf{U}}_{x_{k}}\in\mathbb{C}^{n\times r}$, $\bar{\mathbf{\Sigma}}_{x_{k}} =\text{diag}\{\sigma_1,\dots,\sigma_r\}\in\mathbb{C}^{r\times r}$, $\bar{\mathbf{V}}_{x_{k}}\in\mathbb{C}^{k\times r}$, and $r=n-\text{dim}\left(\text{Ker}\mathbf{X}_k^*\right)$, are defined by the reduced SVD of $\mathbf{X}_{k} \approx \bar{\mathbf{U}}_{x_{k}}\bar{\mathbf{\Sigma}}_{x_{k}}\bar{\mathbf{V}}_{x_{k}}^*$. 
\end{lemma}
% The proof of Lemma~\ref{block DMD lemma} is provided in Appendix~\ref{sec:proof2.1}. We now consider the block computation of the time-varying DMDc in the following lemma. 

% The following lemma considers the time varying DMDc, and can be proven by following the same steps presented in the proof of Lemma \ref{block DMD lemma} argument. 
\begin{lemma}\label{block DMDc lemma}
Consider two sequences of $(w+1)$ measurements $\{ \mathbf{x}_{k-w+1}, \ldots, \mathbf{x}_{k+1} \}$, and $\{ \boldsymbol{\gamma}_{k-w+1}, \cdots, \boldsymbol{\gamma}_{k+1} \}$ at sampling time $t_{k+1}$ that are arranged in the following two matrices 
\begin{eqnarray}
\label{XcY(k)}
\mathbf{X}^c_{k}&=
\begin{bmatrix}
\rho^{k}\mathbf{x}^c_{k-w+1} & \rho^{k-1}\mathbf{x}^c_{k-w+2} &\cdots& \mathbf{x}^c_{k} 
\end{bmatrix} \in \mathbb{R}^{(n+l) \times w}, \\
\mathbf{Y}_{k}&=
\begin{bmatrix}
\rho^{k}\mathbf{y}_{k-w+1} & \rho^{k-1}\mathbf{y}_{k-w+2} &\cdots& \mathbf{y}_{k}
\end{bmatrix}\in \mathbb{R}^{n \times w},\nonumber
\end{eqnarray}  
where $\mathbf{y}_j=\mathbf{x}_{j+1}$, $\mathbf{x}^{c*}_j= \begin{bmatrix} \mathbf{x}_j^* & \boldsymbol{\gamma}_j^*\end{bmatrix} \in \mathbf{R}^{(n+l)}$. Then the following statements hold for the cost function in \eqref{G cost function for control}
\begin{enumerate}
 {\item if $\text{Ker}\mathbf{X}_k^{c*}$ is trivial, then the unique minimizer of \eqref{cost function for control} is 
\begin{align} \label{unique lse Gk}
  \mathbf{G}_k = \mathbf{Y}_{k}\mathbf{X}_{k}^{c*} \left(\mathbf{X}^c_{k}\mathbf{X}_{k}^{c*}\right)^{-1}=\mathbf{Y}_{k}\mathbf{V}^c_{x_{k}} (\mathbf{\Sigma}^c_{x_{k}})^{-1} \mathbf{U}_{x_{k}}^{c*},
\end{align}
where $\mathbf{U}^c_{x_{k}}\in\mathbb{C}^{(n+l)\times (n+l)}$ is a unitary matrix, $\mathbf{\Sigma}^c_{x_{k}} =\text{diag}\{\sigma_1,\dots,\sigma_{n+l}\}\in\mathbb{C}^{(n+l)\times (n+l)}$, $\mathbf{V}^c_{x_{k}}\in\mathbb{C}^{k\times (n+l)}$ has orthonormal columns, and are defined by the reduced SVD of $\mathbf{X}^c_{k} = \mathbf{U}^c_{x_{k}}\mathbf{\Sigma}^c_{x_{k}}\mathbf{V}_{x_{k}}^{c*}$.} 
\item if $\text{Ker}\mathbf{X}_k^*$ is non-trivial, then there exists infinitely many minimizers of \eqref{cost function for control}, and the unique minimizer with the smallest induced two-norm is
 \begin{align}\label{min norm LS G(k)}
    \bar{\mathbf{G}}_k= \mathbf{Y}_{k}^c\bar{\mathbf{V}}^c_{x_{k}} (\bar{\mathbf{\Sigma}}_{x_{k}}^c)^{-1} \bar{\mathbf{U}}_{x_{k}}^{c*}
\end{align}
where $\bar{\mathbf{U}}^c_{x_{k}}\in\mathbb{C}^{(n+l)\times r}$, $\bar{\mathbf{\Sigma}}^c_{x_{k}} =\text{diag}\{\sigma_1,\dots,\sigma_r\}\in\mathbb{C}^{r\times r}$, $\bar{\mathbf{V}}^c_{x_{k}}\in\mathbb{C}^{k\times r}$, and $r=(n+l)-\text{dim}\left(\text{Ker}\mathbf{X}_k^{c*}\right)$, are defined by the reduced SVD of $\mathbf{X}^c_{k} \approx \bar{\mathbf{U}}^c_{x_{k}}\bar{\mathbf{\Sigma}}^c_{x_{k}}\bar{\mathbf{V}}_{x_{k}}^{c*}$. 
\end{enumerate} 
\end{lemma}

% Lemma~\ref{block DMDc lemma} can be proved analogously to Lemma~\ref{block DMD lemma} and its proof is omitted.

Note that using the partitioned form of $\mathbf{U}_k^{c*}=\begin{bmatrix}
\mathbf{U}_k^{c_a*}  & \mathbf{U}_k^{c_b*} \end{bmatrix}$, where $\mathbf{U}_k^{c_a}\in \mathbb{C}^{n\times r}$ and $\mathbf{U}_k^{c_b}  \in \mathbb{C}^{l\times r}$, the matrices $\bar{\mathbf{A}}_k^c$ and $\bar{\mathbf{B}}_k^c$ associated with~\eqref{min norm LS G(k)} can be obtained as follows
% can be obtained from the minimizer $\bar{\mathbf{G}}^*_k= \begin{bmatrix} \bar{\mathbf{A}}_k^{c*} & \bar{\mathbf{B}}_k^{c*}\end{bmatrix}$ as
\begin{align}\label{min norm LS ABc(k)}
    \bar{\mathbf{A}}^c_k= \mathbf{Y}_{k}\bar{\mathbf{V}}^c_{x_{k}} (\bar{\mathbf{\Sigma}}_{x_{k}}^{c})^{-1} \bar{\mathbf{U}}_{x_{k}}^{c_a*},
    \bar{\mathbf{B}}^c_k= \mathbf{Y}_{k}\bar{\mathbf{V}}^c_{x_{k}} (\bar{\mathbf{\Sigma}}_{x_{k}}^{c})^{-1} \bar{\mathbf{U}}_{x_{k}}^{c_b*}.
\end{align}

% The calculations in the block algorithms require the involvement of all accessible data measurements and generate new models at each sampling time, which is considered inefficient from the computational point of view. In the following sections, more efficient algorithms are presented which update the model from the previous sampling time using only the newly accessed measurements. 
}

{Similar to the online DMD~\cite{zhang2019online}}, we focus on two special cases of the cost functions~\eqref{cost function} and \eqref{G cost function for control}, namely, the weighted cost functions and the windowed cost functions. 
% techniques for representing the data measurements in the DMD algorithms. 
The weighted cost functions consider a gradual elimination for the old measurements by assigning the discounting factor to be $\rho \in (0,1)$ and  consider all the available data, i.e., they select $w=k$. 
The windowed cost functions consider a sharp cut-off window and use a non-discounted window of recent measurements, i.e., they select $\rho=1$ and $w<k$. In the remainder of the paper, we will refer to the DMD and the DMDc operators/matrices obtained using the weighted (resp., windowed) cost functions by the weighted (resp., windowed) DMD and DMDc operator/matrices and denote them by $\mathbf{A}_k^\rho$ (resp., $\mathbf{A}_k^w$) and $(\mathbf{A}_k^{\rho_c},\mathbf{B}_k^{\rho_c})$ (resp., $(\mathbf{A}_k^{w_c},\mathbf{B}_k^{w_c})$), respectively.

\section{Background on Incremental SVD Algorithms} \label{Incremental SVD}
At the heart of the DMD framework is the utilization of the SVD to identify an invariant subspace for the low-dimensional dynamics underlying the high-dimensional data. The techniques proposed in this paper for the computation of the time-varying DMD operator require access to the SVD of certain time-varying data matrix at each time. For efficient computation of these SVDs, we resort to incremental SVD techniques proposed in~\cite{brand2002incremental,brand2006fast,oxberry2017limited}. In this section, we present these incremental SVD techniques. The presentation below is adapted from~\cite{brand2002incremental} in order to facilitate better exposition in Section~\ref{Incremental DMD}. 

%  Thus, to reflect the contribution of the incoming measurements to the system dynamics, the left and right singular vectors, and singular values should be updated accordingly. 
 
%  Fortunately, updating the SVD for a dataset of measurements in response to any addition to these measurements is the theme of a popular method known as Incremental SVD. 
 
%  In this section, three algorithms are presented which allow to generate time-varying DMD modes and eigenvalues. Two of them named by weighted and windowed incremental DMD are applied in the absence of any external influence to the system, while the third is devised to include such influence. Throughout this section, 

%  the incremental SVD algorithm presented in~\cite{brand2006fast} is implemented to update the SVD of the dataset after the arrival of a new snapshot for the system states and control inputs. 

%%%%%%%%%%%%%%%%%%%%%%%%%%
%%%%%%%%%%%%%%%%%%%%%%%%%%%
%%%%%%%%%%%%%%%%%%%%%%%%%%%

\subsection{Weighted Incremental SVD Algorithm} \label{Weighed Incremental SVD}
Consider a weighted dataset  
\begin{equation}\label{weighted X(k)}
\mathbf{X}_{k}=
\begin{bmatrix}
\rho^{k-1}\mathbf{x}_{1} & \rho^{k-2}\mathbf{x}_{2} & \cdots & \mathbf{x}_{k}
\end{bmatrix},
\end{equation}   
where $ 0\leq \rho\leq 1$ is a scalar discounting factor and $k\ge n$. Given an additional datum $\mathbf{x}_{k+1}$, let 
% As the new snapshot is available at time $t_{k+2}$, the updated dataset can be defined as:
\begin{eqnarray}\label{weighted X(k+1)}
 {
 \mathbf{X}_{k+1}=
 \begin{bmatrix}
 \rho^{k}\mathbf{x}_{1} & \rho^{k-1}\mathbf{x}_{2} & \cdots & \rho \mathbf{x}_{k}& \mathbf{x}_{k+1}
 \end{bmatrix}
 }
 =\begin{bmatrix}
\rho \mathbf{X}_{k} & \mathbf{x}_{k+1}
\end{bmatrix},
\end{eqnarray}  
be the incremented dataset. Provided that the SVD of $\mathbf{X}_k$ is known, the SVD of $\mathbf{X}_{k+1}$ can be incrementally computed using the following proposition, which is proved in Appendix~\ref{proof of Proposition 1}. 
%= \mathbf{U}_{x_{k+1}} \mathbf{\Sigma}_{x_{k+1}} \mathbf{V}^*_{x_{k+1}}
\begin{proposition}\label{weighted incremental SVD  proposition } 
Let at sampling time $t_k$ the SVD of the dataset in \eqref{weighted X(k)} be defined by $~\boldsymbol{X}_{k}=\mathbf{U}_{x_k} \mathbf{\Sigma}_{x_k} \mathbf{V}_{x_k}^*$, where ${\mathbf{U}_{x_k}} \in\mathbb{C}^{n\times n}$ is a unitary matrix, 
${\mathbf{V}_{x_k}} \in\mathbb{C}^{k\times n}$ has orthonormal columns, 
and $\mathbf{\Sigma}_{x_k}= \text{diag}\{\sigma_1,\dots,\sigma_n\} 
\in\mathbb{C}^{n\times n}$ is a diagonal matrix.
Assume that at sampling time $t_{k+1}$, a new datum $\mathbf{x}_{k+1}$ is accessed. Then, the SVD for the dataset in \eqref{weighted X(k+1)}, defined by $~\mathbf{X}_{k+1}=\mathbf{U}_{x_{k+1}} \mathbf{\Sigma}_{x_{k+1}} \mathbf{V}^*_{x_{k+1}}$, is given by :
\begin{eqnarray}\label{weighted2 updated SVD factors}
\mathbf{U}_{x_{k+1}} = \mathbf{U}_{x_k} \mathbf{U}_{s_k},~
\mathbf{\Sigma}_{k+1}=\rho \mathbf{\Sigma}_{s_k}
,~\text{and}~
\mathbf{V}_{x_{k+1}}=
\begin{bmatrix}
\mathbf{V}_{x_k} \mathbf{V}_{s_{k,1}}\\
\mathbf{v}_{s_{k,2}}
\end{bmatrix},
\end{eqnarray}
where $\mathbf{U}_{s_k}$, $\mathbf{\Sigma}_{s_k}$, and $\mathbf{V}_{s_k}=\begin{bmatrix}\mathbf{V}_{s_{k,1}}\\ \mathbf{v}_{s_{k,2}} \end{bmatrix}$, with $\mathbf{v}_{s_{k,2}} \in \mathbb{C}^{1\times n}$, are defined by the following SVD
\begin{equation}
\mathbf{S}_k \triangleq \begin{bmatrix}\mathbf{\Sigma}_{x_k} & \rho^{-1} \mathbf{U}_{x_k}^* \mathbf{x}_{k+1} \end{bmatrix}=\mathbf{U}_{s_k} \mathbf{\Sigma}_{s_k} \mathbf{V}_{s_k} ^*.\nonumber
\end{equation}
\end{proposition}

The above incremental SVD computation requires the computation of the SVD of matrix $\mathbf{S}_k \in \mathbb{C}^{n \times (n+1)}$. The size of this matrix is smaller than the size of dataset $\mathbf{X}_{k+1}$ and it has the so-called broken arrow structure which enables efficient computation of its SVD~\cite{gu1993stable,huckle1998efficient,chen2014lwi}.

%%%%%%%%%%%%%%%%%%%%%%%%%%
%%%%%%%%%%%%%%%%%%%%%%%%%%%
%%%%%%%%%%%%%%%%%%%%%%%%%%%
\subsection{Windowed Incremental SVD Algorithm}\label{Windowed Incremental SVD}
Consider a windowed dataset
% Assume that the available dataset at time $t_{k+1}$ is an $n \times w$ matrix given by :  
\begin{eqnarray}\label{windowX(k)}
\boldsymbol{\chi}_k=
\begin{bmatrix}
\mathbf{x}_{k-w+1} & \mathbf{x}_{k-w+2} & \cdots & \mathbf{x}_{k} 
\end{bmatrix}\in \real^{n \times w}, 
\end{eqnarray}
where $n$ is the dimension of measurements and $w$ is the length of the time-window of the desired measurements. 
Define $q$ as follows
\begin{equation} \label{generic dim}
  q \triangleq
    \begin{cases}
      n, & \text{if } n < w, \\
      w, & \text{otherwise}. \\
    \end{cases}     
\end{equation} 
Consider another dataset 

\begin{equation}\label{windowX(k+1)}
\boldsymbol{\chi}_{k+1}=
\begin{bmatrix}
\mathbf{x}_{k-w+2} & \mathbf{x}_{k-w+3} &...& \mathbf{x}_{k+1} 
\end{bmatrix}. 
\end{equation}
Then, the incremental SVD computes the SVD of $\boldsymbol{\chi}_{k+1} = \mathbf{U}_{{\chi}_{k+1}} \mathbf{\Sigma}_{{\chi}_{k+1}} \mathbf{V}_{{\chi}_{k+1}}^{*}$ as follows. First, it eliminates $\mathbf{x}_{k-w+1}$ from the dataset to obtain decremented dataset and the associated SVD 
\begin{equation}\label{windowX(k-1)}
\boldsymbol{\acute{\chi}}_{k}=
\begin{bmatrix}
\mathbf{x}_{k-w+2} & \mathbf{x}_{k-w+3} &...& \mathbf{x}_{k} 
\end{bmatrix}. 
\end{equation}
Then, it increments $\boldsymbol{\acute{\chi}}_{k}$ with $\mathbf{x}_{k+1}$ to obtain $\boldsymbol{\acute{\chi}}_{k+1}$ and the associated SVD. The SVD of the dataset $\boldsymbol{\acute{\chi}}_{k}$ can be incrementally computed using the following proposition, which is proved in Appendix~\ref{proof of Proposition 2}.  
\begin{proposition}\label{windowed incremental SVD  proposition}
Let at sampling time $t_k$, the SVD of the dataset in \eqref{windowX(k)} be defined by $~\boldsymbol{\chi}_{k}=\mathbf{U}_{\chi_k} \mathbf{\Sigma}_{\chi_k} {\mathbf{V}_{\chi_k}^*}$, where $\mathbf{U}_{\chi_k} \in \mathbb{C}^{{n} \times {q}}$, $\mathbf{\Sigma}_{\chi_k} = \begin{bmatrix}\text{diag}\{\sigma_1,\dots,\sigma_q\} & \mathbf{0}_{q \times (w-q)}\end{bmatrix} \in \mathbb{C}^{{q} \times {w}}$, and $\mathbf{V}_{\chi_k} \in \mathbb{C}^{{w} \times {w}}$ is a unitary matrix. 
% Assume that at sampling time $t_{k+2}$, the oldest available datum $\mathbf{x}_{k-w+1}$ is eliminated. 
Then, the SVD of the dataset in \eqref{windowX(k-1)} defined as $~\boldsymbol{\acute{\chi}}_{k}= \mathbf{U}_{\boldsymbol{\acute{\chi}}_{k}} \mathbf{\Sigma}_{\boldsymbol{\acute{\chi}}_{k}} \mathbf{V}_{\boldsymbol{\acute{\chi}}_{k}}^*$, is given by :
\begin{eqnarray}
 \mathbf{U}_{\acute{\chi}_{k}}= \mathbf{U}_{\chi_k} \mathbf{U}_{\acute{s}_k}, ~
 {\mathbf{\Sigma}}_{\acute{\chi}_{k}}={\mathbf{\Sigma}}_{\acute{s}_k},~ \text{and} ~
 {\mathbf{V}}_{\acute{\chi}_{k}}= 	\mathbf{V}_{\chi_{k,2}} \mathbf{V}_{\acute{s}_k},
\end{eqnarray}
where $\mathbf{U}_{\acute{s}_k}$, $\mathbf{V}_{\acute{s}_k}$, and $\mathbf{\Sigma}_{\acute{s}_k}$ are defined by the following SVD
\begin{equation}
\mathbf{\acute{S}}_k \triangleq \mathbf{\Sigma}_{\chi_k}- \mathbf{U}_{\chi_k}^{*}\mathbf{x}_{k-w+1} \mathbf{z}_{1}^{\text{T}} \mathbf{V}_{\chi_k} = \mathbf{U}_{\acute{s}_k} \mathbf{\Sigma}_{\acute{s}_k} \mathbf{V}_{\acute{s}_k} ^\ast, \nonumber
\end{equation}
and 
% {\color{red} undefined notation $\mathbf{V}_{\chi_{k,2}}= \mathbf{V}_{\boldsymbol{\acute{\chi}}_{k}(2:w,1:w)}$. 
$\mathbf{V}_{\chi_{k,2}}$ is the submatrix of $\mathbf{V}_{\boldsymbol{{\chi}}_{k}}$ obtained after removing its first row.  
\end{proposition}

In the incremental step, the new datum $\mathbf{x}_{k+1}$ is appended to $\boldsymbol{\acute{\chi}}_{k}$ and the SVD of the resulting  matrix $\boldsymbol{\chi}_{k+1}$ can be obtained using the procedure presented in Section \ref{Weighed Incremental SVD}. In particular, let $\mathbf{\hat{S}}_k \triangleq \begin{bmatrix} {\mathbf{\Sigma}}_{\acute{\chi}_{k}} &  \mathbf{U}_{\acute{\chi}_{k}}^* \mathbf{x}_{k+1} \end{bmatrix}=\mathbf{U}_{\hat{s}_k} \mathbf{\Sigma}_{\hat{s}_k} \mathbf{V}^*_{\hat{s}_k}$, and using equation \eqref{weighted2 updated SVD factors}, the SVD of $\boldsymbol{\chi}_{k+1} = \mathbf{U}_{\chi_{k+1}} \mathbf{\Sigma}_{\chi_{k+1}} \mathbf{V}_{\chi_{k+1}}^{*}$ is 
\begin{eqnarray}\label{windowed updated SVD factors}
\mathbf{U}_{\chi_{k+1}} &=&  \mathbf{U}_{\acute{\chi_k}} \mathbf{U}_{\hat{s}_k} = \mathbf{U}_{\chi_k} \mathbf{U}_{\acute{s}_k} \mathbf{U}_{\hat{s}_k}\nonumber\\ 
\mathbf{\Sigma}_{\chi_{k+1}}&=&\mathbf{\Sigma}_{\hat{s}_k} \\ 
\mathbf{V}_{\chi_{k+1}} &=&
\begin{bmatrix}
\mathbf{V}_{\acute{\chi_k}} \mathbf{V}_{\hat{s}_{k,1}} \\
\mathbf{v}_{\hat{s}_{k,2}}
\end{bmatrix} = 
\begin{bmatrix}
\mathbf{V}_{\chi_{k,2}} \mathbf{V}_{\acute{s}_k} \mathbf{V}_{\hat{s}_{k,1}} \\
\mathbf{v}_{\hat{s}_{k,2}}
\end{bmatrix}.\nonumber
\end{eqnarray}

The windowed incremental SVD computation requires the computation of the SVD of $\mathbf{\acute{S}}_k \in \mathbb{R}^{q \times w}$. In general, $\mathbf{\acute{S}}_k$ may not possess any special structure. However, if $n \gg w$, in which case $q=w$, the above incremental update maybe a lot cheaper than the computation of the SVD of an $n \times w$ matrix. Also, in this context, the above procedure helps in term of storage as it does not require the large matrix $\boldsymbol{\chi}_{k+1}$ to be stored and uses only the first and last column of this matrix along with the SVD at the previous iteration.

%The size of $\acute S$ is the less than the size of the dataset $\mathbf{X}_{k+1}$. Furthermore, in general, $\acute S$ may not possess any special structure. However, if a reduced order representation of rank $r < n$ of $\mathbf{X}_{k+1}$ is of interest, then the above procedure can be applied using a rank $r$ representation of $\mathbf{X}_k$ and in this case $\acute S \in \mathbb{C}^{(r+1) \times (r+1)}$, which may have much smaller size than $\mathbf{X}_{k+1}$. 

% Note that the incremental SVD computations requires the computation of the SVD of matrices $\mathbf{S} \in \mathbb{R}^{r \times (r+1)}$ and $\mathbf{\acute{S}}\in \mathbb{R}^{r \times r}$, where $r \le \min\{n, k\}$ is the targeted dimension of the reduced order system. These SVDs can be computed very efficiently and with a lower storage requirement by comparison with the SVD of $\mathbf{X}_k$ ~\cite{gu1993stable,huckle1998efficient,chen2014lwi}. 

\section{Incremental DMD Algorithms}\label{Incremental DMD}
In this section, we present two algorithms, namely weighted incremental DMD and windowed incremental DMD, respectively, for incremental computation of a time-varying lower-dimensional DMD operator. {Similar to the existing approaches in the literature, the proposed algorithms perform incremental computations either by assigning decaying weights to the received data~\cite{kay1993fundamentals, zhang2019online}, or by using a sliding window of the received data~\cite{grosek2014dynamic, macesic2017koopman,zhang2019online}. The recursive updates derived in this section are similar to those in sequential least square method~\cite[Appendix 8C]{kay1993fundamentals}, recursive least square method~\cite[Section 9.4]{isermann2010identification}, as well as online DMD~\cite{zhang2019online}. }  

% in the absence of any external influence on the system. 

\subsection{Weighted Incremental DMD Algorithm} \label{Weighted Incremental DMD}
In this algorithm,  the time-varying DMD operator is estimated  by assigning a decaying weight to past measurements in order to gradually discount their representation as newer measurements become available.
Assume that at sampling  time $t_{k+1}$, we have access to the following weighted datasets:
\begin{eqnarray}
\label{weighted XY(k)}
\mathbf{X}_{k}=
\begin{bmatrix}
	\rho^{k-1}\mathbf{x}_{1} & \rho^{k-2}\mathbf{x}_{2} &\cdots& \mathbf{x}_{k} 
\end{bmatrix}, 
\mathbf{Y}_{k}=
\begin{bmatrix}
	\rho^{k-1}\mathbf{y}_{1} & \rho^{k-2}\mathbf{y}_{2} &\cdots& \mathbf{y}_{k}
\end{bmatrix},
\end{eqnarray}    
where $\mathbf{y}_{k} = \mathbf{x}_{k+1}$, for each $k\in \mathbb{N}$. Assume that at sampling time $t_{k+2}$, the datasets in \eqref{weighted XY(k)} are updated with a pair of measurements  $(x_{k+1},y_{k+1})$ such that
\begin{align}\label{weighted XY(k+1)}
\begin{split}
\mathbf{X}_{k+1}&=
{\begin{bmatrix}
	\rho^{k}\mathbf{x}_{1} & \rho^{k-1}\mathbf{x}_{2} & \cdots& \rho \mathbf{x}_{k}& \mathbf{x}_{k+1}
\end{bmatrix}}=
\begin{bmatrix}
	\rho \mathbf{X}_{k} & \mathbf{x}_{k+1}
\end{bmatrix}, \\
\mathbf{Y}_{k+1}&=
{\begin{bmatrix}
	\rho^{k}\mathbf{y}_{1} & \rho^{k-1}\mathbf{y}_{2} & \cdots & \rho \mathbf{y}_{k}& \mathbf{y}_{k+1}
\end{bmatrix}}=
\begin{bmatrix}
	\rho \mathbf{Y}_{k} & \mathbf{y}_{k+1}
\end{bmatrix}.
\end{split}
\end{align}    
{Suppose that at sampling time $t_{k+1}$, the DMD operator $\mathbf{A}^{\rho}_{k}$ and the SVD of $\mathbf{X}_{k}$ are known, then at sampling time $t_{k+2}$, the DMD operator can be updated with the new pair of measurements $(\mathbf{x}_{k+1},\mathbf{y}_{k+1})$ using the following theorem, which is proved in Appendix~\ref{proof of Theorem 1}.
\begin{theorem}\label{weighted DMD operator(k+1) theorem}
Let at sampling time $t_{k+1}$ the SVD of $\mathbf{X}_{k}=\mathbf{U}_{x_k} \mathbf{\Sigma}_{x_k} \mathbf{V}_{x_k}^*$ be known, and let the DMD operator $\mathbf{A}^{\rho}_{k} \in \mathbb{R}^{n\times n}$ {minimize} the cost function in \eqref{cost function} with $\mathbf{Y}_k$ and $\mathbf{X}_k$ given in \eqref{weighted XY(k)}.  Assume that at sampling time $t_{k+2}$, a new pair of measurements $(\mathbf{x}_{k+1},\mathbf{y}_{k+1})$ is used to incrementally compute the SVD of $\mathbf{X}_{k+1}$ using Proposition \ref{weighted incremental SVD  proposition }. Then, the DMD operator $\mathbf{A}^{\rho}_{k+1} \in \mathbb{R}^{n\times n}$ which minimizes the cost function in \eqref{cost function} is
\begin{eqnarray}\label{DMD operator(k+1)}
\mathbf{A}^{\rho}_{k+1}=\mathbf{A}^{\rho}_{k}+\left(\mathbf{y}_{k+1}-\mathbf{A}^{\rho}_{k}\mathbf{x}_{k+1}\right)\mathbf{v}_{s_{k,2}}\mathbf{\Sigma}_{x_{k+1}}^{-1} \mathbf{U}_{x_{k+1}}^*.
\end{eqnarray}

\end{theorem}
} If $\mathbf{X}_{k+1}$ is well-approximated by its projection onto its leading $r$ singular vectors given by  $\bar{\mathbf{X}}_{k+1}=\bar{\mathbf{U}}_{x_{k+1}} \bar{\mathbf{\Sigma}}_{x_{k+1}} \bar{\mathbf{V}}^*_{x_{k+1}}$, then the DMD operator update in \eqref{DMD operator(k+1)} takes the  form
\begin{equation}\label{DMD operator(k+1) bar}
\bar{\mathbf{A}}^{\rho}_{k+1}={\bar{\mathbf{A}}^{\rho}}_{k}+\left(\mathbf{y}_{k+1}-{\bar{\mathbf{A}}^{\rho}}_{k}\mathbf{x}_{k+1}\right)\bar{\mathbf{v}}_{s_{k,2}}\bar{\mathbf{\Sigma}}_{x_{k+1}}^{-1} \bar{\mathbf{U}}^*_{x_{k+1}}.
\end{equation}
Moreover,  the DMD operator $\bar{\mathbf{A}}^{\rho}_{k+1} \in \mathbb{R}^{n \times n}$  can be projected onto a subspace spanned by the leading $r$ left singular vectors of $\mathbf{X}_{k+1}$ to obtain a lower-dimensional DMD operator  $\tilde{\mathbf{A}}^{\rho}_{k+1} \in \mathbb{R}^{r \times r}$ given by
\begin{eqnarray}\label{reduced DMD operator(k+1)}
\tilde{\mathbf{A}}^{\rho}_{k+1}=\bar{\mathbf{U}}^*_{x_{k+1}} \bar{\mathbf{A}}^{\rho}_{k+1} {\bar{\mathbf{U}}_{x_{k+1}}}
=\bar{\mathbf{U}}^*_{x_{k+1}} \bar{\mathbf{A}}^{\rho}_{k}\bar{\mathbf{U}}_{x_{k+1}}
+\bar{\mathbf{U}}^*_{x_{k+1}} \left(\mathbf{y}_{k+1}- \bar{\mathbf{A}}^{\rho}_{k}\mathbf{x}_{k+1}\right)\bar{\mathbf{v}}_{s_2}\bar{\mathbf{\Sigma}}_{x_{k+1}}^{-1}.
\end{eqnarray}
The update in equation \eqref{DMD operator(k+1)} requires only the current DMD operator, measurement pair $(\mathbf{x}_{k+1},\mathbf{y}_{k+1})$, and the incremental SVD update. Specifically, these updates do not require the data matrix to be stored. 

\subsection{Windowed Incremental DMD Algorithm} \label{Windowed Incremental DMD}
We now focus on windowed incremental DMD algorithm in which a sliding-window of $w$ most recent measurements is used to estimate the time-varying DMD operator. 
%In this part, an alternative algorithm is presented by which a $w$-length time-window is moved over the datasets to include the most recent snapshots  and abruptly eliminates the snapshots which were sampled before $w$ time samples.
% \subsubsection{Algorithm for Windowed Incremental Dynamic Mode Decomposition}\label{S3.2.2}
Assume that we have access to the following windowed datasets at sampling time $t_{k+1}$:
% the following datasets contain snapshots during the $k^{th}$ time-window:  
\begin{eqnarray}\label{windowed XY(k)}
\boldsymbol{\chi}_k=
\begin{bmatrix}
\mathbf{x}_{k-w+1} & \cdots & \mathbf{x}_{k}
\end{bmatrix}, \quad \text{and} \quad 
\boldsymbol{\Upsilon}_k=
\begin{bmatrix}
\mathbf{y}_{k-w+1} & \cdots & \mathbf{y}_{k}
\end{bmatrix}.
\end{eqnarray}
Accordingly, at sampling time $t_{k+2}$, we have access to
%and for the $(k+1)^{\text{th}}$ time-window, the updated datasets are given by: 
\begin{eqnarray}\label{windowed XY(k+1)}
\boldsymbol{\chi}_{k+1}= 
\begin{bmatrix}
\mathbf{x}_{k-w+2} & \cdots & \mathbf{x}_{k+1}
\end{bmatrix}, \quad \text{and} \quad 
\boldsymbol{\Upsilon}_{k+1}= 
\begin{bmatrix}
\mathbf{y}_{k-w+2} & \cdots & \mathbf{y}_{k+1}
\end{bmatrix}.
\end{eqnarray} 
{Suppose that at sampling time $t_{k+1}$, the DMD operator $\mathbf{A}^{w}_{k}$ and the SVD of $\boldsymbol{\chi}_{k}$ are known, then at sampling time $t_{k+2}$, the DMD operator can be updated with the new pair of measurements $(\mathbf{x}_{k+1},\mathbf{y}_{k+1})$ using the following theorem, which is proved in Appendix~\ref{proof of Theorem 2}.
\begin{theorem}\label{windowed DMD operator(k+1) theorem}
Let at sampling time $t_{k+1}$ the SVD of $\boldsymbol{\chi}_{k}=\mathbf{U}_{\chi_{k}} \mathbf{\Sigma}_{\chi_{k}}\mathbf{V}^*_{\chi_{k}}$ be known, and let the DMD operator $\mathbf{A}^{w}_{k} \in \mathbb{R}^{n\times n}$ {minimize} the cost function in \eqref{cost function} with $\boldsymbol{\Upsilon}_{k}$ and $\boldsymbol{\chi}_k$ given in \eqref{windowed XY(k)}.  
Assume that at sampling time $t_{k+2}$, a new pair of measurements $(\mathbf{x}_{k+1},\mathbf{y}_{k+1})$ is used to incrementally compute 
% the oldest pair of measurements $(\mathbf{x}_{k-w+1},\mathbf{y}_{k-w+1})$ are replaced by the newest pair of measurements $(\mathbf{x}_{k+1},\mathbf{y}_{k+1})$ in order to generate the datasets in \eqref{windowed XY(k+1)}, and
 the SVD of $\boldsymbol{\chi}_{k+1}$ using Proposition \ref{windowed incremental SVD  proposition} and equation~\eqref{windowed updated SVD factors}. Then, the DMD operator $\mathbf{A}^{w}_{k+1} \in \mathbb{R}^{n\times n}$ which minimizes the cost function in \eqref{cost function} is
\begin{equation}\label{windowed DMD operator(k+1)}
\mathbf{A}^w_{k+1}=\mathbf{A}^w_{k}+\left(\mathbf{y}_{k+1}-\mathbf{A}^w_{k}\mathbf{x}_{k+1}\right)\mathbf{v}_{\hat{s}_{k,2}}\mathbf{\Sigma}_{\chi_{k+1}}^{-1} \mathbf{U}_{\chi_{k+1}}^*.
\end{equation}
\end{theorem}

}If $\boldsymbol{\chi}_{k+1}$ is well-approximated by its projection onto its leading $r$ singular vectors given by $\bar{\boldsymbol{\chi}}_{k+1}=\bar{\mathbf{U}}_{\chi_{k+1}} \bar{\mathbf{\Sigma}}_{\chi_{k+1}} \bar{\mathbf{V}}^*_{\chi_{k+1}}$, then the DMD operator takes the form
\begin{equation}\label{windowed DMD operator(k+1) bar}
\bar{\mathbf{A}}^w_{k+1}=\bar{\mathbf{A}}^w_{k}+\left(\mathbf{y}_{k+1}-\bar{\mathbf{A}}^w_{k}\mathbf{x}_{k+1}\right)\bar{\mathbf{v}}_{\hat{s}_{k,2}}\bar{\mathbf{\Sigma}}_{\chi_{k+1}}^{-1} \bar{\mathbf{U}}_{\chi_{k+1}}^*.
\end{equation}
Moreover,  the DMD operator $\bar{\mathbf{A}}^w_{k+1} \in \mathbb{R}^{n \times n}$  can be projected onto a subspace spanned by the leading $r$ left singular vectors of $\boldsymbol{\chi}_{k+1}$ to obtain a lower-dimensional DMD operator  ${\tilde{\mathbf{A}}}^w_{k+1} \in \mathbb{R}^{r \times r}$ given by
\begin{align}\label{reduced windowed DMD operator(k+1)}
\tilde{\mathbf{A}}^w_{k+1}&=\bar{\mathbf{U}}_{\chi_{k+1}}^* \bar{\mathbf{A}}^w_{k+1} \bar{\mathbf{U}}_{\chi_{k+1}}\nonumber \\
&=\bar{\mathbf{U}}_{\chi_{k+1}}^* \bar{\mathbf{A}}^w_{k} \bar{\mathbf{U}}_{\chi_{k+1}}
+\bar{\mathbf{U}}_{\chi_{k+1}}^*\left(\mathbf{y}_{k+1}-\bar{\mathbf{A}}^w_{k}\mathbf{x}_{k+1}\right)\bar{\mathbf{v}}_{\hat{s}_{k,2}}\bar{\mathbf{\Sigma}}_{\chi_{k+1}}^{-1}.
\end{align}
 Similar to the weighted incremental DMD, {the} update in equation \eqref{windowed DMD operator(k+1)} requires only the current DMD operator, measurement pair $(\mathbf{x}_{k+1},\mathbf{y}_{k+1})$, and the windowed incremental SVD update. 
 
{For a fixed dimension $r$ of the reduced order system, the computational complexity of rank one incremental SVD update is $\mathcal{O}(nr +r^3)$~\cite[Section 4.3]{brand2006fast}. In comparison, computational complexities of each iteration in the streaming DMD  and the online DMD are  $\mathcal{O}(nr^2)$ and  $\mathcal{O}(n^2)$, respectively. When data matrices are non-singular, both online and incremental DMD approaches can be used to compute time-varying DMD operators. In such cases, if a reduced order DMD operator of dimension $r$ is of interest, then the proposed incremental DMD approach has a better computational complexity, when $r< n^{2/3}$.  
}
 
%  \textcolor{green}{Like the streaming DMD, the computational cost of the presented approaches in this section is $\mathcal{O}(nr)$ without the calculations of DMD modes and eigenvalues and $\mathcal{O}(nr^2)$ with them , and it is dominated by the computational cost of the incremental SVD. By comparison with the computational cost of online DMD which is $\mathcal{O}(n^2)$, the proposed approach is faster only if $r<\sqrt{n}$.}    
 
 \section{Dynamic Mode Decomposition for Error-Related Potentials in EEG data}\label{ErrPs}
\begin{figure}[ht!]
  \centering
  {
  \includegraphics[width=0.5\linewidth]{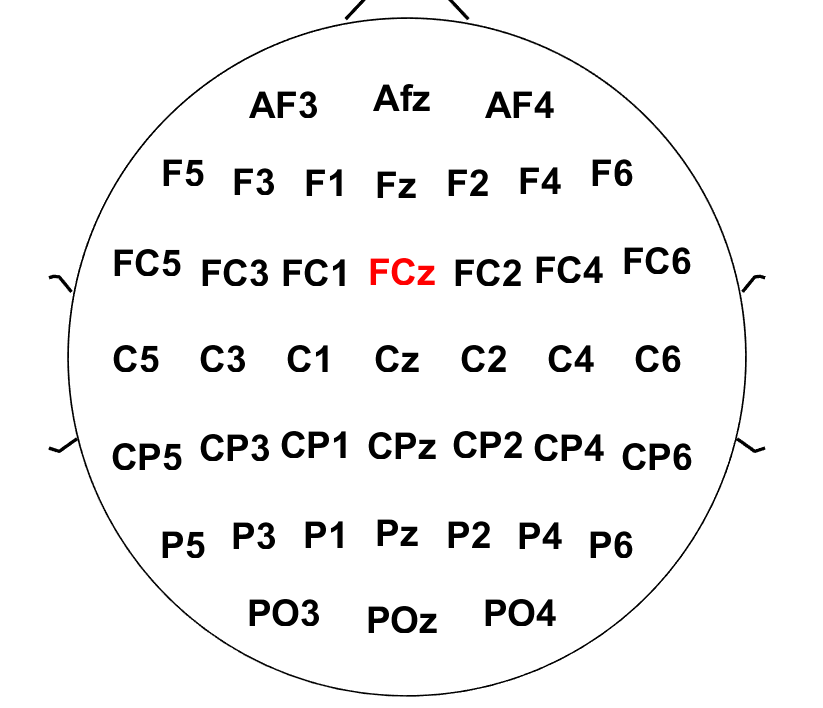}
  }
 \caption{Topographical view for EEG channels with the channel FCz, where the ErrPs can be characterized, marked in red bold font }
 \label{$10/20$ biosemi}
 \end{figure}

Error‐ Related Potentials (ErrPs) are slow cortical potentials seen in the EEG signal of human subjects that {are} elicited by an unexpected (erroneous) outcome. For example, within the context of human-machine interaction, such signals are observed when the machine takes an unexpected action~\cite{ferrez2008error,chavarriaga2010learning,chavarriaga2014errare}. 
%is induced in the EEG of human operators as a result of monitoring an action which is in contrast with their intentions.
In this section, we illustrate the efficacy of the proposed DMD algorithms on EEG recordings taken from Monitoring Error-Related Potentials {database} \cite{chavarriaga2010learning,chavarriaga2014errare}.
 These EEG recordings were collected from six subjects while they were observing the movement of a cursor between two labeled targets on a screen. The subjects had no control on the motion of the cursor. However, they had a priori knowledge about the intended directions of movement. The cursor movements were set to elicit ErrPs by generating two types of events, namely, correct events and erroneous events. The correct events correspond to the cursor movement in the intended direction, while  the erroneous events correspond to the cursor movement in any unintended direction. EEG signals were recorded at sampling rate of $512$ Hz, using Biosemi ActiveTwo system with $64$ electrodes distributed according to the standard $10/20$ international system as it is shown in Figure \ref{$10/20$ biosemi}.
 
\begin{figure}[ht!]
  \centering
  \subfloat[EEG pattern during correct event]
  {
   \label{ErrPs database:a}
  \includegraphics[width=0.3\linewidth]{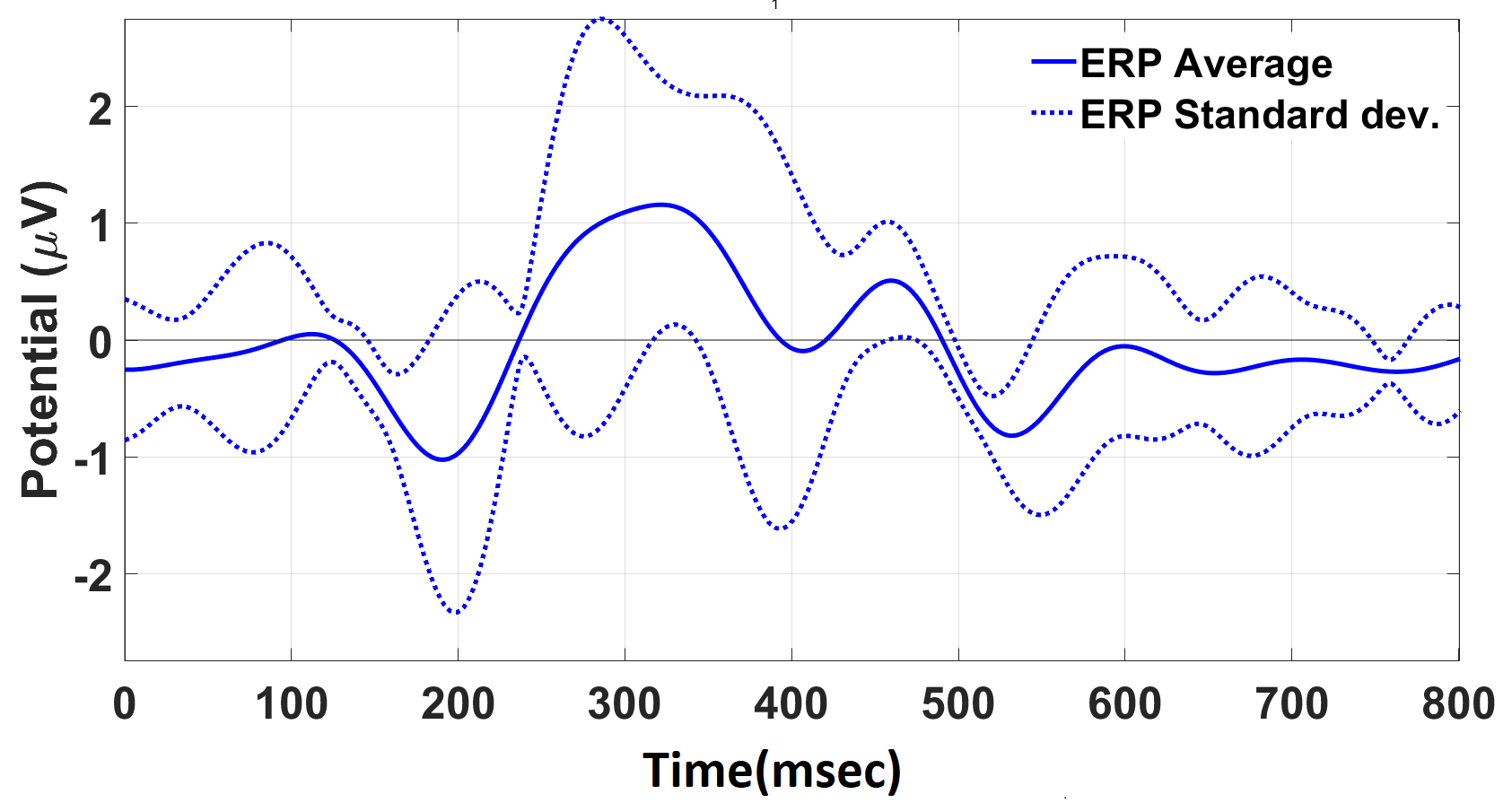}
  \includegraphics[width=0.6\linewidth]{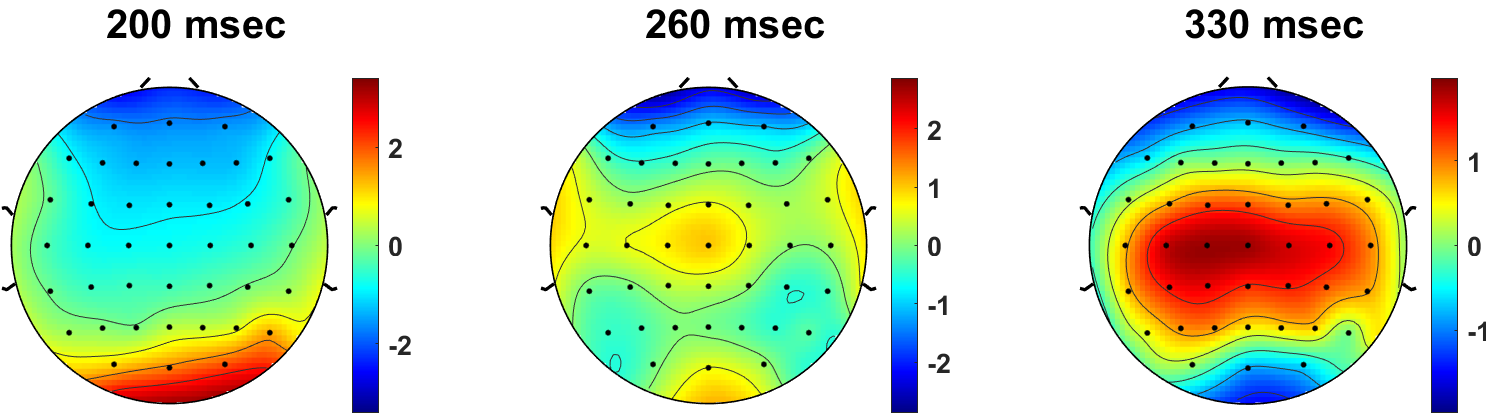}
  } \\
  \subfloat[EEG pattern during erroneous event]
  {
   \label{ErrPs database:b}
  \includegraphics[width=0.3\linewidth]{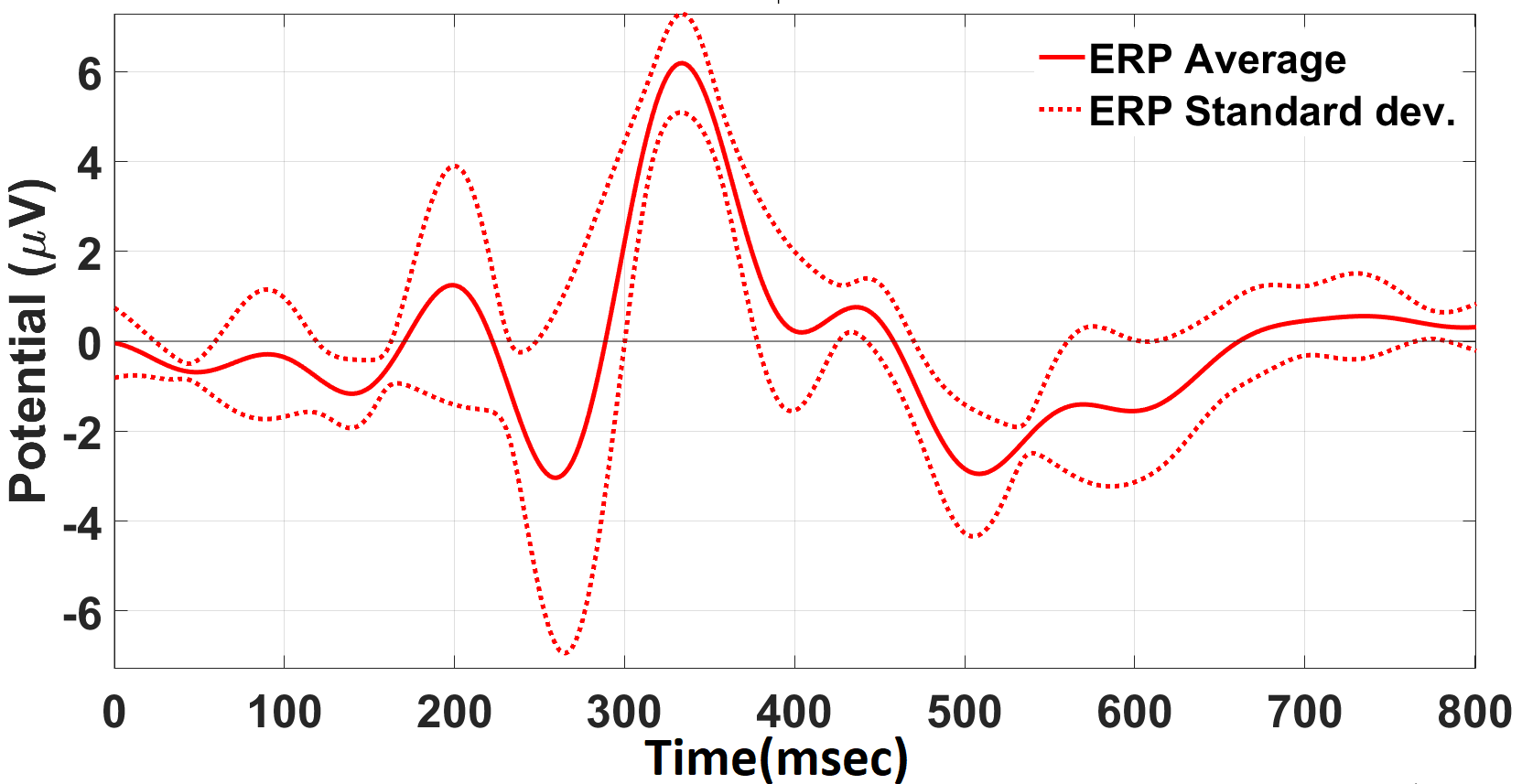}
  \includegraphics[width=0.6\linewidth]{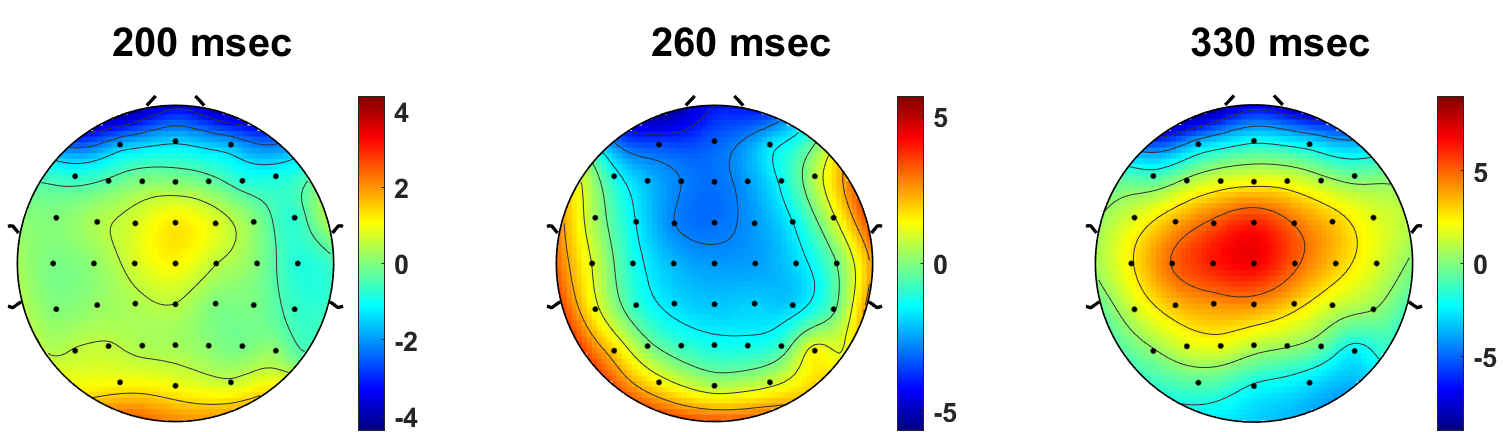}
  }\\
  \subfloat[Differential EEG pattern]
  {
   \label{ErrPs database:c}
  \includegraphics[width=0.3\linewidth]{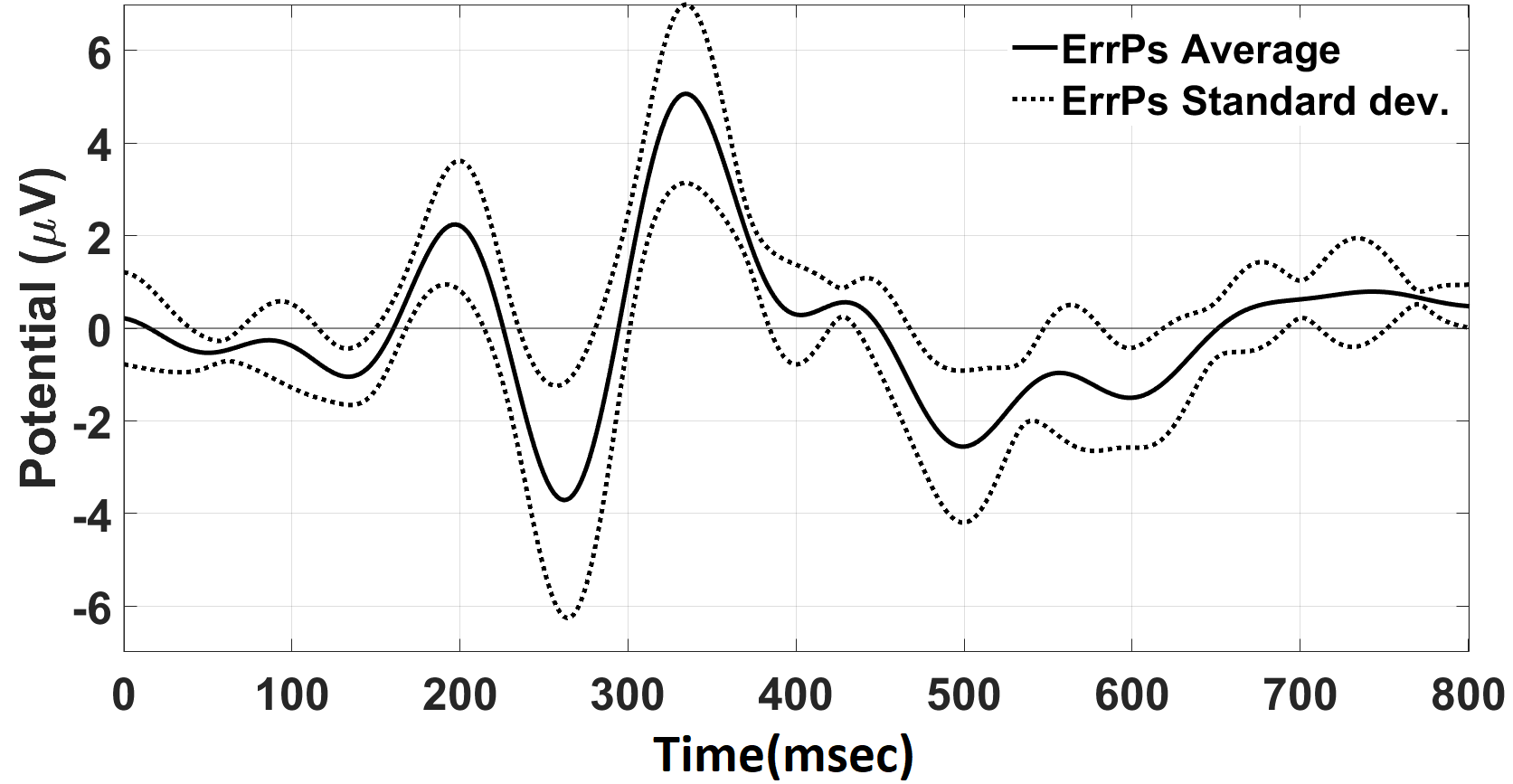}
  \includegraphics[width=0.6\linewidth]{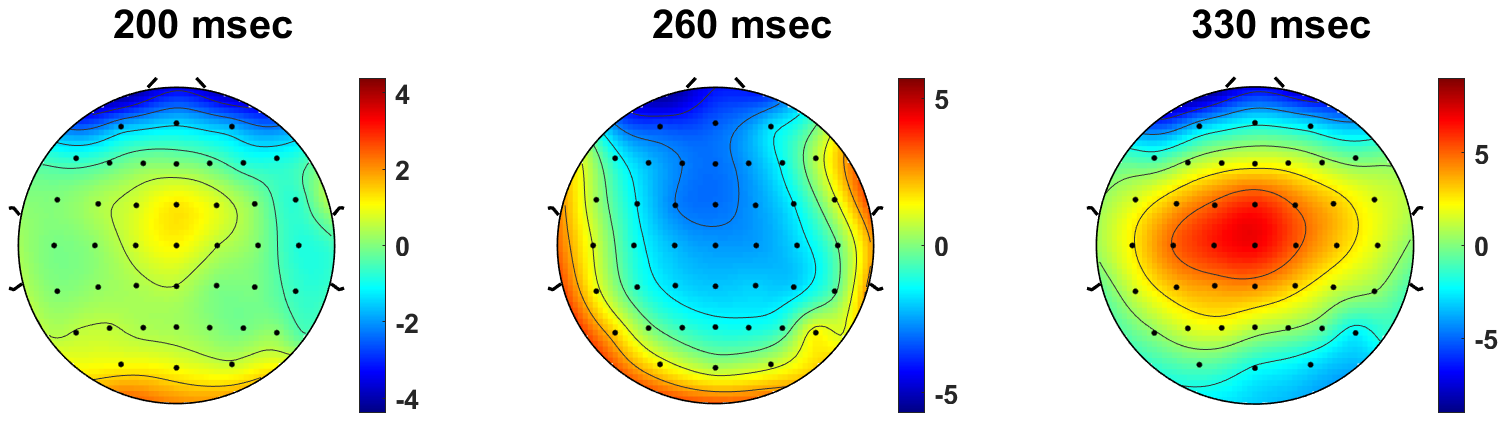}
  }
  \caption{The average response to an event at $t=0$: the mean ERP (confidence level = $95\%$) at the FCz channel (left panel) and the topographical view for brain activity across all channels (right panel). The top and middle panels show the patterns during the correct event and the erroneous event, respectively.  The bottom panel shows that ErrP obtained by subtracting the signal associated with the correct event from that of the erroneous event.  The topographical views are shown at the three characterizing peaks that occur at $200$ \text{msec}, $260$ \text{msec}, and $360$ \text{msec}, respectively.
%  for (a) correct events ERP and (b) erroneous events ERP, and (c) differential ErrPs $=$ erroneous events ERP $-$correct events ERP.
  }
 \label{ErrPs database}
 \end{figure}

 A two step standard EEG preprocessing for ErrPs which consists of Common Average Reference (CAR) filtering, and then $(1-10)$ Hz Band Pass (BP) filtering is performed on the data. Segments of EEG signals, from all the participants, corresponding to the correct and erroneous events were extracted into two separate groups. Then, the segments within each group were averaged to obtain a signal each for the correct and erroneous events.  Finally, the ErrPs are calculated by subtracting the signal for correct event from the signal for the erroneous event. 
Figure \ref{ErrPs database} shows the ERP signals and topographical views corresponding to the correct and erroneous events, and the calculated ErrPs. EEGLAB  was used for the processing of EEG signals and generating the topographical views for the related brain activities \cite{delorme2004eeglab}. Note that for consistency with the literature, we investigate the averaged ERP signal. However, the confidence regions around the averaged signal shown in Figure \ref{ErrPs database} reveal that despite the variability in these signals, {the key features} are also seen in the signal associated with a single event. Thus, the following analysis can be performed on a single subject real time data as well. 

% {\color{red} Need to resolve ERP/ErrP}

As reported in \cite{chavarriaga2010learning,chavarriaga2014errare}, the ErrPs are characterized by a sequence of three peaks/troughs after the onset of the event in the EEG signal measured at  channel FCz (see Figure~\ref{$10/20$ biosemi}). A sample ErrP is shown in Figure \ref{ErrPs database:c} where the first small positive peak is observed at 200 ms, followed by a negative trough at 260 ms, and finally a larger positive peak at 330 ms.

Two observations can be made by comparing the temporal behavior and the topographical views during the correct and erroneous events in Figures \ref{ErrPs database:a} and \ref{ErrPs database:b}, respectively. First, the ERP signal during the correct event tends to have broader peaks in comparison with the ERP signal during erroneous events, which {indicates that it} has slower dynamics, i.e., has smaller growth or decay rates. Second,
 by comparing the topographical views during the three characterizing peaks at FCz channel, it can be noticed that the ERP peaks during erroneous events have higher amplitudes than the peaks during correct events in both negative and positive directions. Note that the colormap bar values are different for the topographical views for the correct ERP, the erroneous ERP, and the ErrP in Figure~\ref{ErrPs database}. 

%Let assume that the dynamics, which underlie the brain activity during the two investigated events, can be represented in a $n$-dimensional space $\mathbb{U}$, where $n=64$ EEG channels. Then as a result of the local characterization of ErrPs around FCz channel, which can be observed from the topographical views in figure \ref{ErrPs database}, it is necessary to identify an $r$-dimensional subspace $\hat{\mathbb{U}} \subset \mathbb{U}$ with $r \ll n$ to represent these dynamics. The necessity to identify a lower-dimensional subspace is also supported by the ill-conditioned EEG datasets $(\text {condition number} \gg 1)$ resulted from CAR and BP filtering. So in this section, the incremental DMD algorithms presented in section \ref{Incremental DMD} are used to identify a subspace $\hat{\mathbb{U}}$ and an $r$-dimensional DMD operator to analyze the dynamics and predict future EEG states during the correct and erroneous events.     

As shown in Figure~\ref{ErrPs database}, the EEG patterns show coherent activities within the brain suggesting the possibility of a low-dimensional dynamics underlying this high dimensional data. Furthermore, these patterns change significantly with time and can possibly be better explained using a time-varying underlying model. To investigate these hypotheses, we apply the incremental DMD algorithms to the EEG recordings corresponding to  correct and erroneous events. The weighted incremental DMD algorithm is applied with discounting factors $\rho =\left\lbrace 0.1,0.2,0.4,0.8 \right\rbrace$, and the windowed incremental DMD algorithm with a time-window of width $w=512$ samples. All singular values  greater than $\sigma_{\text{thr}}=\{0.01,0.001\}$ are used for computing the reduced order DMD operator.  An initial window of $512$ samples before the event occurrence is used to initialized the DMD models for both algorithms.   

We compare the proposed algorithms with the online DMD approach proposed in~\cite{zhang2019online}. To this end, we use the same parameters (discounting factor and width of time-window) in the online DMD algorithms as the incremental DMD algorithms. Recall that the online DMD algorithms rely on {the Sherman-Morrison identity to compute the DMD operator} and do not have access to the (time-varying) singular values of the data matrix and hence, the reduction of the DMD operator to leading singular vectors does not apply to online DMD algorithms.

%
%A comparison between the two models in term of prediction accuracy indicates that the incremental DMD models have a higher accuracy, which reflects the effectiveness of identifying a lower-dimensional subspace for dynamics representation.

In order to compute {the time-varying DMD operator}, we first compute a standard DMD operator using  EEG data from a one-second window ($512$ samples) just before  the correct or erroneous event. 
%Note that this data corresponds to a $512 \times 64$ matrix, i.e., $512$ measurements at $64$ electrodes. 
Subsequently, the EEG measurements are streamed with a rate of $1$ $\text{sample}/\text{iteration}$ to update the initial DMD operator. At each iteration, both weighted and windowed versions of the incremental DMD and online DMD algorithms are used to predict $64$  future samples of the EEG signal at channel FCz. We compare the performance of these algorithms using normalized Root Mean Square (RMS) prediction error, denoted as ${e}_{\text{nrms}}(k)$ at each $t_k \in \{0, \cdots, 0.6\}$ using the following formula:
\begin{equation}
 {e}_{\text{nrms}}(k) = \frac{\sqrt{ \frac{1}{64}\sum_{i=k+1}^{k+64}(\hat{\mathbf{y}}_i- \mathbf{y}_i)^2}}{\overline{\mathbf{y}}_k-\underline{\mathbf{y}}_k},
\end{equation}
where $\hat{\mathbf{y}}_i$ and $\mathbf{y}_i$ are the estimated and recorded values, respectively, $\overline{\mathbf{y}}_k= \max\limits_{k+1 \le i \le k+64} \mathbf{y}_i$, and $\underline{\mathbf{y}}_k= \min\limits_{k+1 \le i \le k+64} \mathbf{y}_i$.

\begin{figure}[ht!]
\centering
\subfloat[Correct Event]
{
\includegraphics[width=0.5\linewidth]{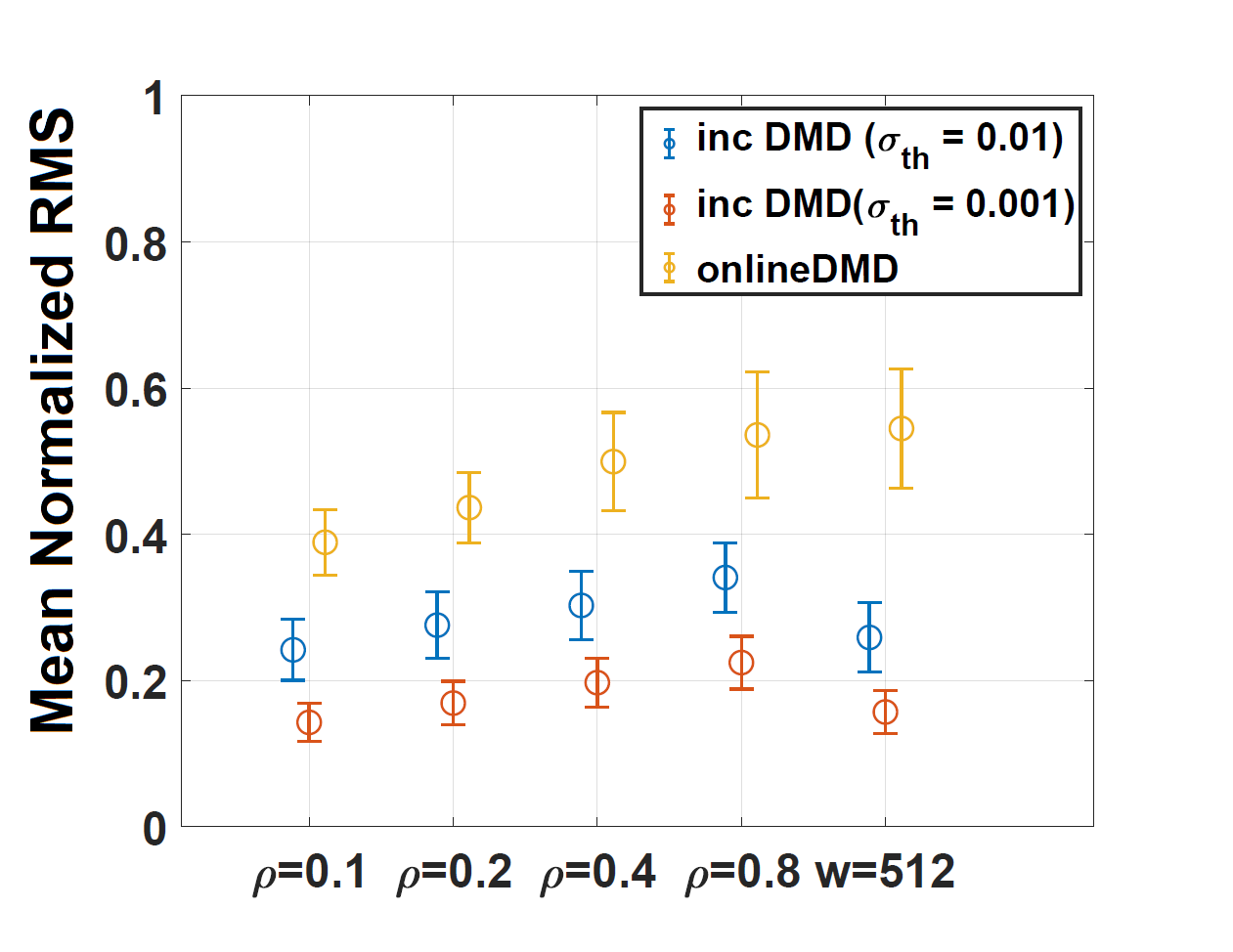}
} 
\subfloat[Erroneous Event]
{
\includegraphics[width=0.5\linewidth]{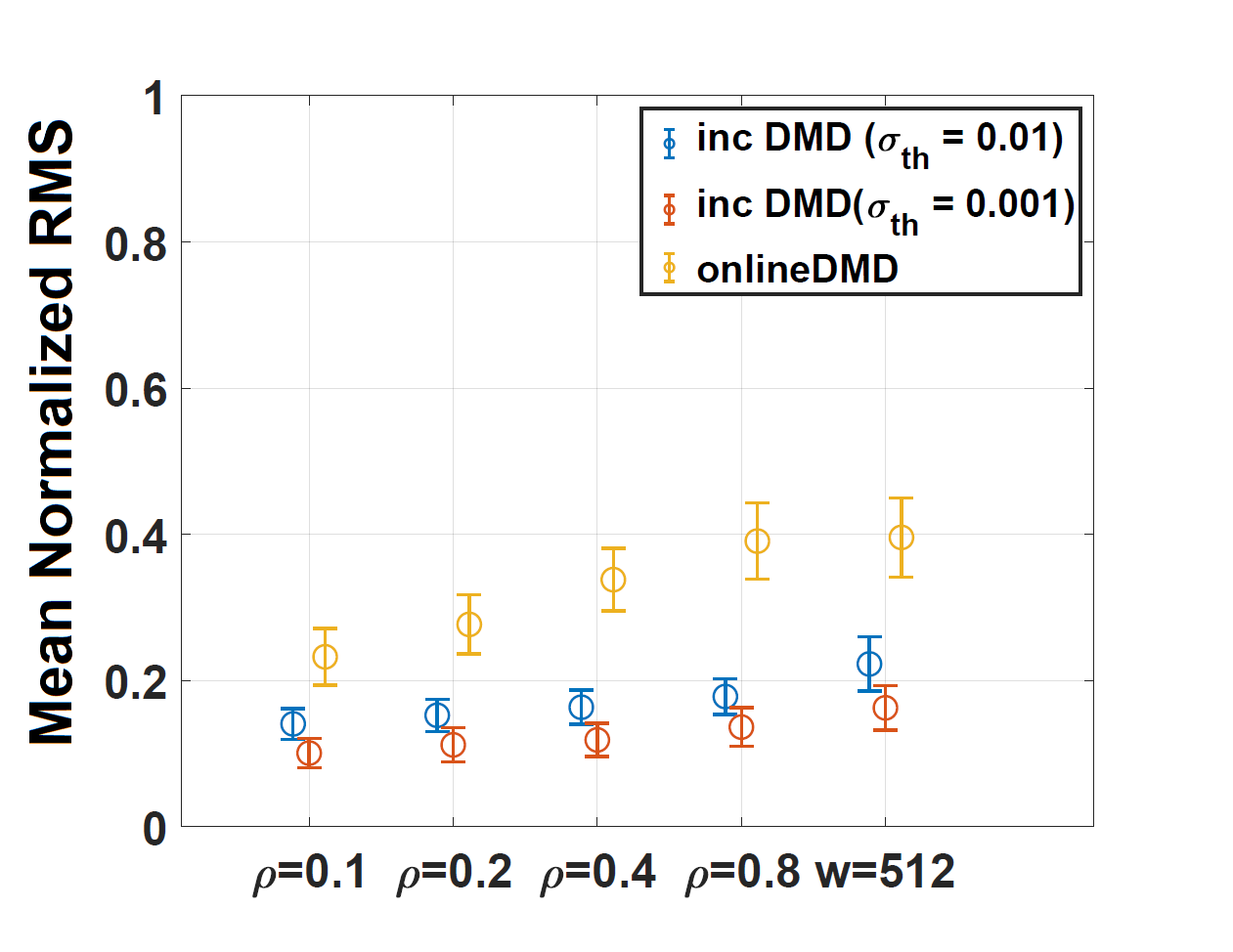}
} 
\caption{The mean of the normalized RMS prediction error computed over all iterations as well as the associated 95\% confidence sets for (a) correct events and (b) erroneous events.}
\label{Pred_SEM}
\end{figure}

Figure \ref{Pred_SEM} shows the mean of the normalized RMS prediction error computed over all iterations as well as the associated 95\% confidence sets for weighted and windowed incremental DMD algorithms and different choices of parameters. 
% The mean  (confidence level $= \% 95$) of the calculated normalized RMS error values for all iterations per discounting factor per threshold value for each event is presented in Figure \ref{Pred_SEM}. 
The performance of the incremental DMD does not appear to vary much with the {weighting} factor and the two choices of the threshold on singular values. The performance of the windowed incremental DMD algorithm is similar to the weighted incremental DMD. However, incremental DMD algorithms seem to outperform online DMD algorithms. This suggests that having a small threshold $\subscr{\sigma}{thr}$ is more beneficial than having no threshold as in the case of online DMD algorithms. Figure \ref{EEG pred rms} in Appendix~\ref{EEG_Pred_nrmse} presents the evolution of the normalized RMS prediction error for each event.    

\begin{figure}[ht!]
\centering
\subfloat[Correct event]
{
\includegraphics[width=0.33\linewidth]{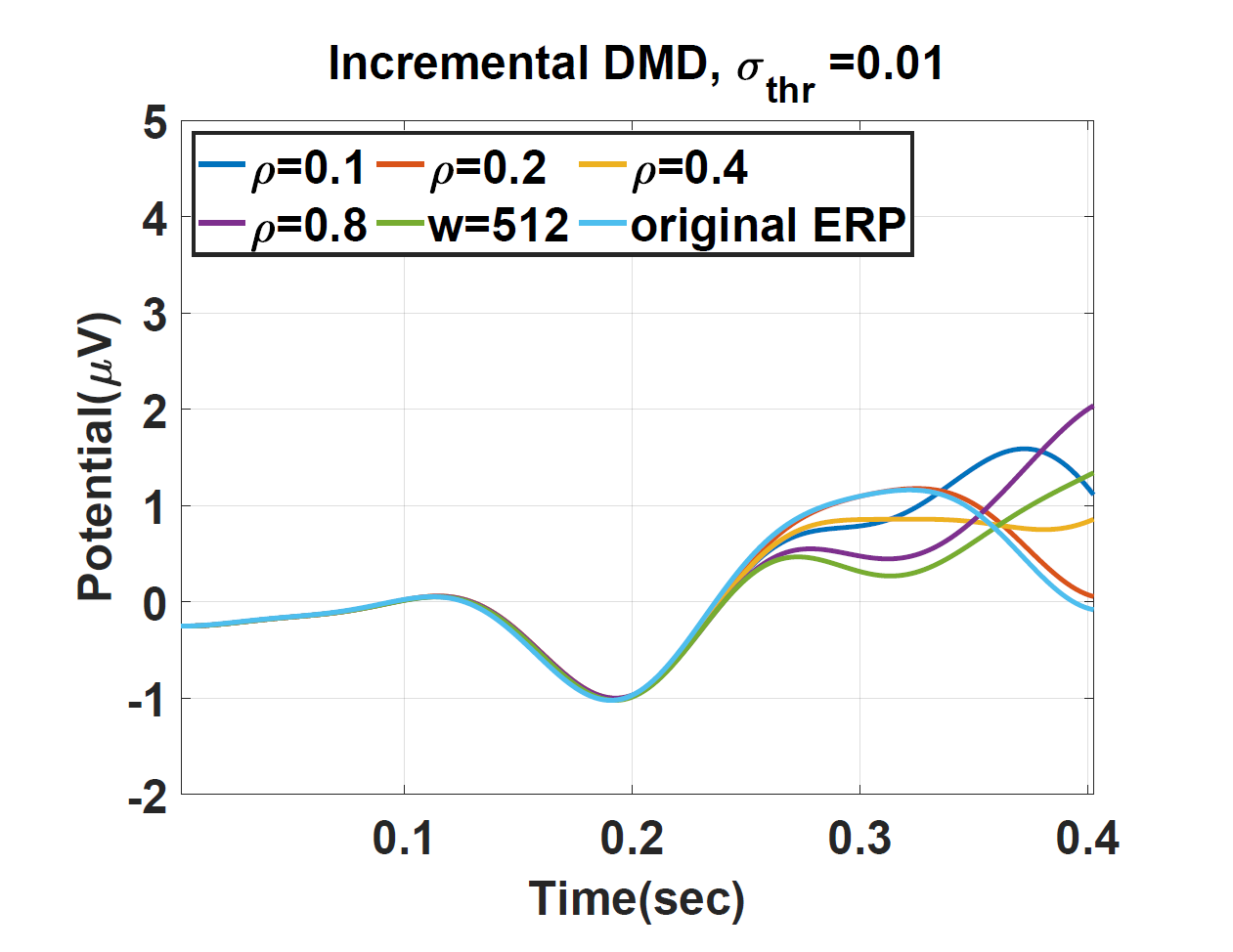}
\includegraphics[width=0.33\linewidth]{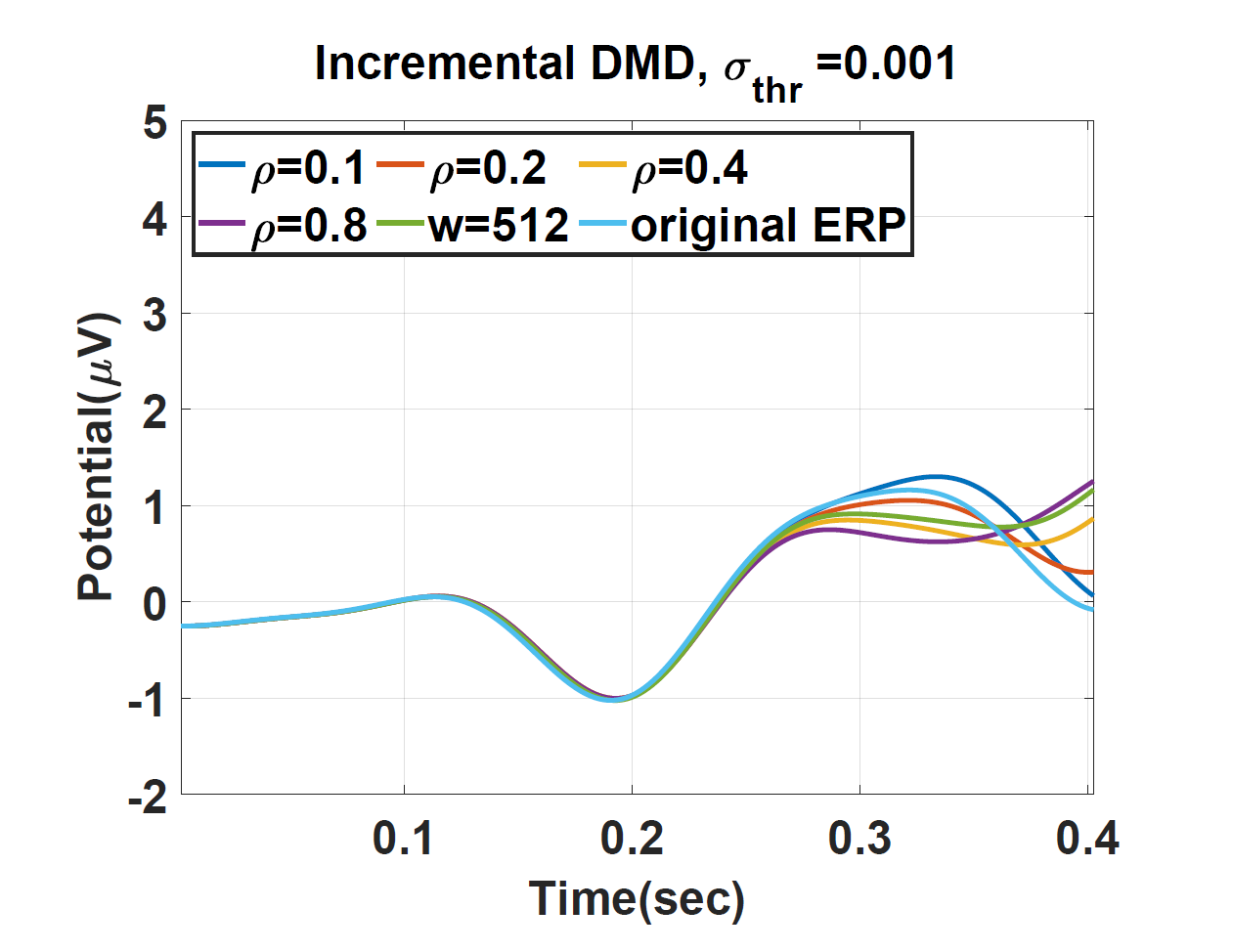}
\includegraphics[width=0.33\linewidth]{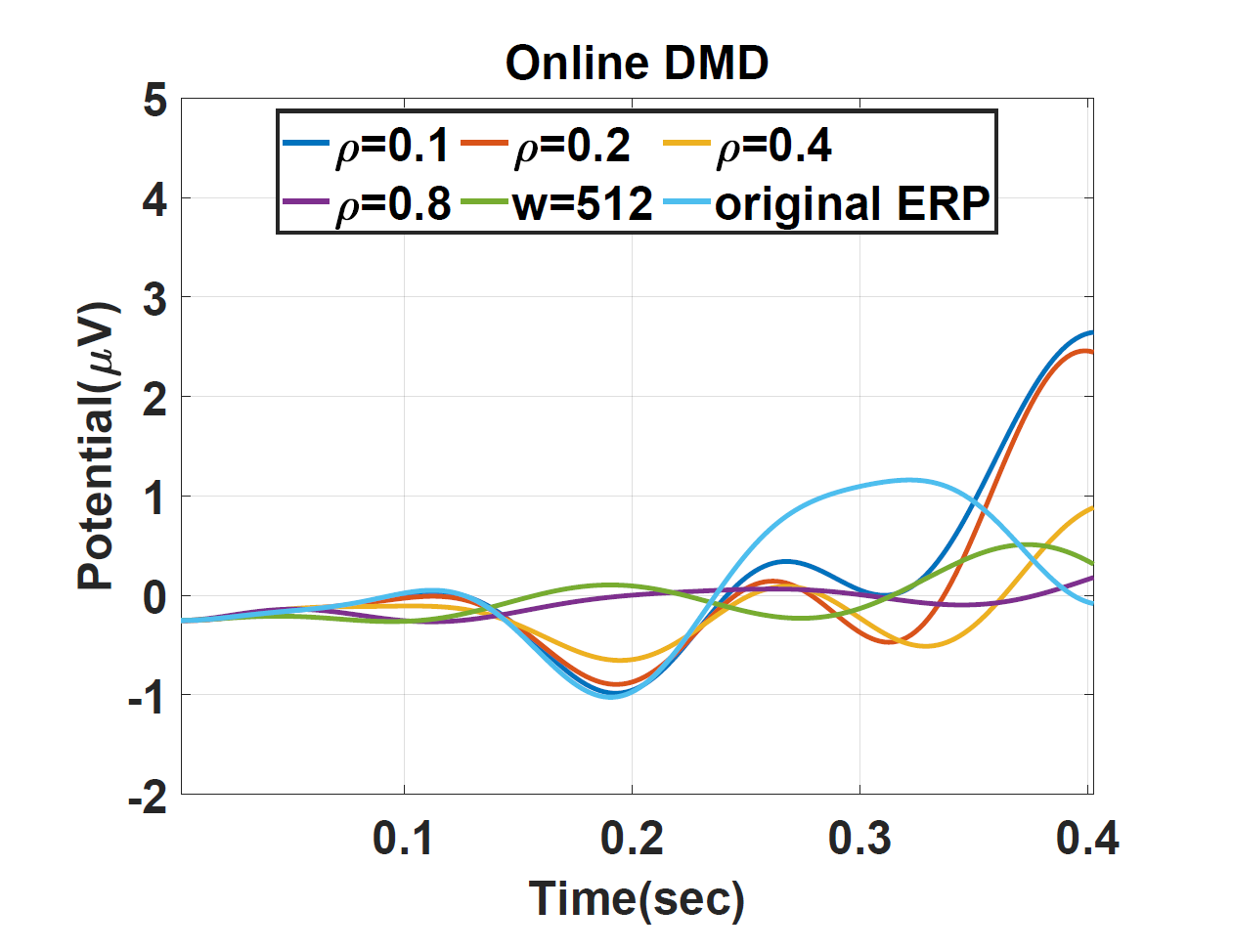}
} \\
\subfloat[Erroneous event]
{
\includegraphics[width=0.33\linewidth]{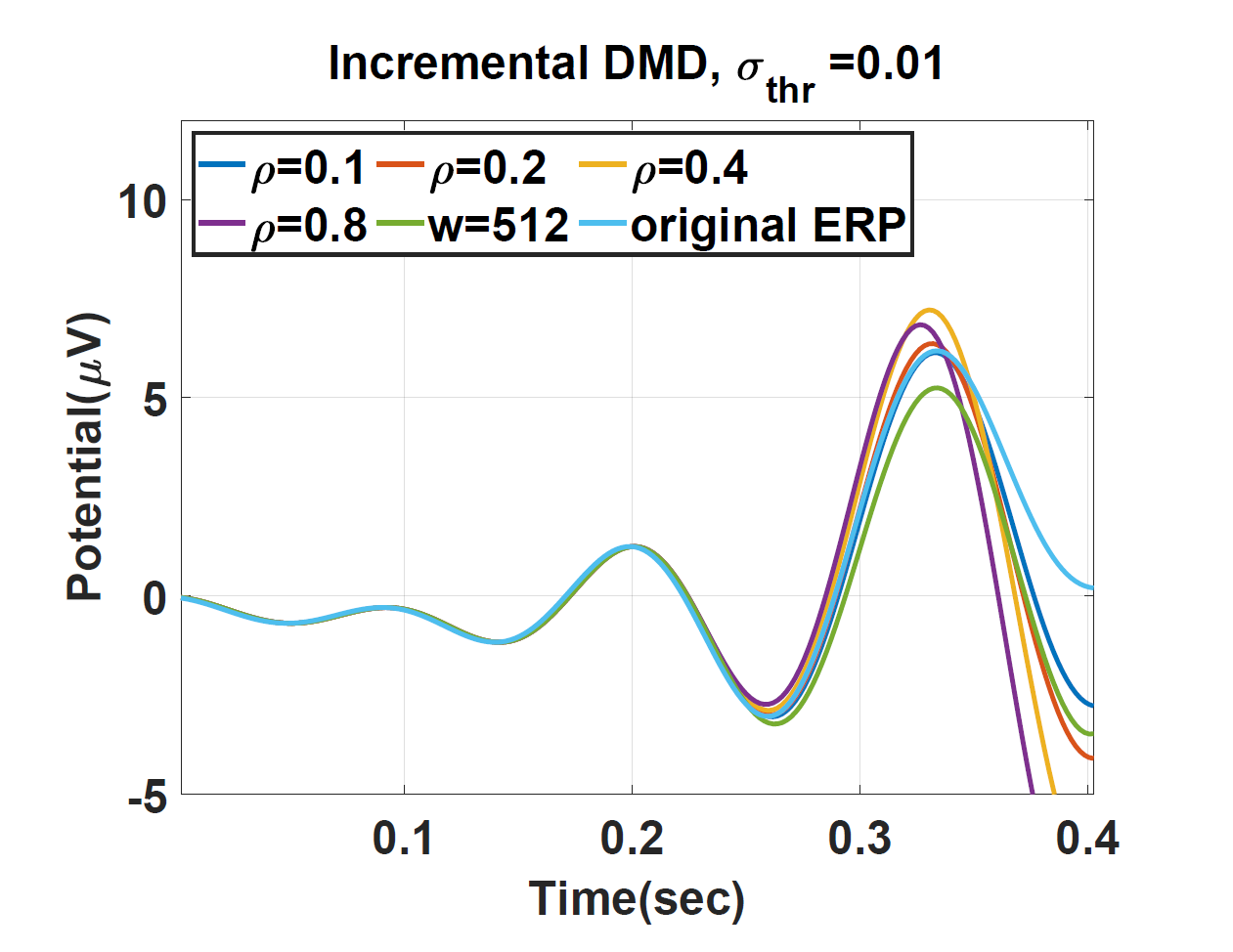}
\includegraphics[width=0.33\linewidth]{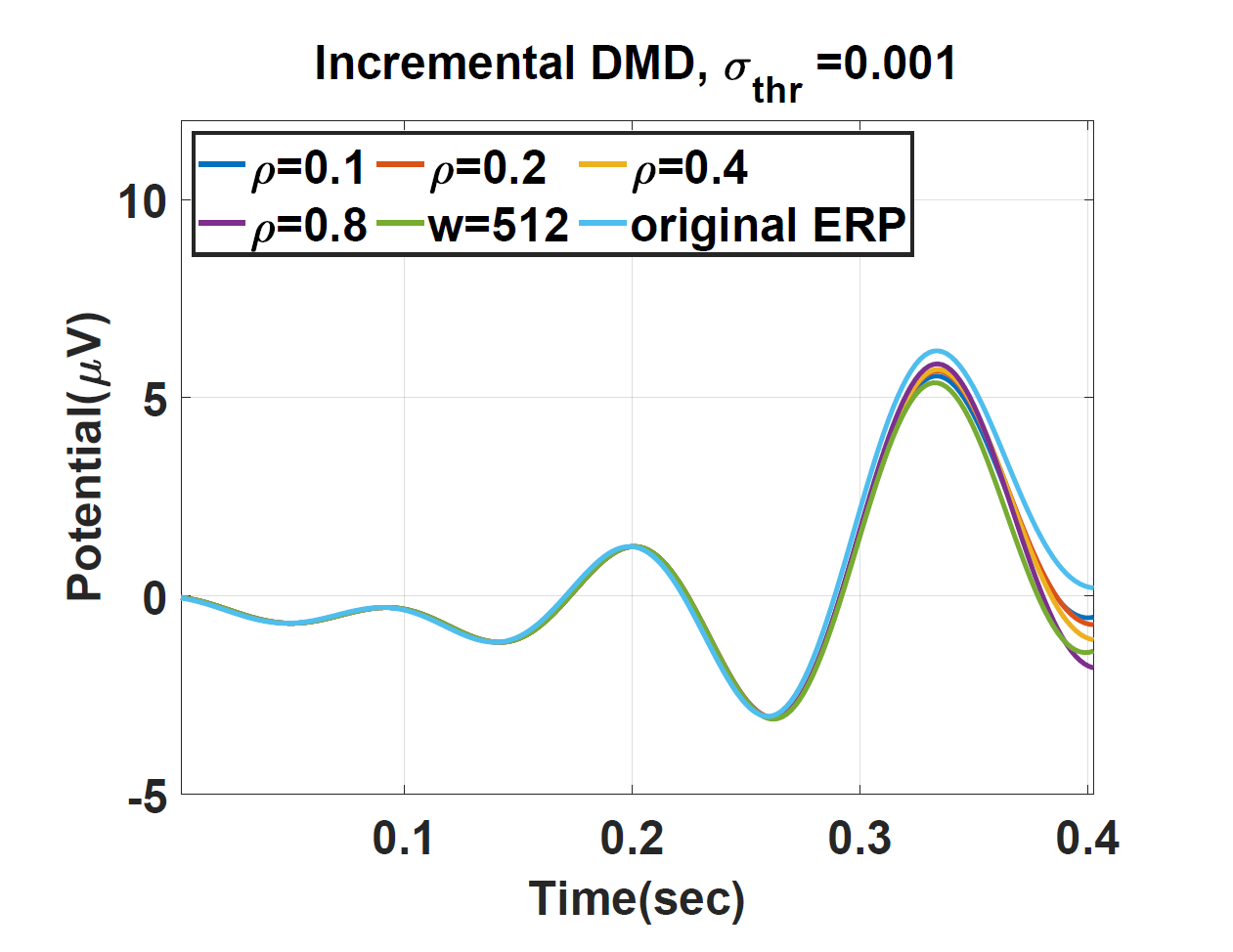}
\includegraphics[width=0.33\linewidth]{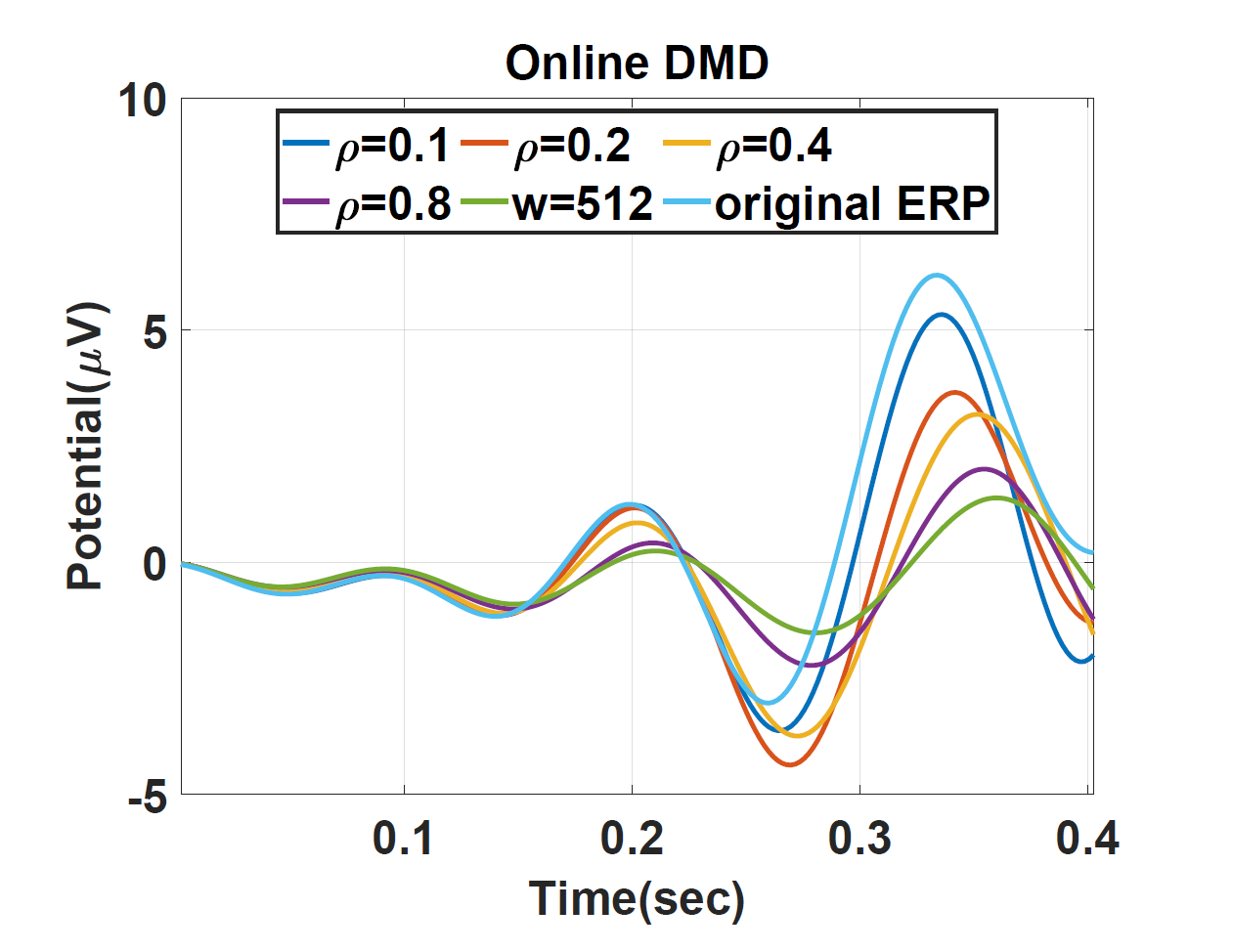}
}
\caption{Predicted ERP signal at channel FCz using incremental DMD with $\sigma_{\text{thr}}=0.01$ (left panel), incremental DMD with $\sigma_{\text{thr}}=0.001$ (middle panel), and online DMD (right panel) for (a) correct event, and (b) erroneous event.}
\label{ERP pred}
\end{figure}

\begin{figure}[ht!]
\centering
\subfloat[Incremental DMD]
{
\includegraphics[width=0.45\linewidth]{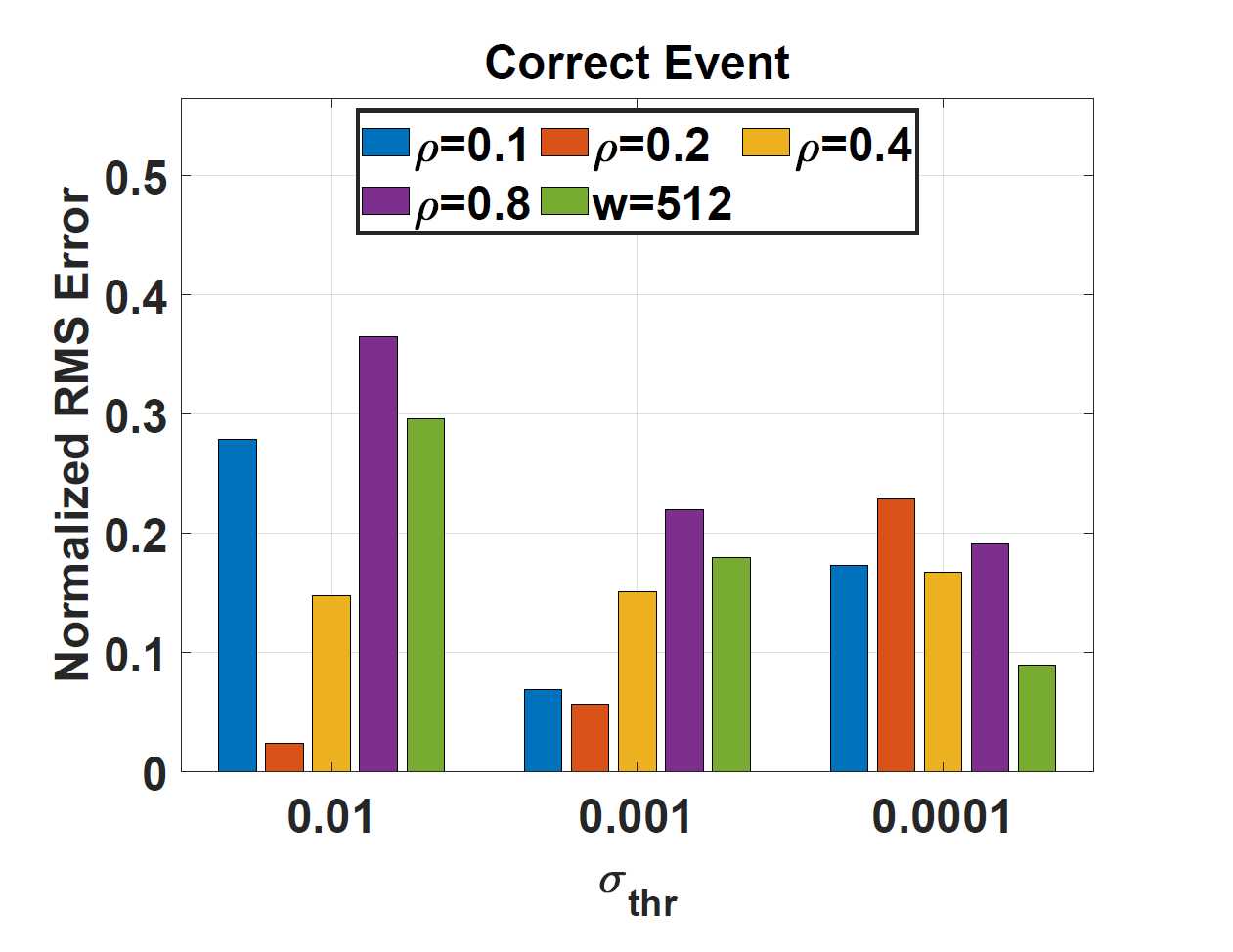}
\includegraphics[width=0.45\linewidth]{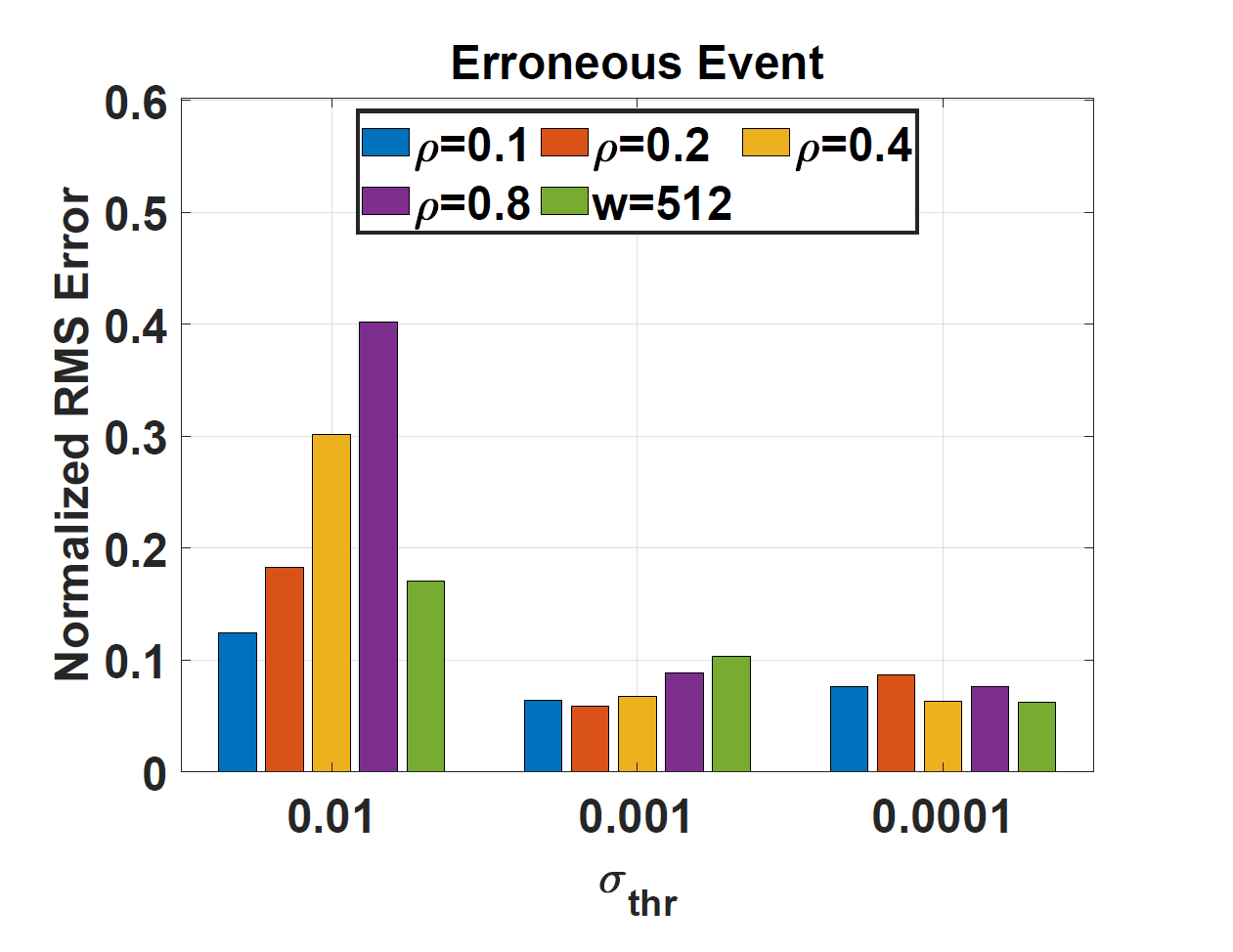}
} \\
\subfloat[Online DMD]
{
\includegraphics[width=0.45\linewidth]{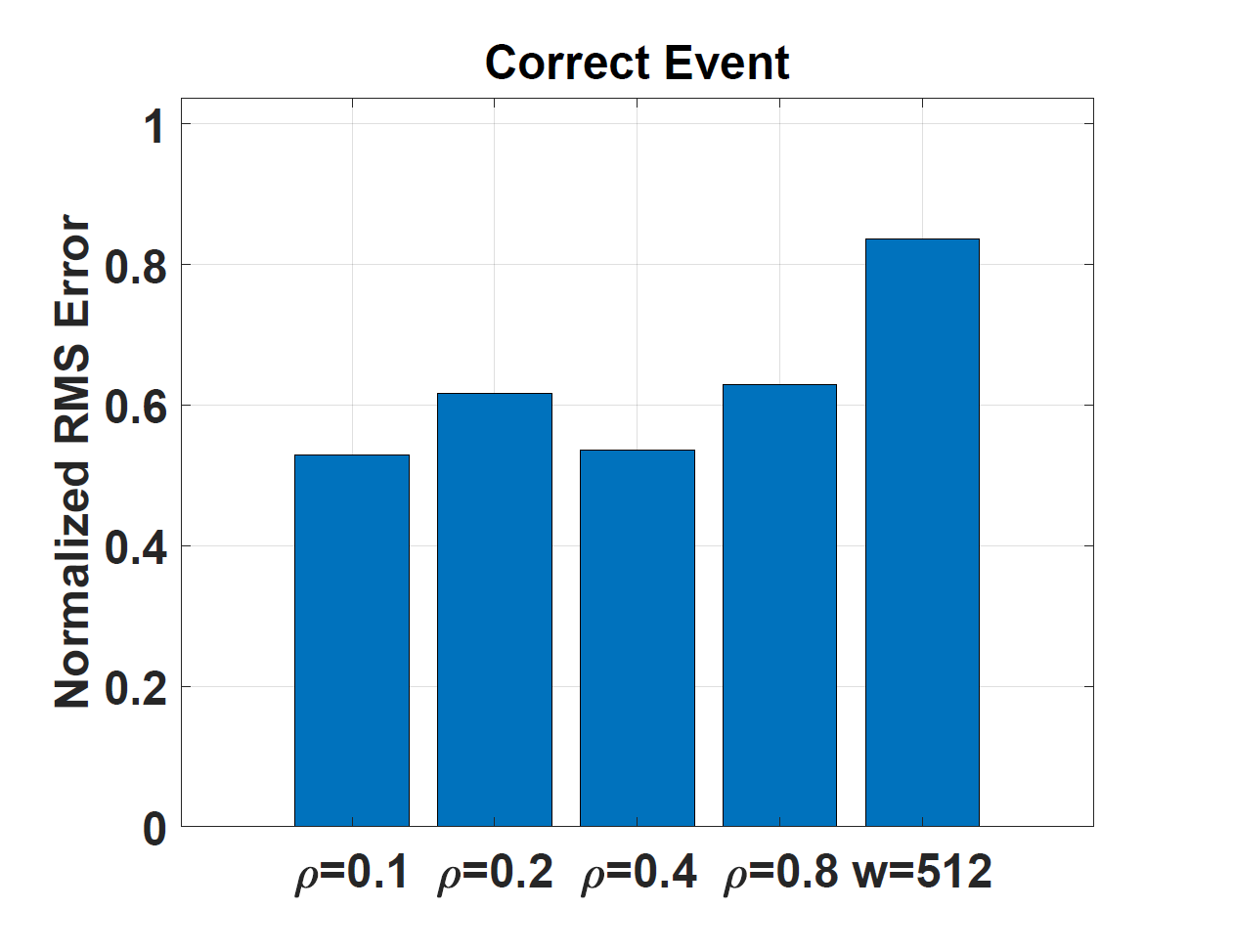}
\includegraphics[width=0.45\linewidth]{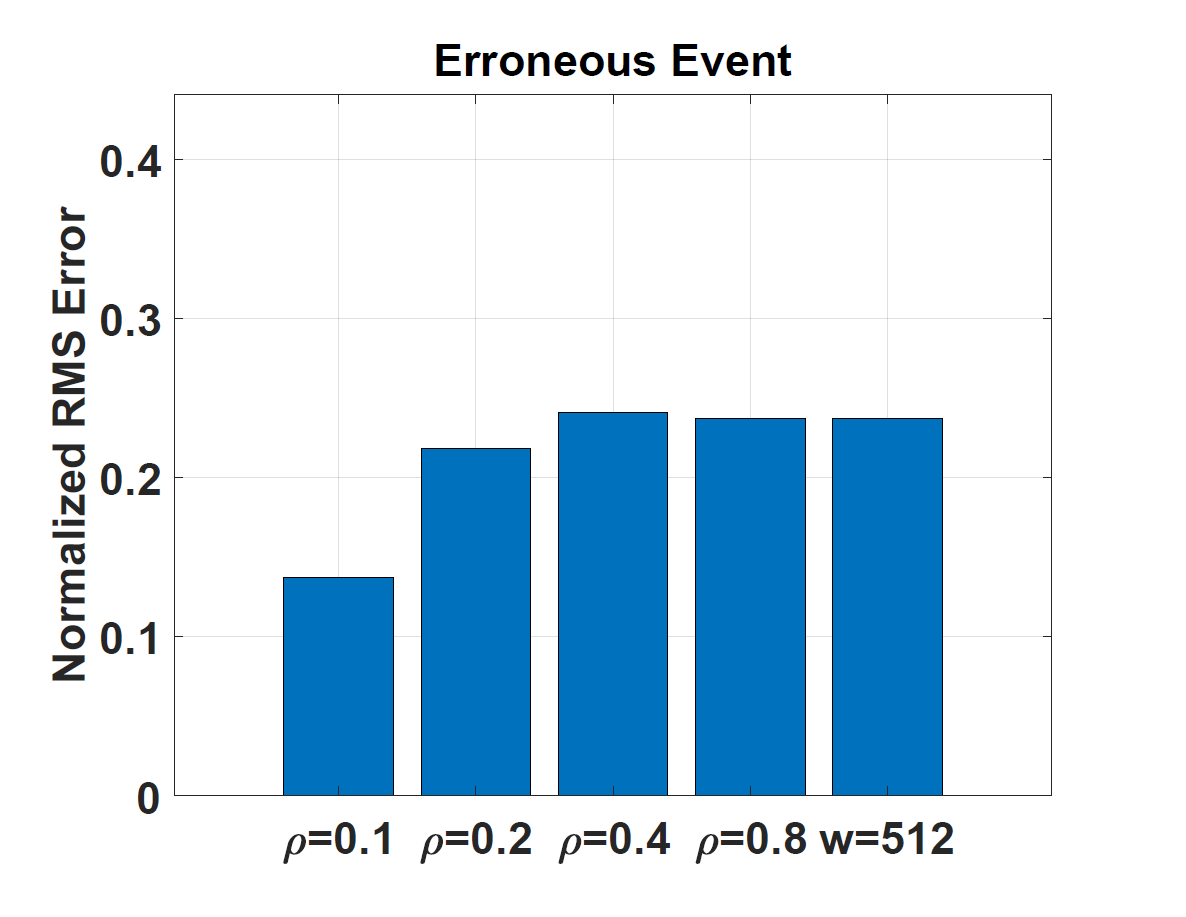}
} 
\caption{Normalized RMS error for the predicted ERP signal at channel FCz for correct events (left panel) and erroneous events (right panel), using (a) incremental DMD, and (b) online DMD.}
\label{ERP pred rms}
\end{figure}

In order to gain further insight into the influence of different thresholds $\sigma_{\text{thr}}$ and {weighting} factors $\rho$, we investigate how accurately the incremental DMD predicts the signature ERPs at channel FCz. To this end, we updated the initial DMD operator using incremental DMD algorithms until the first peak at 200 msec. We then used the updated lower-dimensional DMD operator to predict the future ERP until the third peak at 330 msec. 
%  prior to the onset of the first peak at $200$ msec, these models are used to predict the ERP states until the end of the third peak at $330$ msec during the two events. 
 The predicted signal and the associated normalized RMS are shown in Figures \ref{ERP pred} and \ref{ERP pred rms}, respectively. For the erroneous event, the predicted signal from incremental DMD algorithms capture the trend of the original signal, while for the correct event, most incremental DMD algorithms fail to capture the final dip in the original signal. The only exception is the windowed incremental DMD with $\subscr{\sigma}{thr}=0.001$. For $\subscr{\sigma}{thr}=0.01$, the number of DMD modes used by the incremental DMD algorithms range from {26 to 31} for correct events and {28 to 32} for erroneous events. Similarly, for $\subscr{\sigma}{thr}=0.001$, the number of DMD modes used by the incremental DMD algorithms range from {26 to 31} for correct events and {28 to 33} for erroneous events. 
 In comparison, the predicted signal from the online DMD algorithms performs poorly. The key reason for this poor performance, is that the data matrix obtained after prepossessing is not full rank even if all the measurements are included. Thus, the Sherman-Morrison update cannot be applied in the standard form. Recall that the online DMD algorithm starts by initializing the DMD operator $\mathbf{A}_k$ and the inverse of covariance matrix  $\mathbf{P}_k=\left(\mathbf{X}_k \mathbf{X}_k^{\text{T}} \right)^{-1}$. For initialization of online DMD algorithms, we adopted the suggestion in~\cite{zhang2019online}: in the event that $\mathbf{X}_k \mathbf{X}_k^{\text{T}}$ is not full rank,  $\mathbf{A}_k$ can be initialized by an $n \times n$ zero matrix, and $\mathbf{P}_k$ can be initialized by $\mathbf{P}_{\text{init}}=\alpha \text{I}$ where $\alpha$ is a large positive scalar.


 During the correct events, there appears to be an interplay between the threshold $\sigma_{\text{thr}}$ and {weighting} factor $\rho$. As shown in Figure~\ref{ERP pred rms} for correct events, the normalized RMS decreases with decreasing the {weighting} factor until $\rho=0.2$ and increases with further decrease in $\rho$. For $\rho=0.2$, the higher dimension of the reduced system ($\subscr{\sigma}{thr}=0.001$) leads to higher prediction error compared with the lower dimension reduced system  ($\subscr{\sigma}{thr}=0.01$). This suggests {that for a weighting factor} that leads to the smallest prediction error, including too many modes is not beneficial for prediction performance, since it leads to over-fitting.  
%  {\color{red}
 %The same observation can not be seen for incremental DMD models with smaller initial window sizes of $256$ and $128$ samples, as it can be seen in Figure \ref{ERP_RMS_256_windows} and Figure \ref {ERP_RMS_128_windows}, respectively in Appendix \ref{app_wind DMD with different sizes}. 
 A similar interplay is not seen for the erroneous event in Figure~\ref{ERP pred rms}. However, if we choose a smaller initial window to initialize the DMD model, we observe similar effects as shown in Figure~\ref {ERP_RMS_128_windows} in Appendix~\ref{app_wind DMD with different sizes}. 
%  In contrast, for erroneous events, the aforementioned observation is detected when using smaller initial window size of $128$ samples which can be seen in Figure \ref {ERP_RMS_128_windows}}.
 Similar effects are observed for the windowed DMD as shown in Figure~\ref{ERP_RMS_windows} in Appendix \ref{app_wind DMD with different sizes}.

\begin{figure}[ht!]
\centering
\subfloat[correct events] 
{
\label{ERP pred noisy:a}
\includegraphics[width=0.45\linewidth]{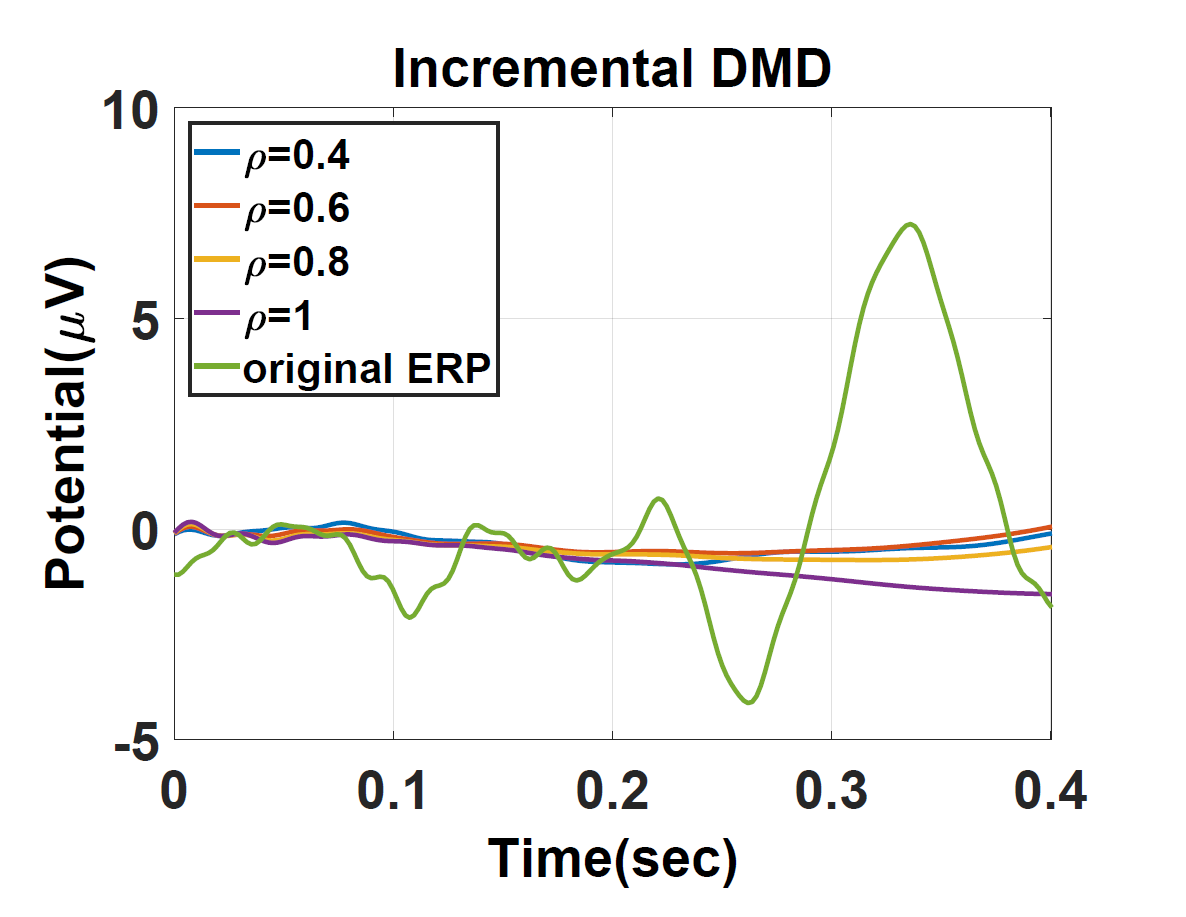}
\includegraphics[width=0.45\linewidth]{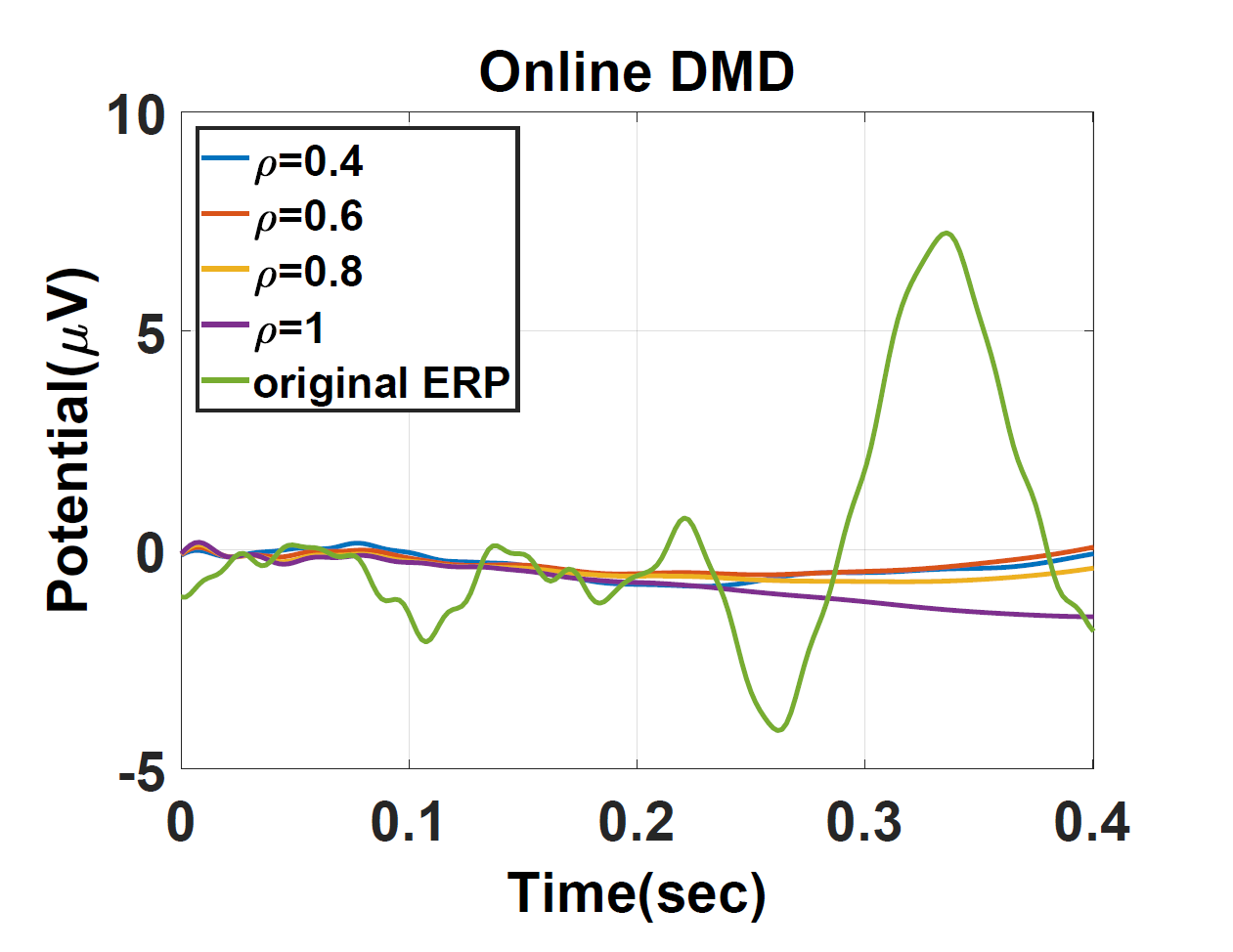}
} \\
\subfloat[{erroneous events}] 
{
\label{ERP pred noisy:b}
\includegraphics[width=0.45\linewidth]{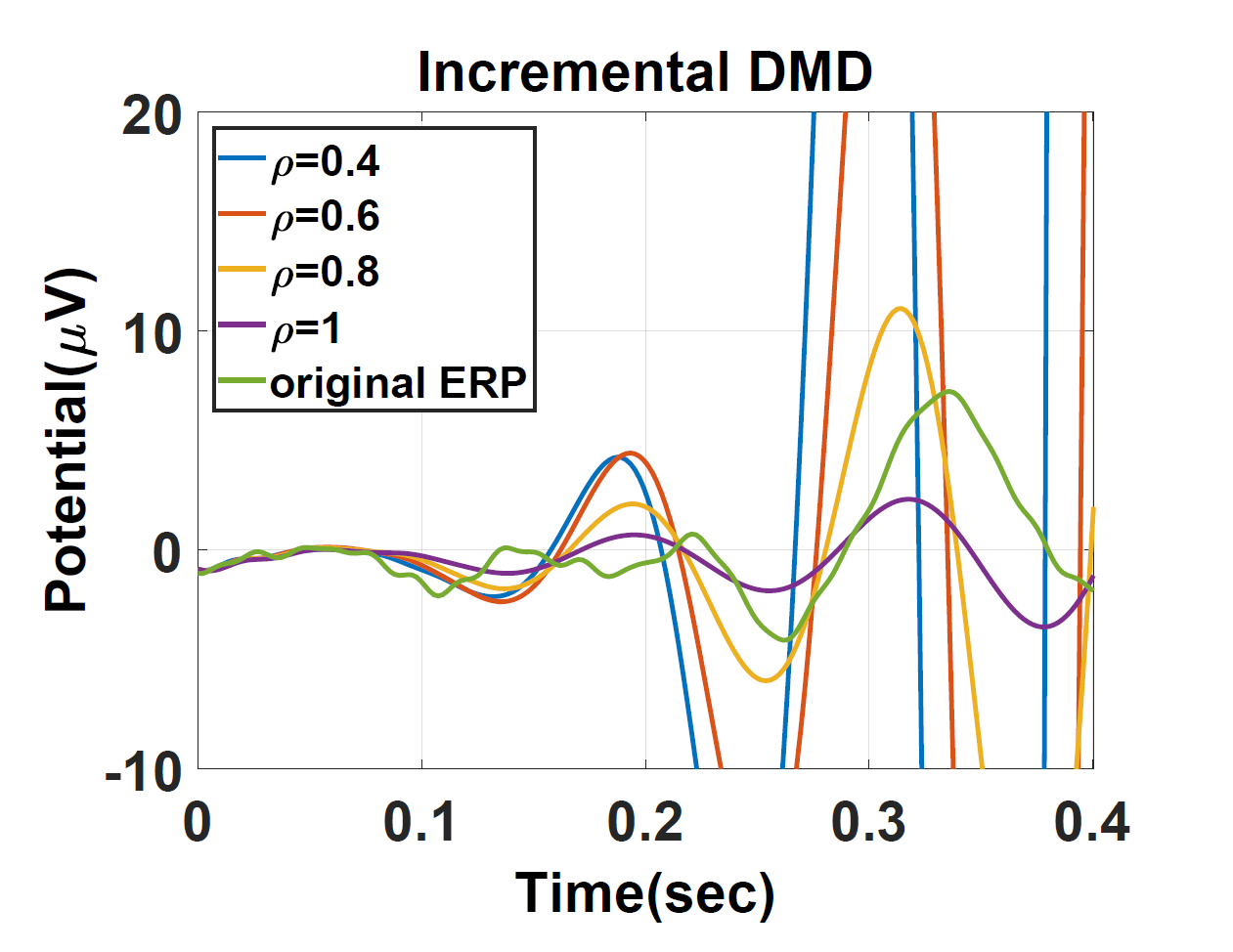}
\includegraphics[width=0.45\linewidth]{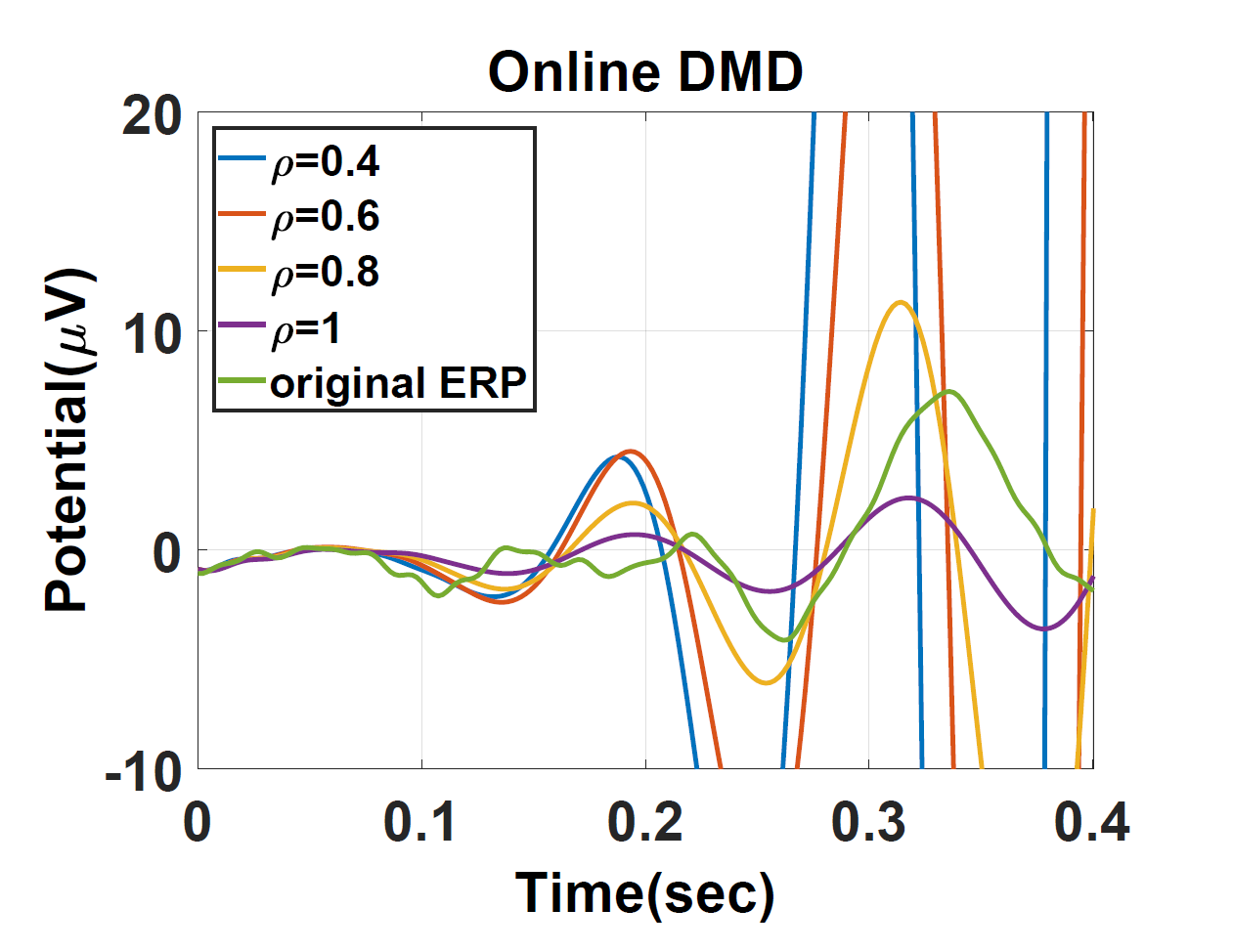}
} 
\caption{The predicted ERP signal at channel FCz based on  well conditioned EEG datasets using incremental DMD model (left panel) and online DMD model (right panel) during (a) correct events and (b) erroneous events.}
\label{ERP pred noisy}
\end{figure}

To further compare the online DMD with the incremental DMD, we took the raw EEG data, i.e., data without any preprocessing  (CAR and BP filtering). In this case, the data matrix is well-conditioned and the Sherman-Morrison identity is well defined. The performance of online DMD and the incremental DMD in terms of predicting future signal is shown in Figure \ref{ERP pred noisy}. In this case, both incremental and online DMD algorithms have similar poor performance. This suggests that the difference in the performance of these algorithms is primarily due to ill-conditioned dataset. This also highlights the utility of applying appropriate band-pass filtering to the raw EEG data. Without such filtering, the key activities seem to be lost in the background noise resulting in a poor prediction performance.

% The incremental DMD model emulates the online DMD model when the EEG dataset is well conditioned. For the sake of illustration, we used a special EEG dataset for which the preprocessing part (CAR and BP filtering) is skipped. 

% This dataset are used to build incremental and online DMD models to predict the same signals in figure \ref{ERP pred}. The identical predicted signals for the two models with different $\rho$ values and $64$ DMD modes are shown in figures \ref{ERP pred noisy:a} and figure \ref{ERP pred noisy:b} for correct and erroneous events, respectively.  

\begin{figure}[ht!]
\centering
\subfloat[DMD modes $\sigma_{\text{thr}}=0.01$.]
{
\includegraphics[width=0.9\linewidth]{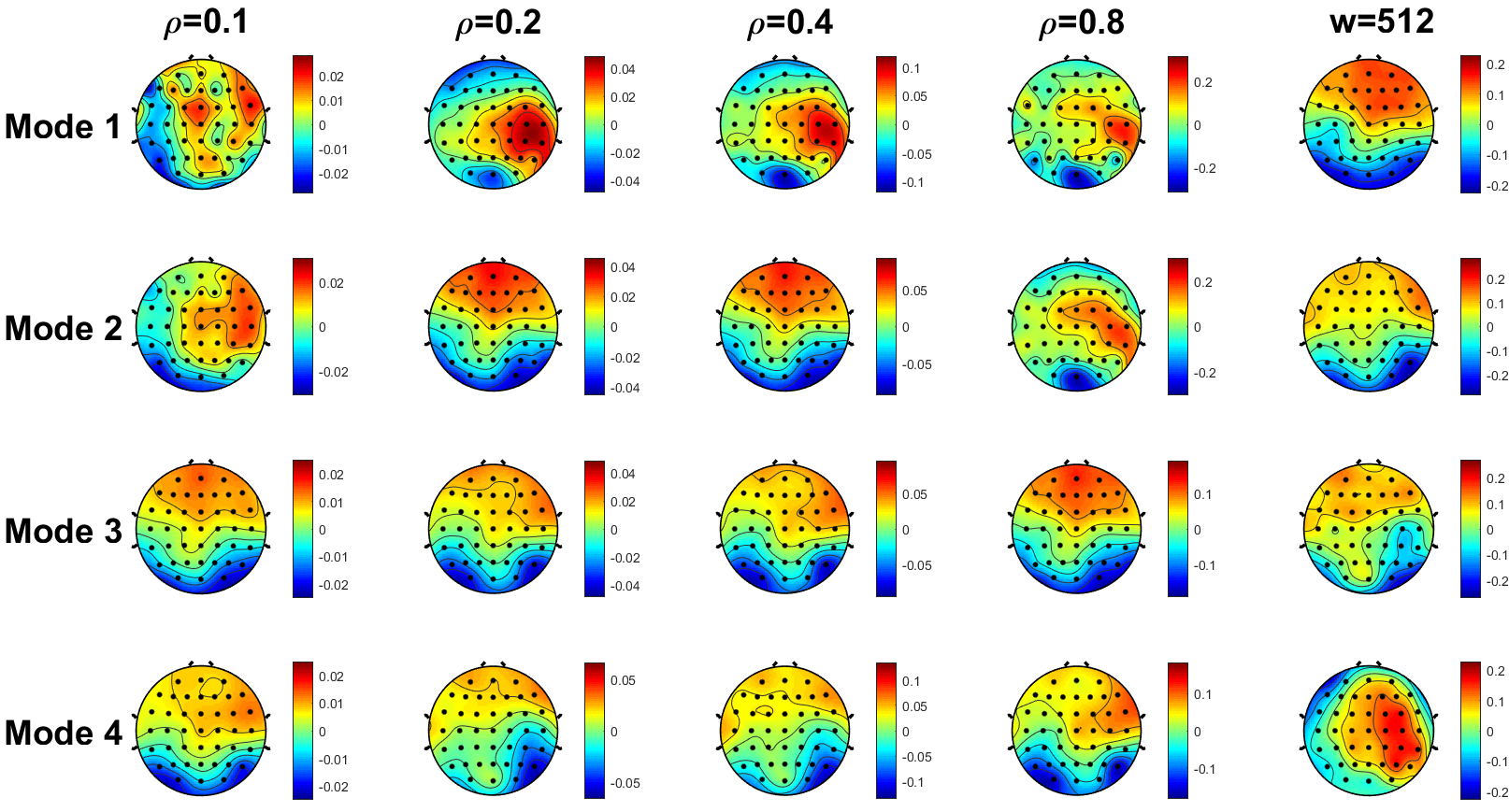}
} 
\\
\subfloat[DMD modes $\sigma_{\text{thr}}=0.001$.]
{
\includegraphics[width=0.9\linewidth]{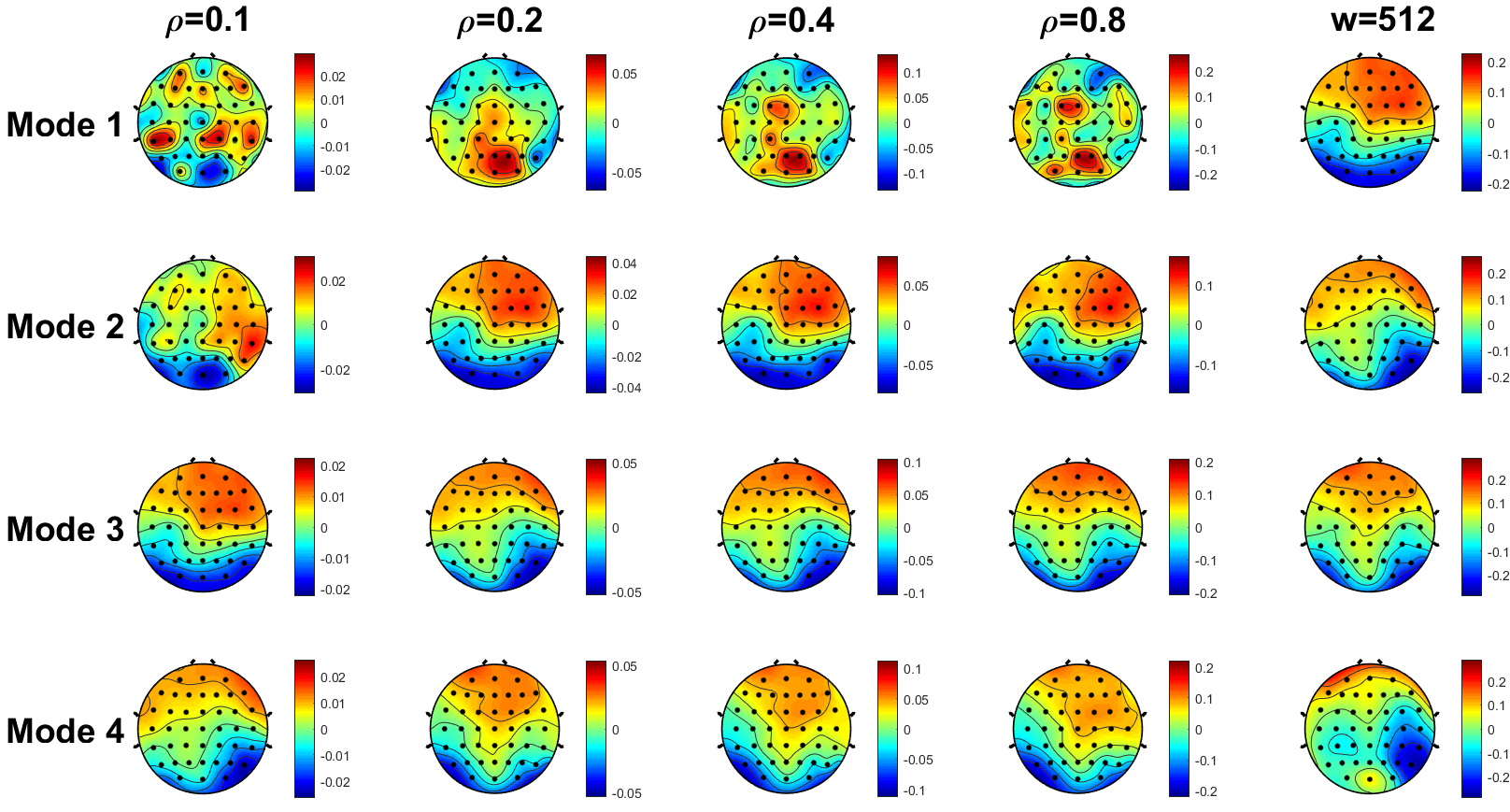}
} 
\caption{Topographical views for the real part part of the $4$ dominant DMD modes during correct events using threshold values of (a) $\sigma_{\text{thr}}=0.01$, and (b)     $\sigma_{\text{thr}}=0.001$.}
\label{ERP pred topodmd correct}
\end{figure}

\begin{figure}[ht!]
\centering
\subfloat[DMD modes $\sigma_{\text{thr}}=0.01$.]
{
\includegraphics[width=0.9\linewidth]{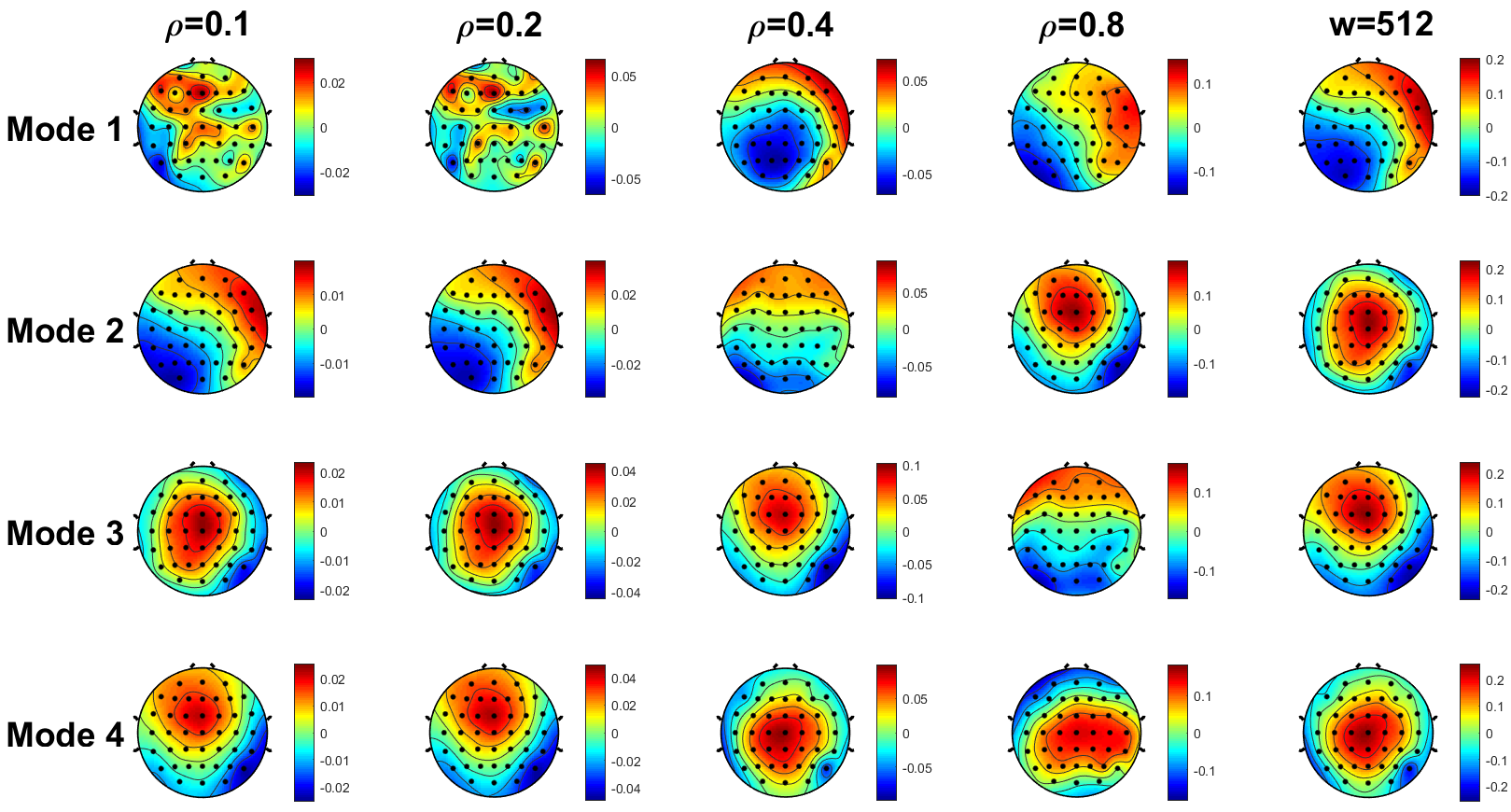}
} 
\\
\subfloat[DMD modes $\sigma_{\text{thr}}=0.001$.]
{
\includegraphics[width=0.9\linewidth]{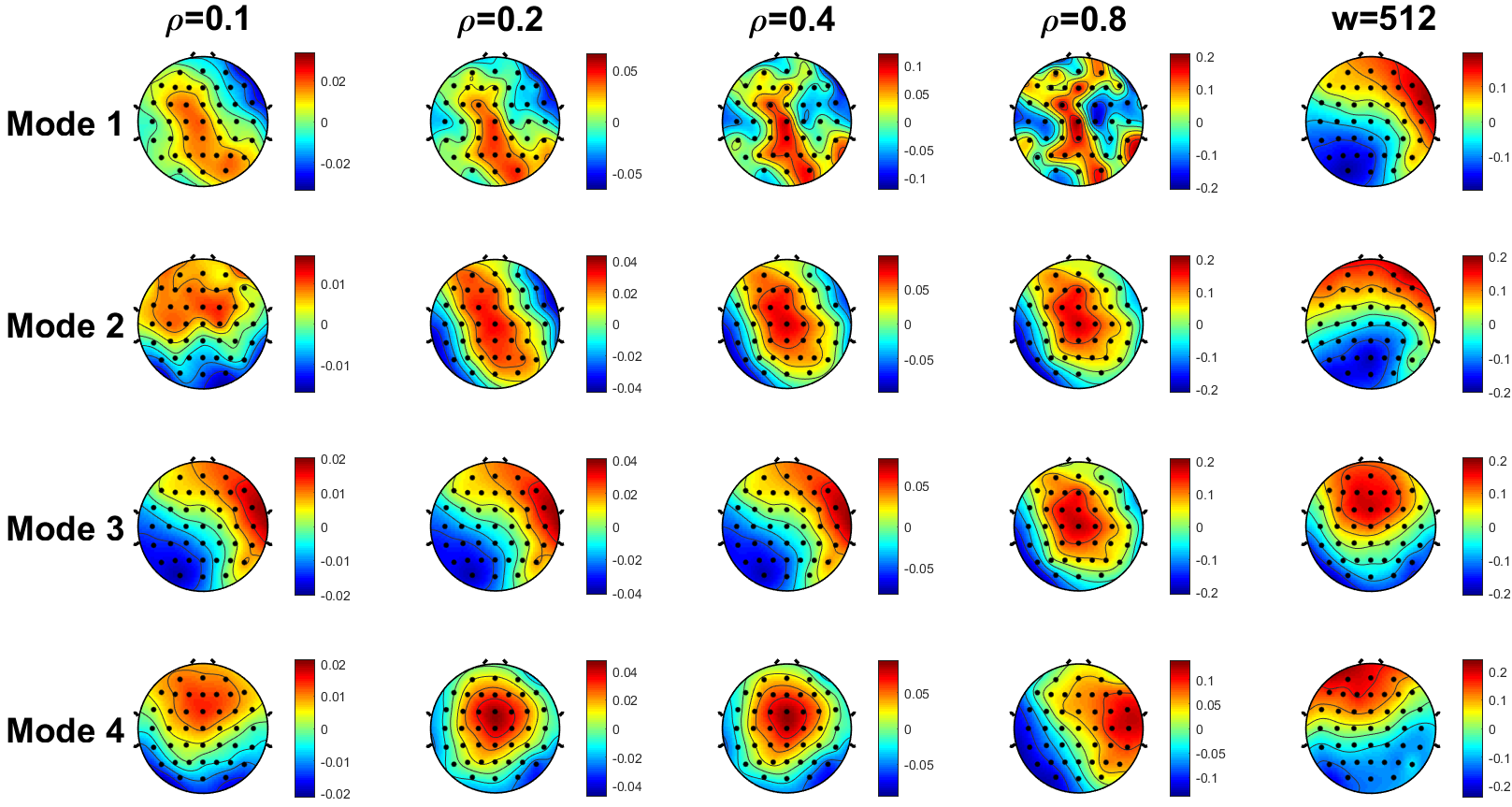}
} 
\caption{Topographical views for the real part of the $4$ dominant DMD modes during {erroneous} events using threshold values of (a) $\sigma_{\text{thr}}=0.01$, and (b) $\sigma_{\text{thr}}=0.001$.}
\label{ERP pred topodmd error}
\end{figure}

Topographical views for the real part of the four dominant DMD modes at 200 msec obtained from the incremental DMD algorithms for the correct and erroneous events are shown in Figure \ref{ERP pred topodmd correct} and Figure \ref{ERP pred topodmd error}, respectively. A comparison between the DMD modes during the erroneous and correct events shows that there is a stronger activity in the frontal lobe during the erroneous events especially around the FCz channel. This is consistent with the topographical views from EEG recordings shown in Figures \ref{ErrPs database:a} and \ref{ErrPs database:b} that show stronger activity in the frontal lobe during an erroneous event. This illustrates that a few principal DMD modes are able to capture the dominant activity in the brain during these experiments.
%the peaks during the erroneous events have larger amplitudes in both negative and positive directions. 

\begin{figure}[ht!]
\centering
\subfloat[Correct events]
{
\includegraphics[width=0.45\linewidth]{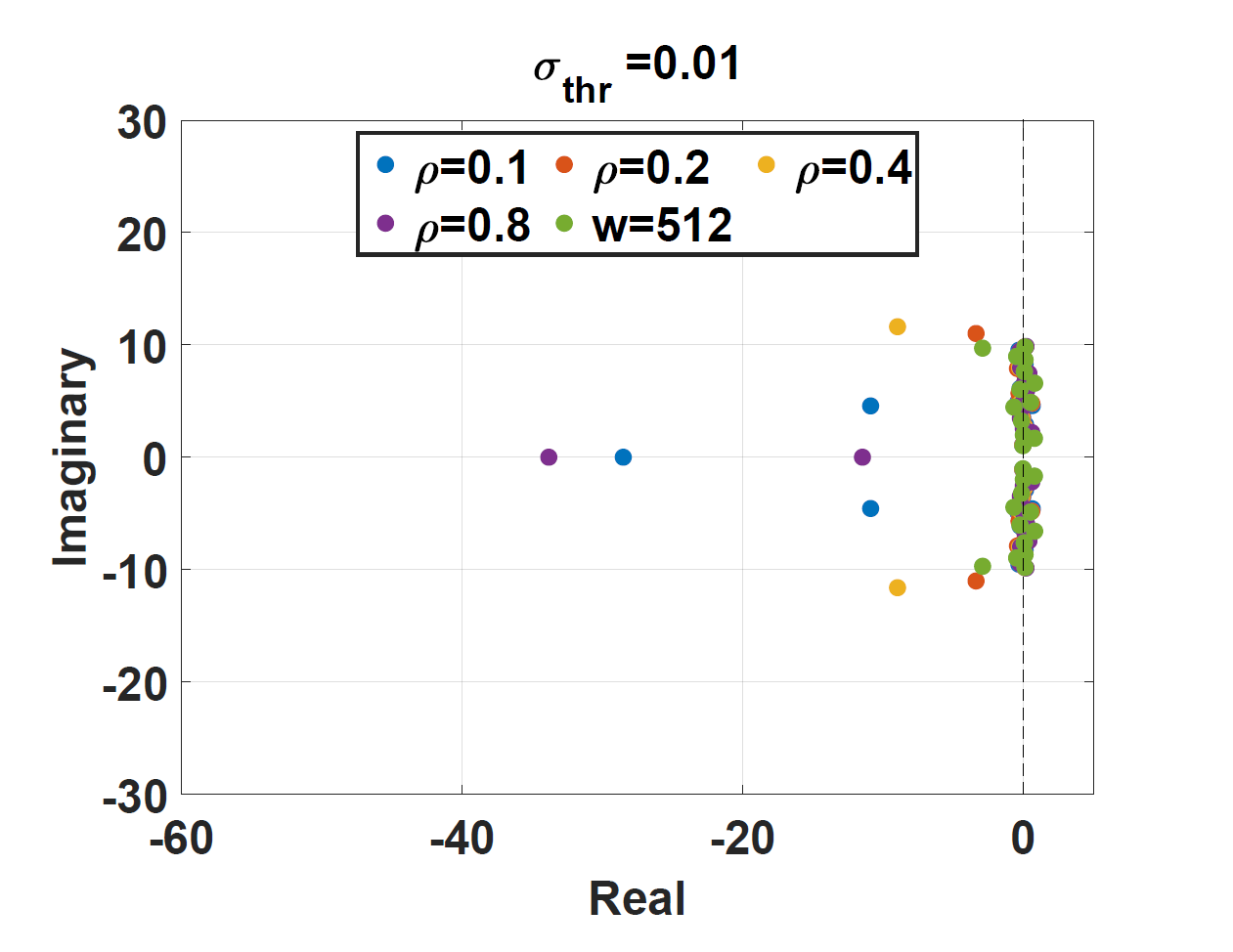}
\includegraphics[width=0.45\linewidth]{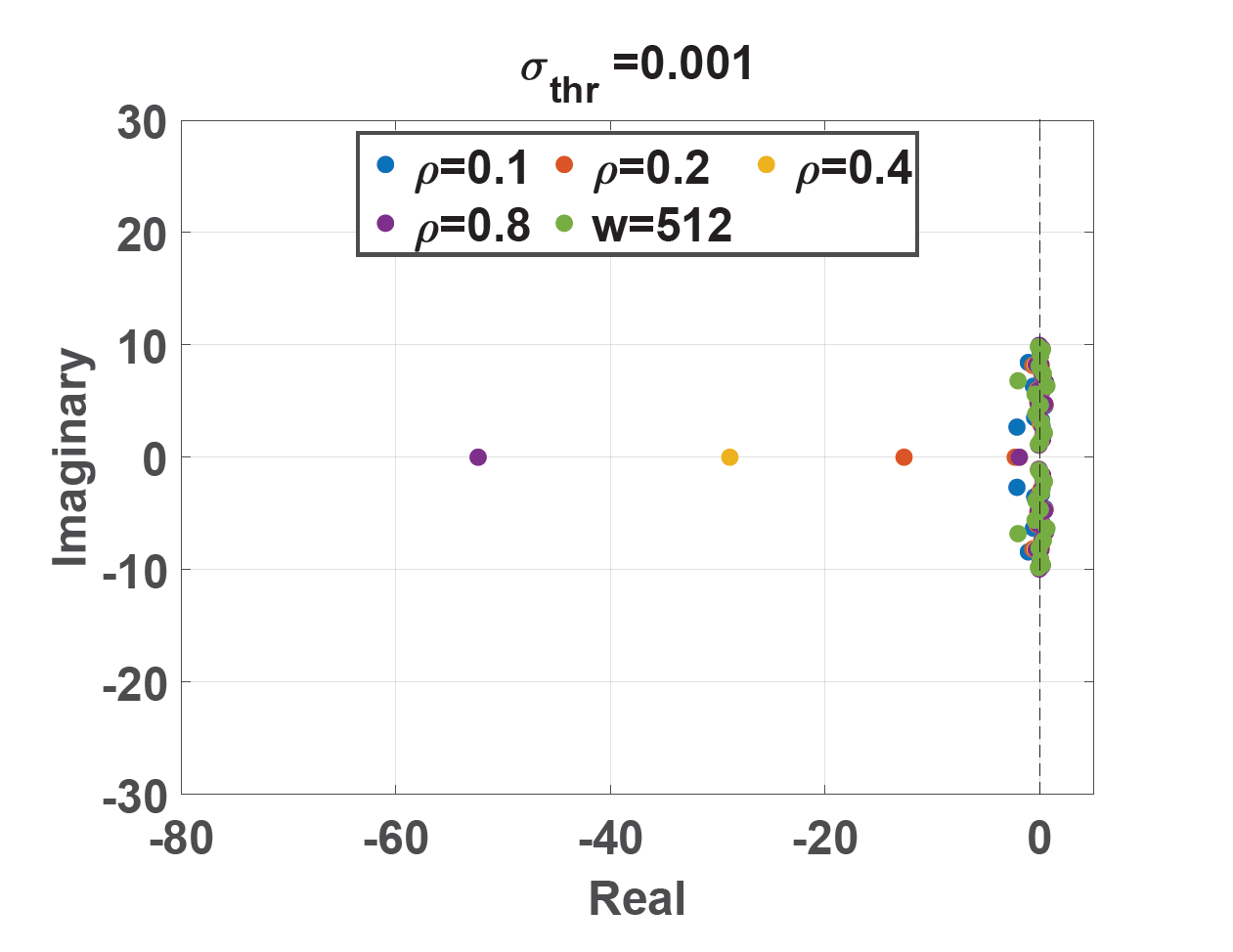}
} \\
\subfloat[Erroneous events]
{
\includegraphics[width=0.45\linewidth]{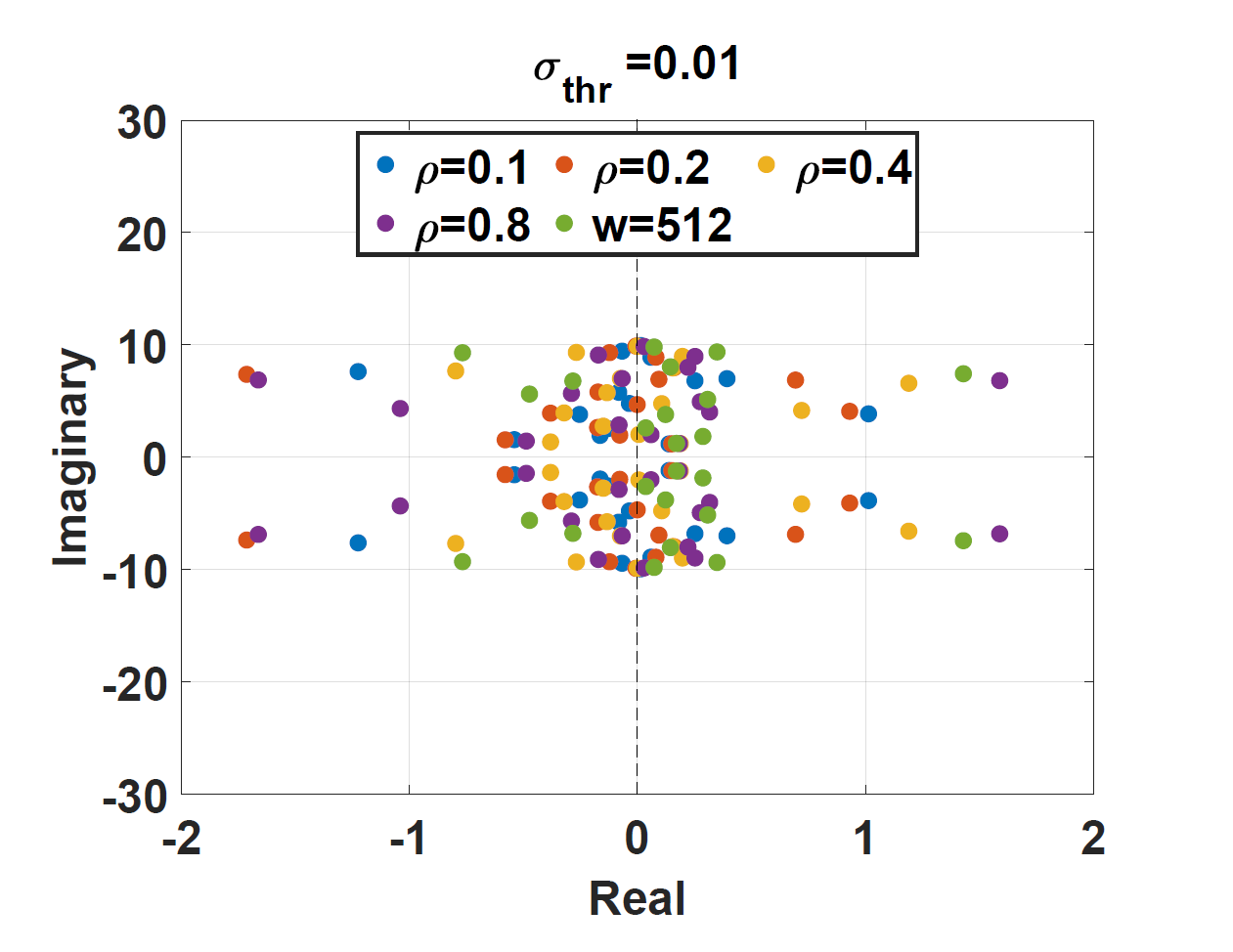}
\includegraphics[width=0.45\linewidth]{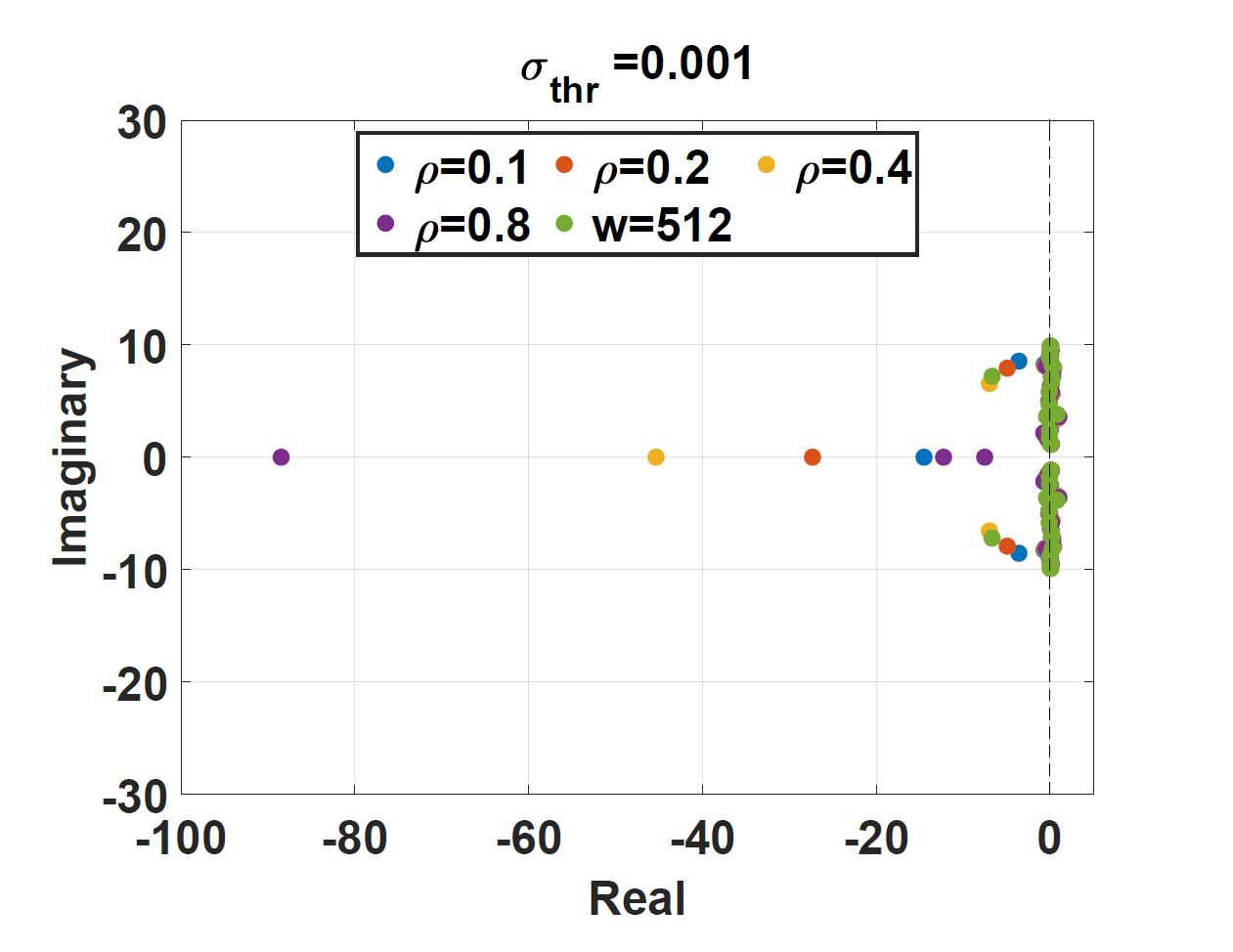}
} 
\caption{The left panel show the {continuous time DMD eigenvalues} for $\sigma_{\text{thr}}=0.01$ and the right panel shows the {continuous time DMD eigenvalues} for $\sigma_{\text{thr}}=0.001$  during (a) correct events (b) erroneous events}
\label{number of modes}
\end{figure}

Figure \ref{number of modes} shows the logarithm of eigenvalues of the DMD operator at 200 msec associated with different thresholds $\sigma_{\text{thr}}$, weight factor $\rho$ and time-window width. For smaller value of $\subscr{\sigma}{thr}$, the range of eigenvalues in the left half plane is bigger. This is consistent with the fact that smaller $\sigma_{\text{thr}}$ implies a larger reduced order model in which some states converge to zero much faster than other states. For the erroneous event, some eigenvalues are spread in the right half plane, which suggest a locally unstable dynamics underlie the evolution of ERP signal during erroneous trials. This is also consistent with the faster dynamics (sharper peaks) in the EEG signal during the {erroneous} event (see Figure~\ref{ErrPs database}).

% It can be seen that the discovered decay rates during erroneous events with best prediction accuracy models (incremental models with $\sigma_{\text{thr}}=0.001$) are faster than the the rates discovered during correct events (incremental models with $\sigma_{\text{thr}}=0.01$), which is consistent with our observation in figure \ref{ErrPs database} in which the ERP during the erroneous events tends to have a faster dynamics (decay rates) than the ERP during correct events.   

\begin{figure}[ht!]
\centering
\subfloat[Correct event]
{
\label{EEG recon rms:a}
\includegraphics[width=0.33\linewidth]{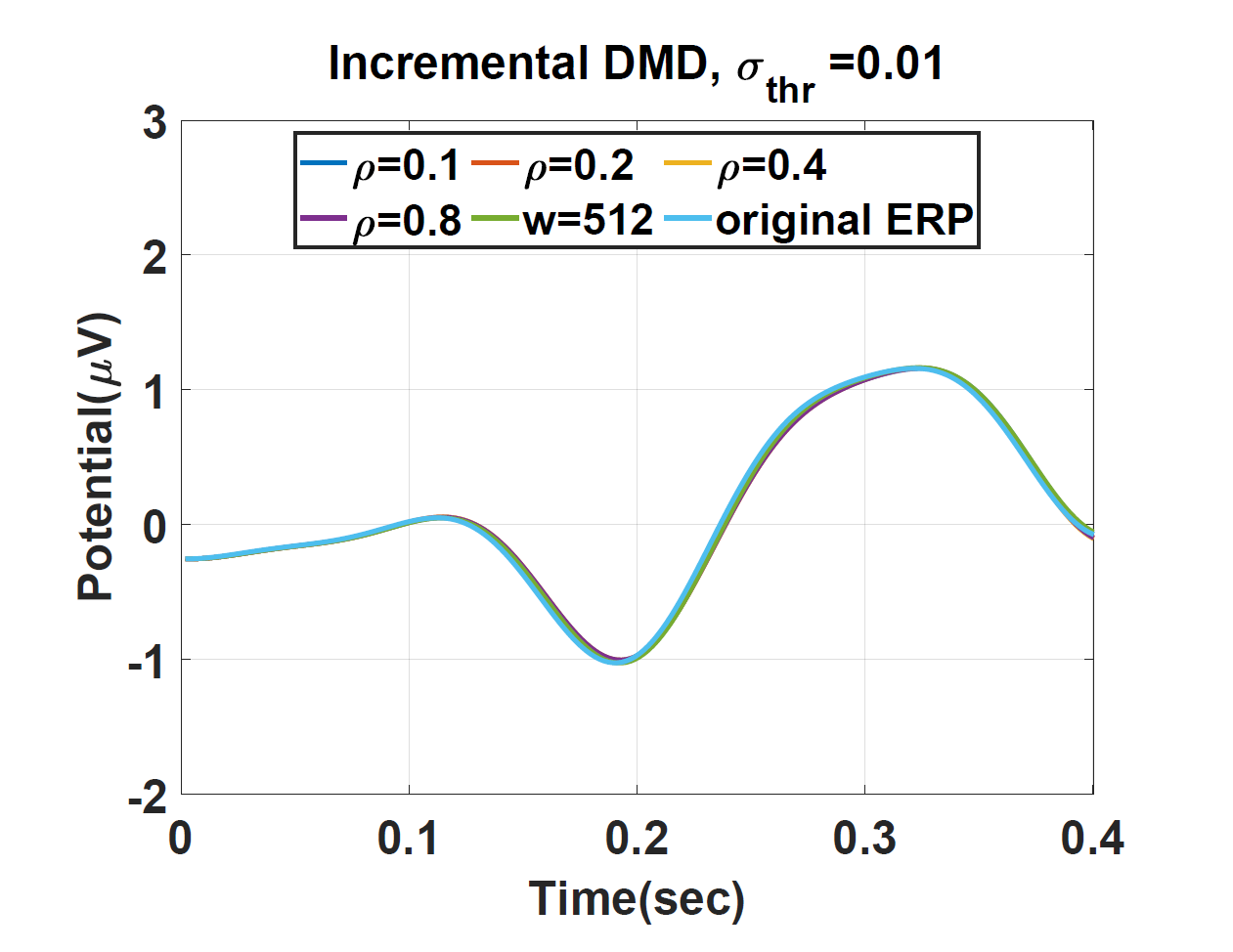}
\includegraphics[width=0.33\linewidth]{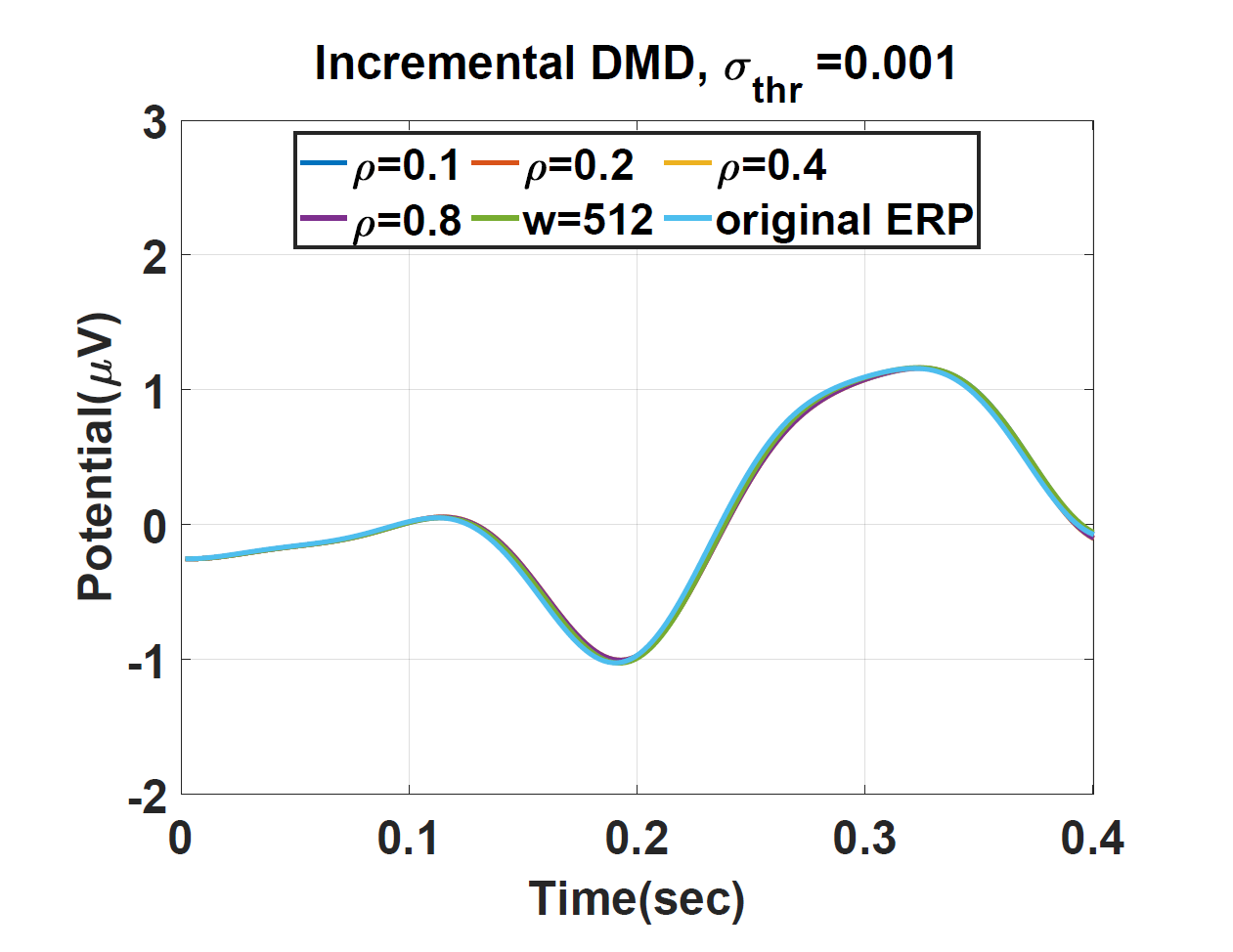}
\includegraphics[width=0.33\linewidth]{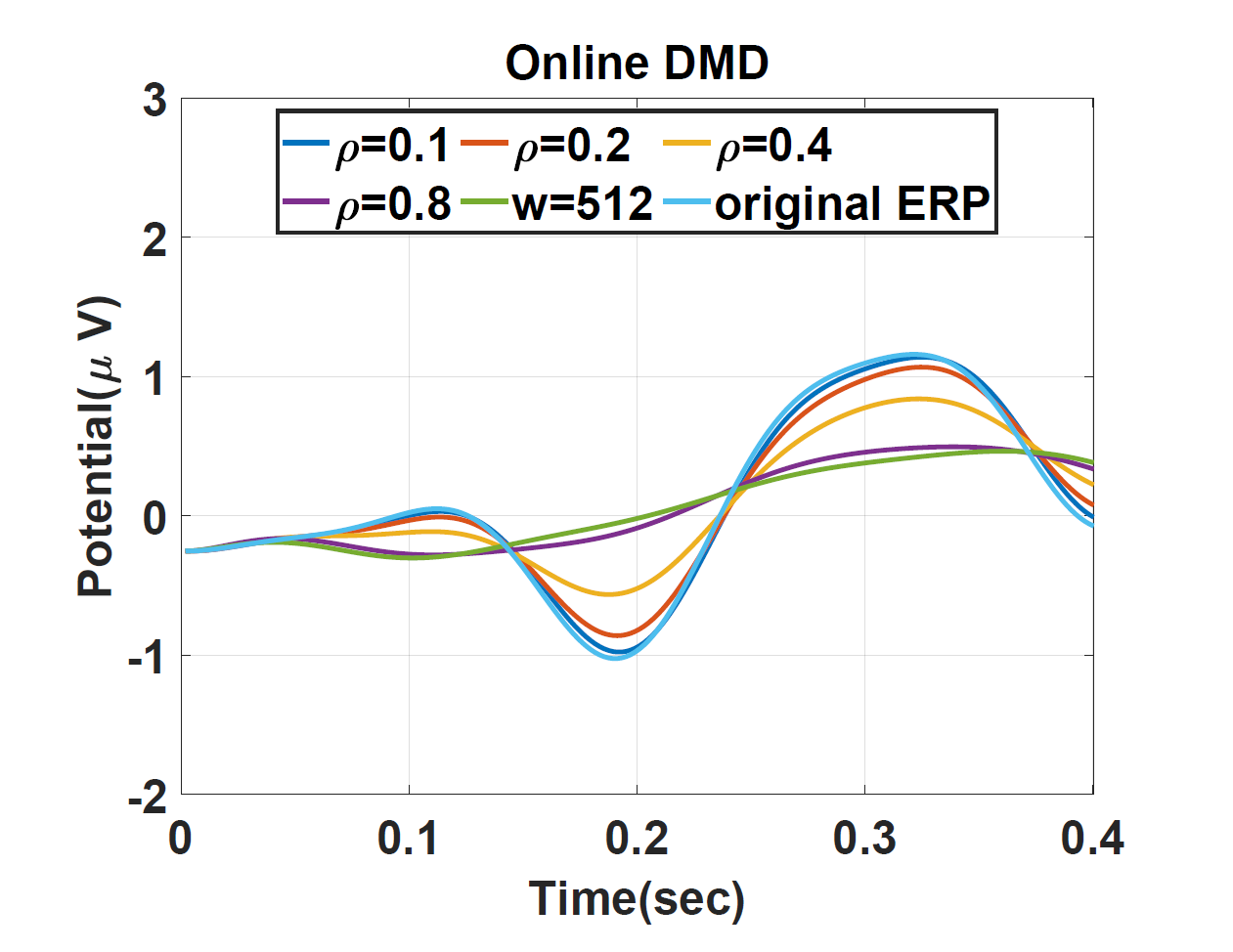}
} \\
\subfloat[Erroneous event]
{
\label{EEG recon rms:b}
\includegraphics[width=0.33\linewidth]{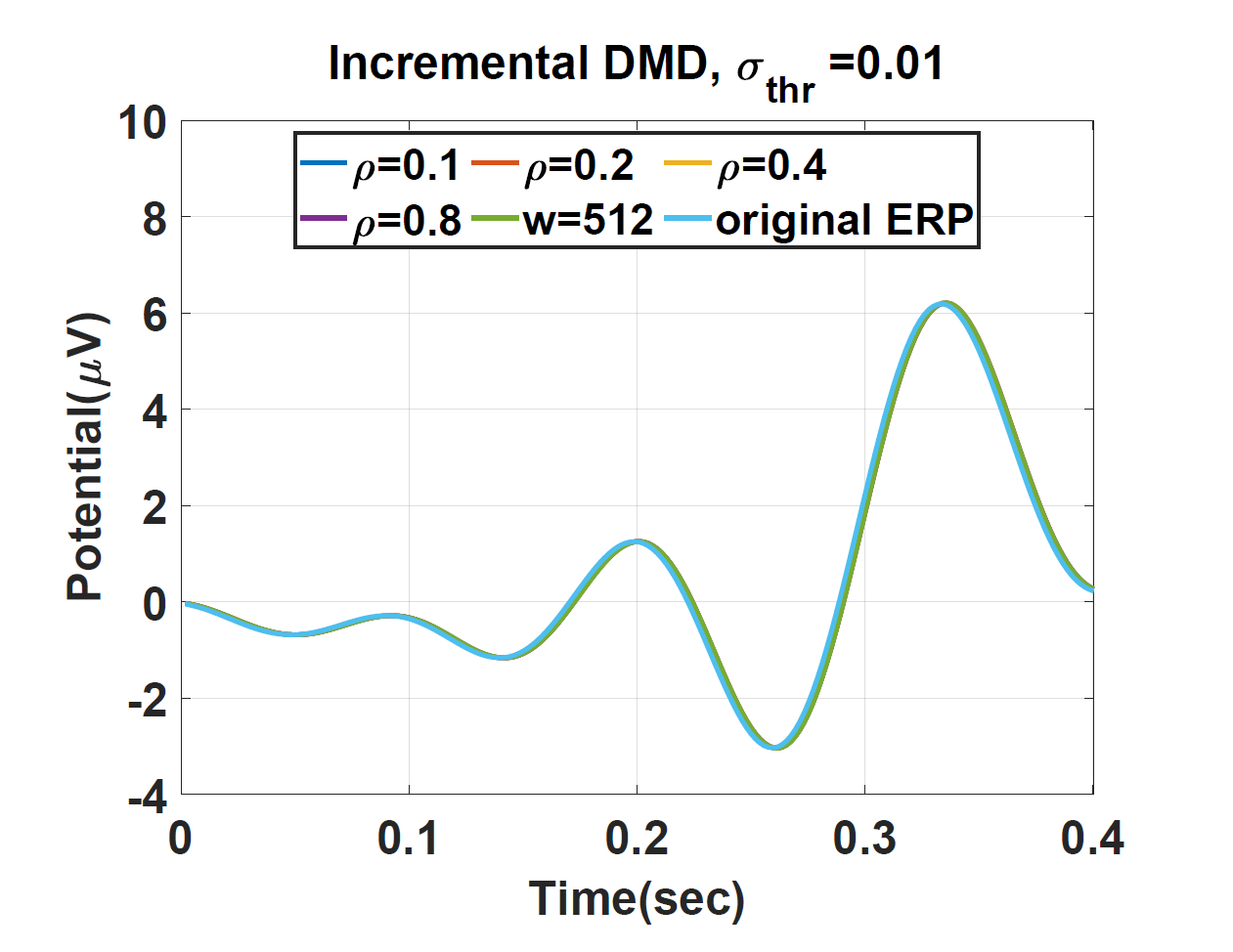}
\includegraphics[width=0.33\linewidth]{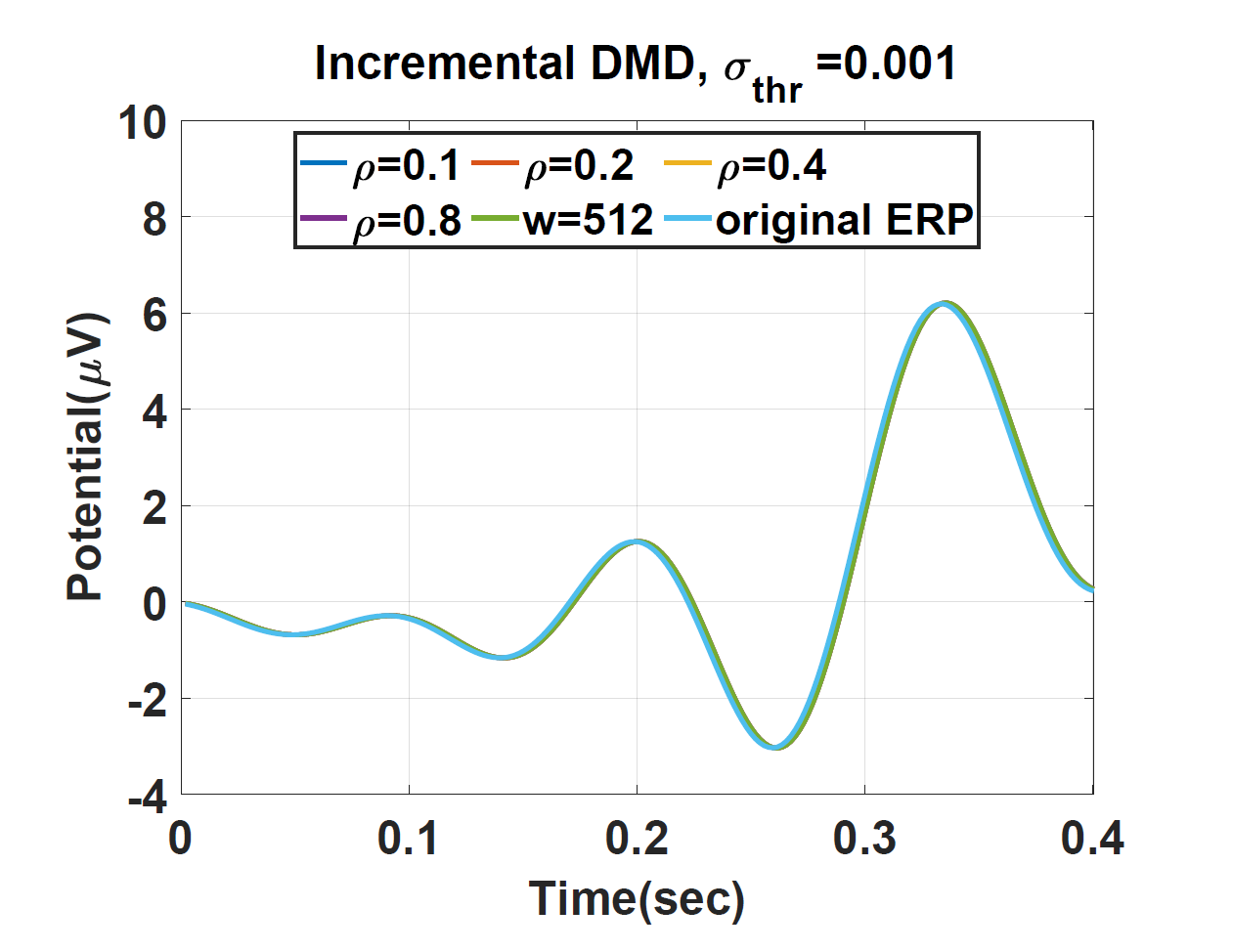}
\includegraphics[width=0.33\linewidth]{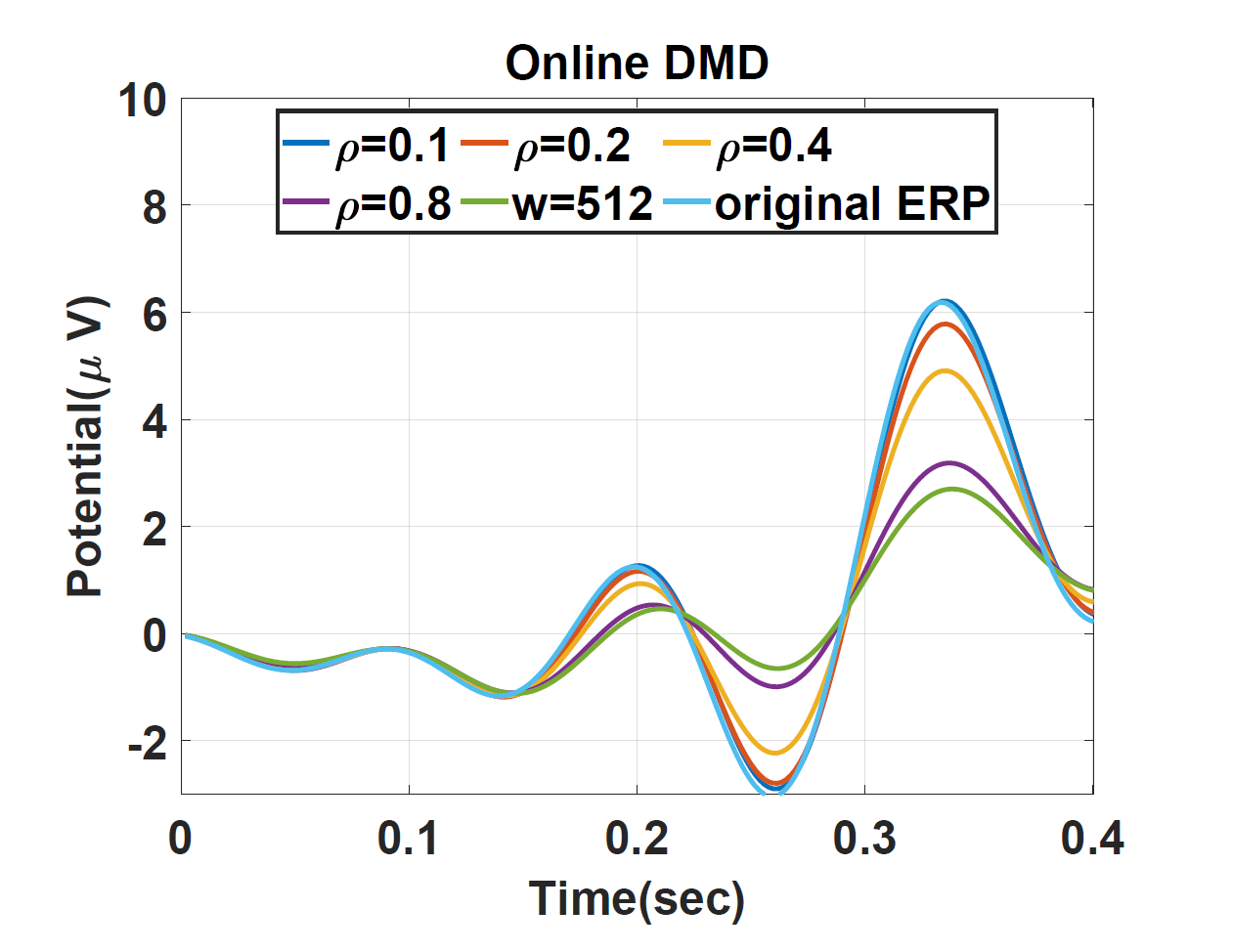}
} 
\caption{Reconstructed ERP signal at channel FCz {using incremental DMD with $\sigma_{thr}=0.01$ (left panel), incremental DMD with $\sigma_{thr}=0.001$ (middle panel), and online DMD (right panel) for (a) correct events and (b) erroneous events.}}
\label{EEG recon}
\end{figure}

\begin{figure}[ht!]
\centering
\subfloat[Incremental DMD]
{
\includegraphics[width=0.45\linewidth]{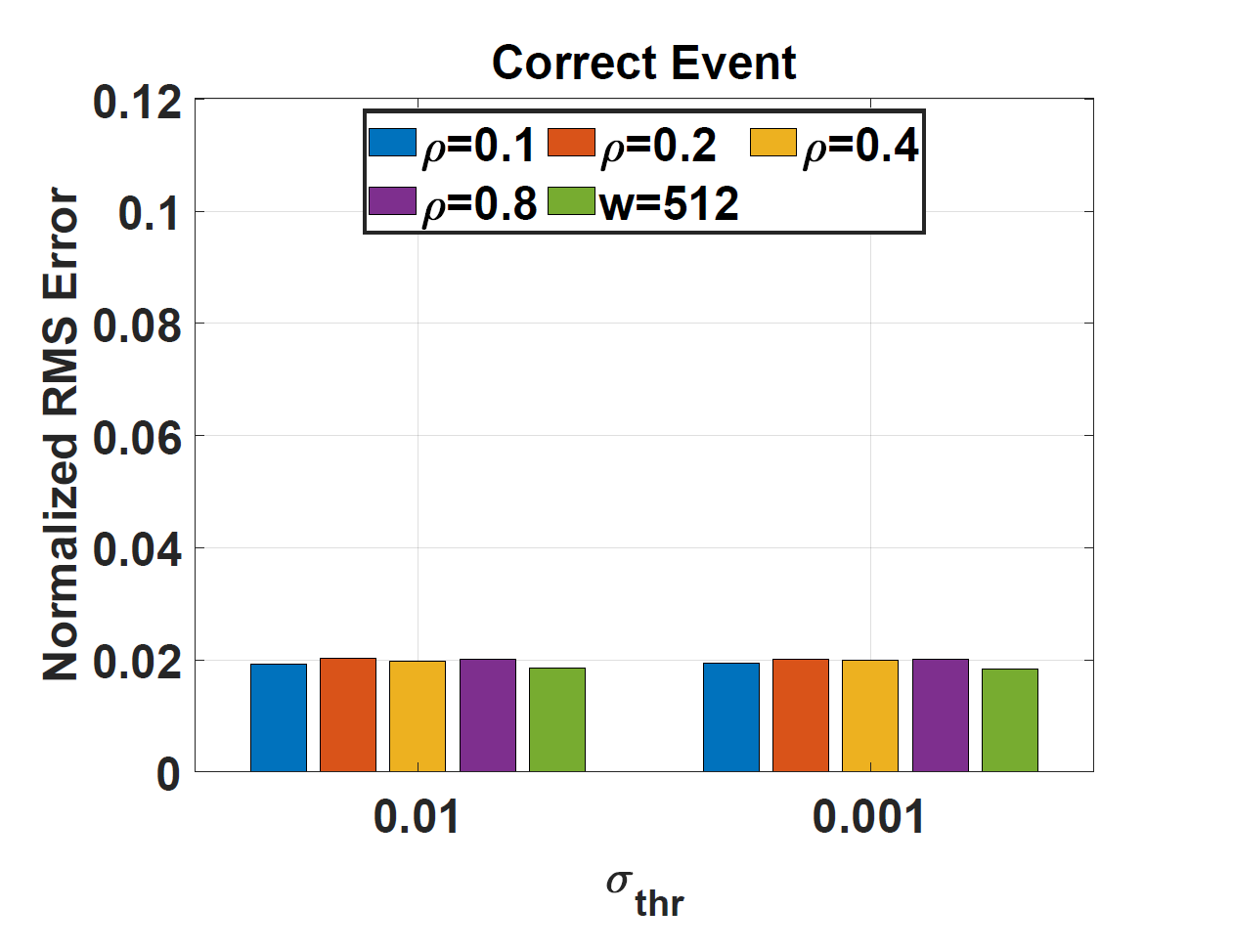}
\includegraphics[width=0.45\linewidth]{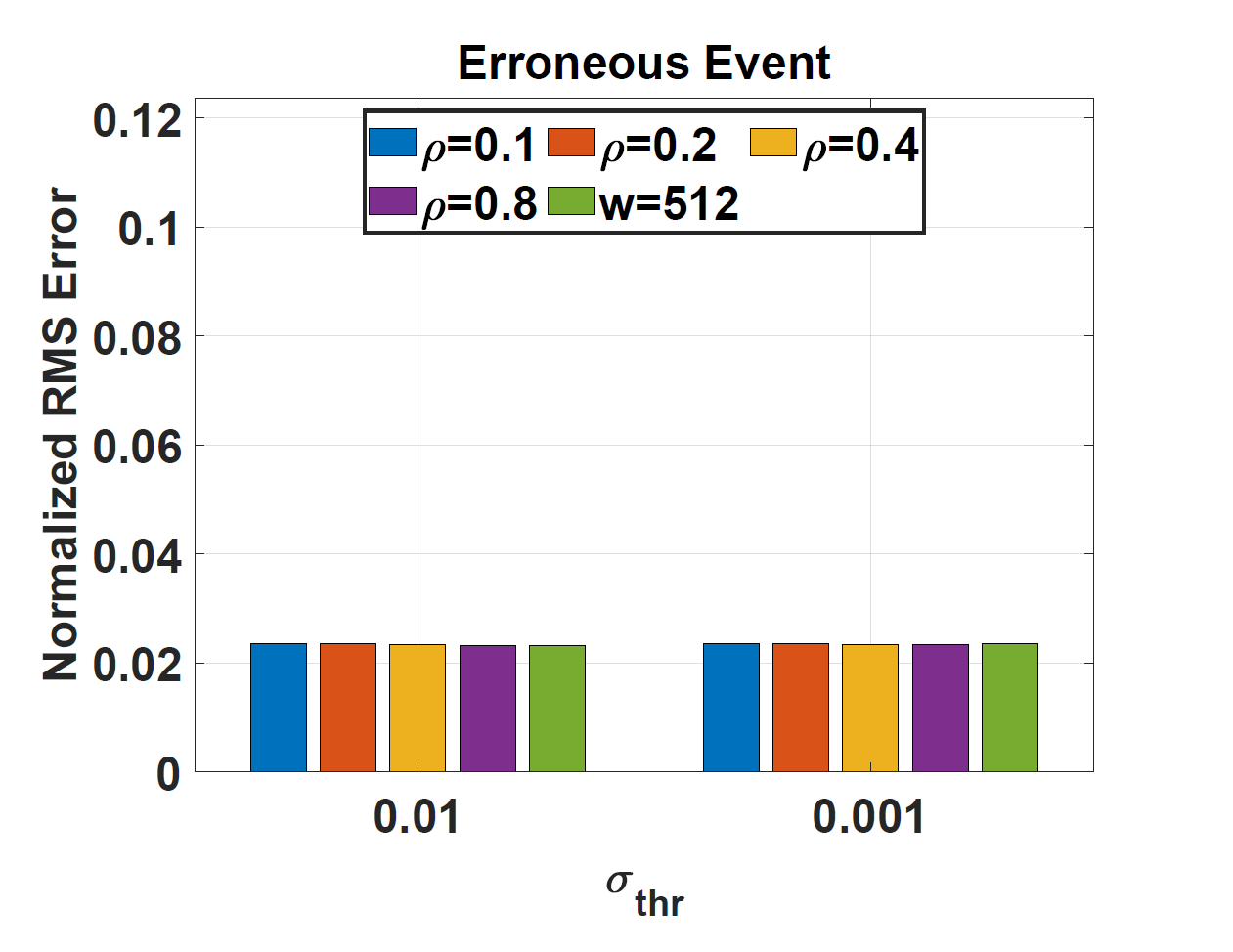}
} \\
\subfloat[Online DMD]
{
\includegraphics[width=0.45\linewidth]{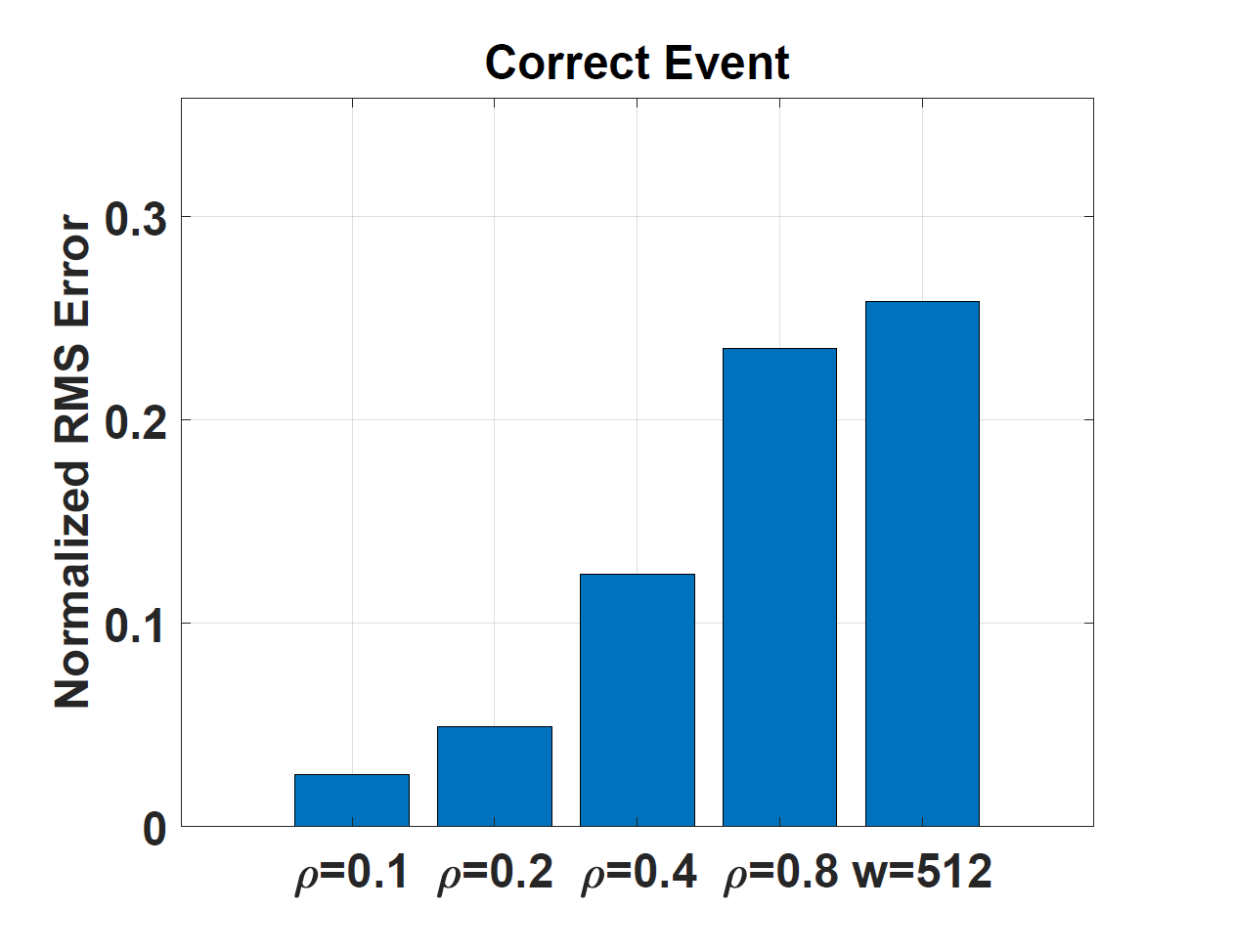}
\includegraphics[width=0.45\linewidth]{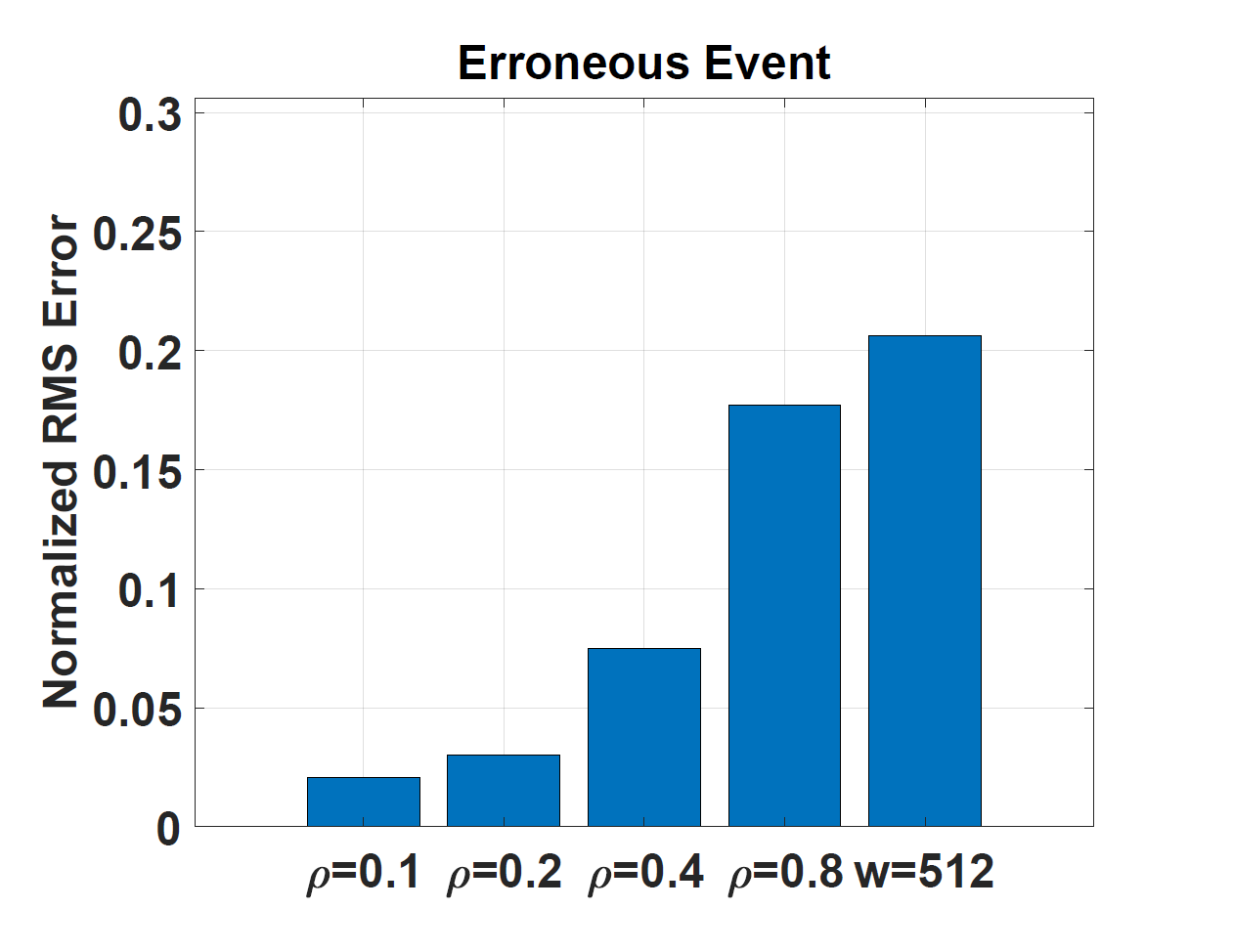}
} 
\caption{Normalized RMS error for the reconstructed ERP signal at channel FCz for correct events (left panel) and erroneous events (right panel), using (a) incremental DMD, and (b) online DMD.}
\label{EEG recon rms}
\end{figure}

We now investigate the effectiveness of the incremental DMD algorithms in terms of reconstructing the  ERP signals during the correct and erroneous events at FCz channel. To this end, we reconstruct the signal starting from the onset of the events until the end of the third peak at $330$ msec. The reconstructed signals and the normalized RMS error for the reconstruction are shown in Figures \ref{EEG recon} and \ref{EEG recon rms}, respectively. These results indicate  that the incremental DMD models incur smaller reconstruction error than the online DMD models. This result is counter-intuitive, since online DMD uses more modes and should lead to smaller reconstruction error. This discrepancy happens due to the fact that the data matrix never achieves full rank and the online DMD is not able to overcome the heuristic initialization discussed above. 
% It also shows that representing the EEG dynamics in a subspace not only improves the prediction accuracy, but also improves the reconstruction of the ERP signals during the two events. 
There is no significant difference in the reconstruction error for incremental DMD models with different $\rho$ and $\sigma_{\text{thr}}$ values as shown in Figure \ref{EEG recon rms:a}.

Finally, we summarize the above investigation into the utility of incremental and online DMD in modeling EEG data. It appears that the pre-processing of the EEG data that requires domain-specific knowledge such as the band of frequency in which the event of interest is observed is vital to obtain sensible DMD-based data-driven models. However, after such pre-processing, the data matrix may become ill-conditioned and under such scenarios incremental DMD techniques proposed in this paper appear more promising that the online DMD techniques. There exists a trade-off between the threshold on the singular values used for model reduction and the window-width or the discount factor used in the incremental DMD algorithms. We also observed that the principal DMD modes are consistent with the activity in the brain during the studied experiments. This suggests that the incremental DMD algorithms can compute an efficient basis for describing the evolution of the EEG activity. Finally, we observed that under erroneous events, some DMD eigenvalues moved towards the right-half complex plane suggesting that certain events trigger (small time) unstable dynamics.


 \section{Incremental Dynamic Mode Decomposition for Systems with Control Input}\label{Incremental DMD with input}
In this section, we extended the incremental DMD algorithms to the case of non-autonomous dynamical systems. We first recall the Dynamic Mode Decomposition with Control (DMDc) algorithm proposed in \cite{proctor2016dynamic} that estimates time-invariant non-autonomous system dynamics underlying high-dimensional data.

% take into consideration the case in which the dynamics are influenced by external input by approximating approximating a time-varying DMD operator and input matrix to represent the external influence on the system states. 

For the scenarios in which the non-autonomous system underlying the measurements is time-varying, the DMDc algorithm has been extended to the online DMDc algorithm \cite{zhang2019online}. The online DMDc algorithm operates similarly to the online DMD algorithm and yields time-varying matrices $\mathbf{A}^c_k \in \mathbb{R}^{n \times n} $ and $\mathbf{B}^c_k \in \mathbb{R}^{n \times l} $, $k \in \{1,2,\cdots\}$. Similar to the online DMD algorithm, the DMDc algorithm also does not allow for incremental computation of the lower dimensional non-autonomous system underlying high dimensional measurements. In the following, we extend the DMDc algorithm to the weighted and windowed incremental DMDc algorithms that enable us to obtain a lower-dimensional time-varying  approximation for the underlying dynamics.

\subsection{Weighted Incremental DMDc Algorithm}\label{Weighted Incremental DMDc}
Assume that at sampling time $t_{k+1}$, the measurements and the associated exogeneous inputs are arranged in the following datasets 

\begin{equation}\label{XGAmma(k)}
\begin{aligned}
    \mathbf{X}_k &= \begin{bmatrix} \rho^{k-1}\mathbf{x}_1 &\rho^{k-2}\mathbf{x}_2 &\cdots & \mathbf{x}_k\end{bmatrix} ,
    \mathbf{Y}_k = \begin{bmatrix} \rho^{k-1}\mathbf{y}_1 &\rho^{k-2}\mathbf{y}_2 &\cdots & \mathbf{y}_k\end{bmatrix} \\
    \mathbf{\Gamma}_{k} &= \begin{bmatrix} \rho^{k-1}\boldsymbol{\gamma}_1 & \rho^{k-2}\boldsymbol{\gamma}_2  & \cdots & \boldsymbol{\gamma}_k\end{bmatrix},
    \mathbf{X}_k^c = \begin{bmatrix}\mathbf{X}_k^*& \mathbf{\Gamma}_k^* \end{bmatrix}^*
\end{aligned}
\end{equation}
respectively, where $\mathbf{X}_k,\mathbf{Y}_k \in \mathbb{R}^{n\times k}$, $\mathbf{\Gamma}_{k} \in \mathbb{R}^{l \times k}$, and $\mathbf{X}_k^c \in \mathbb{R}^{(n+l)\times k} $. 
At sampling time $t_{k+1}$, the datasets are updated with a new set of measurements $\left(\mathbf{x}_{k+1},\mathbf{y}_{k+1}, \boldsymbol{\gamma}_{k+1} \right)$ which results in the following
\begin{equation}\label{XGAmma(k+1)}
\begin{aligned}
    \mathbf{X}_{k+1}&=\begin{bmatrix} \rho \mathbf{X}_{k} & \mathbf{x}_{k+1}\end{bmatrix},
    \mathbf{Y}_{k+1}=\begin{bmatrix} \rho \mathbf{Y}_{k} & \mathbf{y}_{k+1}\end{bmatrix},\\
    \mathbf{\Gamma}_{k+1}&=\begin{bmatrix} \rho \mathbf{\Gamma}_{k} & \boldsymbol{\gamma}_{k+1}\end{bmatrix}, 
    \mathbf{X}_{k+1}^c = \begin{bmatrix}
\mathbf{X}_{k+1}^*& \mathbf{\Gamma}_{k+1}^*
\end{bmatrix}^*.
\end{aligned}
\end{equation}
{
Suppose that the weighted DMDc operator and input matrix at $t_{k+1}$, and the SVD of both $\mathbf{X}_{k}$ and $\mathbf{X}_k^c$ are known, then according to the new incoming measurements at $t_{k+2}$, the updates for the weighted DMDc operator and input matrix can be obtained using the following theorem, which is proved in Appendix~\ref{proof of Theorem 3}.
\begin{theorem}\label{weighted incremental DMD with controlinput}
Let at sampling time $t_{k+1}$ the SVD of $\mathbf{X}^{c}_k= \mathbf{U}^{c}_{x_k} \mathbf{\Sigma}^{c}_{x_k} \mathbf{V}^{c*}_{x_k}$ and $\mathbf{X}_k= \mathbf{U}_{x_k} \mathbf{\Sigma}_{x_k} \mathbf{V}^{*}_{x_k}$ be known. Assume that the pair  $(\mathbf{A}^{\rho_c}_k, \mathbf{B}^{\rho_c}_k)$ minimizes the cost function in \eqref{cost function for control} with $\mathbf{X}_k$ and $\mathbf{\Gamma}_{k}$ defined in \eqref{XGAmma(k)}.
Consider that at sampling time $t_{k+2}$, a new triplet of measurements $(\mathbf{x}_{k+1},\mathbf{y}_{k+1},\boldsymbol{\gamma}_{k+1})$ is used to incrementally compute
% obtain the datasets in \eqref{XGAmma(k+1)}, and 
the SVD of both $\mathbf{X}_{k+1}^c$ and $\mathbf{X}_{k+1}$ using Proposition \ref{weighted incremental SVD  proposition }. Then, the pair $(\mathbf{A}^{\rho_c}_{k+1}, \mathbf{B}^{\rho_c}_{k+1})$ which minimizes the cost function in \eqref{cost function for control} is
\begin{align}\label{inc weighted DMDc thm}
\begin{split}
{\mathbf{A}}^{\rho_c}_{k+1}&={\mathbf{A}}^{\rho_c}_{k}+\left(\mathbf{y}_{k+1}-{\mathbf{A}}^{\rho_c}_{k}\mathbf{x}_{k+1}-{\mathbf{B}}^{\rho_c}_{k}\boldsymbol{\gamma}_{k+1}\right){\mathbf{v}}^{c}_{s_{k,2}} ({\mathbf{\Sigma}}^{c}_{x_{k+1}})^{-1} {\mathbf{U}}^{c_a*}_{x_{k+1}},   \\  {\mathbf{B}}^{\rho_c}_{k+1}&={\mathbf{B}}^{\rho_c}_{k}+\left(\mathbf{y}_{k+1}-{\mathbf{A}}^{\rho_c}_{k}\mathbf{x}_{k+1}-{\mathbf{B}}^{\rho_c}_{k}\boldsymbol{\gamma}_{k+1}\right){\mathbf{v}}^{c}_{s_{k,2}} ({\mathbf{\Sigma}}^{c}_{x_{k+1}})^{-1} {\mathbf{U}}^{c_b*}_{x_{k+1}}.   
\end{split}
\end{align}
\end{theorem}
}
Assume that $\mathbf{X}^{c}_{k+1}$ can be well-approximated by its projection onto its leading $p$ singular vectors such that $\mathbf{X}^{c}_{k+1} \approx \bar{\mathbf{X}}^{c}_{k+1}=\bar{\mathbf{U}}^{c}_{x_{k+1}} \bar{\mathbf{\Sigma}}^{c}_{x_{k+1}} \bar{\mathbf{V}}^{c*}_{x_{k+1}}$, where $\bar{\mathbf{U}}_{x_{k+1}}^{c} \in \mathbb{C}^{(n+l) \times p}$, $\bar{\mathbf{\Sigma}}_{x_{k+1}}^{c} \in \mathbb{C}^{p \times p}$ ,and $\bar{\mathbf{V}}_{x_{k+1}}^{c} \in \mathbb{C}^{(k+1) \times p }$. Let $\bar{\mathbf{U}}_{x_{k+1}}^{c*}$ be partitioned such that $\bar{\mathbf{U}}_{x_{k+1}}^{c*} = [\bar{\mathbf{U}}^{c_a*}_{x_{k+1}} \bar{\mathbf{U}}^{c_b*}_{x_{k+1}}]$,  where $\bar{\mathbf{U}}^{c_a}_{x_{k+1}} \in \mathbb{C}^{n \times p}$ and $\bar{\mathbf{U}}^{c_b}_{x_{k+1}} \in \mathbb{C}^{l \times p}$. Then, the update in \eqref{inc weighted DMDc thm} yields the following updates for $ \bar{\mathbf{A}}^{\rho_c}_{k+1} \in \mathbb{R}^{n \times n}$ and $\bar{\mathbf{B}}^{\rho_c}_{k+1} \in \mathbb{R}^{n \times l}$:
\begin{align}
\bar{\mathbf{A}}^{\rho_c}_{k+1}&=\bar{\mathbf{A}}^{\rho_c}_{k}+\left(\mathbf{y}_{k+1}-\bar{\mathbf{A}}^{\rho_c}_{k}\mathbf{x}_{k+1}-\bar{\mathbf{B}}^{\rho_c}_{k}\boldsymbol{\gamma}_{k+1}\right)\bar{\mathbf{v}}^{c}_{s_{k,2}} (\bar{\mathbf{\Sigma}}^{c}_{x_{k+1}})^{-1} \bar{\mathbf{U}}^{c_a*}_{x_{k+1}}  \nonumber \\                 
\bar{\mathbf{B}}^{\rho_c}_{k+1}&=\bar{\mathbf{B}}^{\rho_c}_{k}+\left(\mathbf{y}_{k+1}-\bar{\mathbf{A}}^{\rho_c}_{k}\mathbf{x}_{k+1}-\bar{\mathbf{B}}^{\rho_c}_{k}\boldsymbol{\gamma}_{k+1}\right)\bar{\mathbf{v}}^{c}_{s_{k,2}} (\bar{\mathbf{\Sigma}}^{c}_{x_{k+1}})^{-1} \bar{\mathbf{U}}^{c_b*}_{x_{k+1}}. \nonumber                   
\end{align}
Assume that $\mathbf{X}_{k+1}$ is well-approximated by its projection onto its leading $r$ singular vectors such that $\mathbf{X}_{k+1} \approx\bar{\mathbf{X}}_{k+1}=\bar{\mathbf{U}}_{x_{k+1}} \bar{\mathbf{\Sigma}}_{x_{k+1}} \bar{\mathbf{V}}^*_{x_{k+1}}$ with $r \le p$. Then a reduced order model can be represented as:
\begin{eqnarray}
\tilde{\mathbf{x}}_{k+1}= \tilde{\mathbf{A}}^{\rho_c}_{k}\tilde{\mathbf{x}}_{k}+\tilde{\mathbf{B}}^{\rho_c}_{k} {\gamma}_k, \nonumber
\end{eqnarray}
where the lower-dimensional approximation  $\tilde{\mathbf{A}}^{\rho_c}_{k+1} \in \mathbb{R}^{r \times r}$ and $\tilde{\mathbf{B}}^{\rho_c}_{k+1} \in \mathbb{R}^{r \times l}$ can obtained by projecting $\bar{\mathbf{A}}^{\rho_c}_{k+1}$ and $\bar{\mathbf{B}}^{\rho_c}_{k+1}$ onto a subspace spanned by the columns of $\bar{\mathbf{U}}_{x_{k+1}}$ as follows:
\begin{align}\label{tilde AB weighted DMDc(k+1)} 
\tilde{\mathbf{A}}^{\rho_c}_{k+1}&=\bar{\mathbf{U}}^*_{x_{k+1}}\bar{\mathbf{A}}^{\rho_c}_{k}\bar{\mathbf{U}}_{x_{k+1}}+\bar{\mathbf{U}}^*_{x_{k+1}}\left(\mathbf{y}_{k+1}-\bar{\mathbf{A}}^{\rho_c}_{k}\mathbf{x}_{k+1}-\bar{\mathbf{B}}^{\rho_c}_{k}\boldsymbol{\gamma}_{k+1}\right)\bar{\mathbf{v}}^{c}_{s_{k,2}} (\bar{\mathbf{\Sigma}}^{c}_{x_{k+1}})^{-1} \bar{\mathbf{U}}^{c_a*}_{x_{k+1}}\bar{\mathbf{U}}_{x_{k+1}} \nonumber \\                 
\tilde{\mathbf{B}}^{\rho_c}_{k+1}&=\bar{\mathbf{U}}^*_{x_{k+1}} \bar{\mathbf{B}}^{\rho_c}_{k}+\bar{\mathbf{U}}^*_{x_{k+1}}\left(\mathbf{y}_{k+1}-\bar{\mathbf{A}}^{\rho_c}_{k}\mathbf{x}_{k+1}-\bar{\mathbf{B}}^{\rho_c}_{k}\boldsymbol{\gamma}_{k+1}\right)\bar{\mathbf{v}}^{c}_{s_{k,2}} (\bar{\mathbf{\Sigma}}^{c}_{x_{k+1}})^{-1} \bar{\mathbf{U}}^{c_b*}_{x_{k+1}}. \nonumber      \end{align}             
The update in equation \eqref{inc weighted DMDc thm}   requires only the current DMD operator $(\mathbf{A}^{\rho_c}_{k}, \mathbf{B}^{\rho_c}_{k})$, measurements set of  $(\mathbf{y}_{k+1},\mathbf{x}_{k+1},{\gamma}_{k+1})$, and the incremental SVD update. Specifically, these updates do not require the large data matrix to be stored.

\subsection{Windowed Incremental DMDc Algorithm}\label{Windowed Incremental DMDc}
Assume that at sampling time $t_{k+1}$, the past $w$ states and control input measurements are arranged in the following datasets 
\begin{equation}\label{windowedXGamma(k)}
    \begin{aligned}
    \boldsymbol{\chi}_k &= \begin{bmatrix} \mathbf{x}_{k-w+1}&\cdots &\mathbf{x}_k\end{bmatrix},
    \boldsymbol{\Upsilon}_{k}=\begin{bmatrix} \mathbf{y}_{k-w+1}&\cdots &\mathbf{y}_k\end{bmatrix},\\
    \mathbf{\Gamma}_{k} &= \begin{bmatrix} \boldsymbol{\gamma}_{k-w+1} & \cdots & \boldsymbol{\gamma}_k\end{bmatrix},
    \boldsymbol{\chi}^c_k=\begin{bmatrix} \boldsymbol{\chi}_k^* &\mathbf{\Gamma}_{k}^*\end{bmatrix}^*,
    \end{aligned}
\end{equation}
where $\boldsymbol{\chi}_k, \boldsymbol{\Upsilon}_{k} \in \mathbb{R}^{n\times w}$, $\mathbf{\Gamma}_{l}  \in \mathbb{R}^{l \times w}$, and $\boldsymbol{\chi}^c_k \in \mathbb{R}^{(n+l)\times w} $, respectively. 
At sampling time $t_{k+1}$, the datasets are updated by adding a new set of measurements $\left(\mathbf{x}_{k+1}, \mathbf{y}_{k+1},\boldsymbol{\gamma}_{k+1} \right)$, and removing the oldest set of measurements $\left(\mathbf{x}_{k-w+1},\mathbf{y}_{k-w+1},\boldsymbol{\gamma}_{k-w+1}\right)$, such that:
\begin{equation}\label{windowedXGamma(k+1)}
    \begin{aligned}
    \boldsymbol{\chi}_{k+1} &= \begin{bmatrix} \mathbf{x}_{k-w+2}&\cdots &\mathbf{x}_{k+1}\end{bmatrix},
    \boldsymbol{\Upsilon}_{k+1}=\begin{bmatrix} \mathbf{y}_{k-w+2}&\cdots &\mathbf{y}_{k+1}\end{bmatrix},\\
    \mathbf{\Gamma}_{k+1} &= \begin{bmatrix} \boldsymbol{\gamma}_{k-w+2} & \cdots & \boldsymbol{\gamma}_{k+1}\end{bmatrix},
    \boldsymbol{\chi}^c_{k+1}=\begin{bmatrix} \boldsymbol{\chi}_{k+1}^* &\mathbf{\Gamma}_{k+1}^*\end{bmatrix}^*.
    \end{aligned}
\end{equation}
{
Suppose that the windowed DMDc operator and input matrix at $t_{k+1}$, and the SVD of both $\boldsymbol{\chi}_k$ and $\boldsymbol{\chi}^c_k$ are known. Then, according to the new incoming measurements at $t_{k+2}$, the updates for the windowed DMDc operator and input matrix can be obtained using the following theorem, which is proved in Appendix~\ref{proof of Theorem 4}.
\begin{theorem}\label{windowed incremental DMD with controlinput}
Let at sampling time $t_{k+1}$ the SVD of $\boldsymbol{\chi}_{k}=\mathbf{U}_{\chi_{k}} \mathbf{\Sigma}_{\chi_{k}}\mathbf{V}^*_{\chi_{k}}$ and $ \boldsymbol{\chi}_k^{c}= \mathbf{U}^{c}_{\chi_k} \mathbf{\Sigma}^{c}_{\chi_k} \mathbf{V}^{c*}_{\chi_k}$ be known, and let the pair $(\mathbf{A}^{w_c}_{k}, \mathbf{B}^{w_c}_{k})$ {minimize} the cost function in \eqref{cost function for control} with $\boldsymbol{\Upsilon}_{k}$ and $\boldsymbol{\chi}_k$ given in \eqref{windowedXGamma(k)}.  Assume that at sampling time $t_{k+2}$, a new triplet of measurements $(\mathbf{x}_{k+1},\mathbf{y}_{k+1},\boldsymbol{\gamma}_{k+1})$ is used to incrementally compute
% added and the oldest triplet of measurements $(\mathbf{x}_{k-w+1},\mathbf{y}_{k-w+1},\boldsymbol{\gamma}_{k-w+1})$ is eliminated, such that the datasets in \eqref{windowedXGamma(k+1)} are obtained, and 
the SVD of $\boldsymbol{\chi}^c_{k+1}$ and $\boldsymbol{\chi}_{k+1}$ using Proposition \ref{windowed incremental SVD  proposition} and equation~\eqref{windowed updated SVD factors}. Then, the pair $(\mathbf{A}^{w_c}_{k+1}, \mathbf{B}^{w_c}_{k+1})$  which minimizes the cost function in \eqref{cost function for control} is
\begin{align}\label{inc windowed DMDc thm}
\begin{split}
  {\mathbf{A}}^{w_c}_{k+1}&={\mathbf{A}}^{w_c}_{k}+\left(\mathbf{y}_{k+1}-{\mathbf{A}}^{w_c}_{k}\mathbf{x}_{k+1}-{\mathbf{B}}^{w_c}_{k}\boldsymbol{\gamma}_{k+1}\right)
{\mathbf{v}}_{\hat{s}_{k,2}}^{c}({\mathbf{\Sigma}}^{c}_{\chi_{k+1}})^{-1} \mathbf{U}^{c_a*}_{\chi_{k+1}},  \\                 
{\mathbf{B}}^{w_c}_{k+1}&={\mathbf{B}}^{w_c}_{k}+\left(\mathbf{y}_{k+1}-{\mathbf{A}}^{w_c}_{k}\mathbf{x}_{k+1}-{\mathbf{B}}^{w_c}_{k}\boldsymbol{\gamma}_{k+1}\right)
{\mathbf{v}}_{\hat{s}_{k,2}}^{c}({\mathbf{\Sigma}}^{c}_{\chi_{k+1}})^{-1} \mathbf{U}^{c_b*}_{\chi_{k+1}}.   
\end{split}
\end{align}
\end{theorem}
}

If $\boldsymbol{\chi}^{c}_{k+1}$ can be well-approximated by its projection onto {its} leading $p$ singular vectors such that $\bar{\boldsymbol{\chi}}^{c}_{k+1}=\bar{\mathbf{U}}^{c}_{\chi_{k+1}} \bar{\mathbf{\Sigma}}^{c}_{\chi_{k+1}} \bar{\mathbf{V}}^{c*}_{\chi_{k+1}}$, where $\bar{\mathbf{U}}_{\chi_{k+1}}^{c} \in \mathbb{C}^{(n+l) \times p}$, $\bar{\mathbf{\Sigma}}_{\chi_{k+1}}^{c} \in \mathbb{C}^{p \times p}$, and $\bar{\mathbf{V}}_{\chi_{k+1}}^{c} \in \mathbb{C}^{w\times p}$, then equation \eqref{inc windowed DMDc thm} yields the following updates for $\bar{\mathbf{A}}^{w_c}_{k+1} \in \mathbb{R}^{n \times n}$ and $\bar{\mathbf{B}}^{w_c}_{k+1} \in \mathbb{R}^{n \times n}$: 
\begin{align}
\bar{\mathbf{A}}^{w_c}_{k+1}&=\bar{\mathbf{A}}^{w_c}_{k}+\left(\mathbf{y}_{k+1}-\bar{\mathbf{A}}^{w_c}_{k}\mathbf{x}_{k+1}-\bar{\mathbf{B}}^{w_c}_{k}\boldsymbol{\gamma}_{k+1}\right)
\bar{\mathbf{v}}_{\hat{s}_{k,2}}^{c}(\bar{\mathbf{\Sigma}}^{c}_{\chi_{k+1}})^{-1} \bar{\mathbf{U}}^{c_a*}_{\chi_{k+1}},  \nonumber \\                 
\bar{\mathbf{B}}^{w_c}_{k+1}&=\bar{\mathbf{B}}^{w_c}_{k}+\left(\mathbf{y}_{k+1}-\bar{\mathbf{A}}^{w_c}_{k}\mathbf{x}_{k+1}-\bar{\mathbf{B}}^{w_c}_{k}\boldsymbol{\gamma}_{k+1}\right)
\bar{\mathbf{v}}_{\hat{s}_{k,2}}^{c}(\bar{\mathbf{\Sigma}}^{c}_{\chi_{k+1}})^{-1} \bar{\mathbf{U}}^{c_b*}_{\chi_{k+1}}, \nonumber 
\end{align}
where  $\bar{\mathbf{U}}^{c_a}_{\chi_{k+1}} \in \mathbb{C}^{n \times p}$ and $\bar{\mathbf{U}}^{c_b}_{\chi_{k+1}} \in \mathbb{C}^{l \times p}$ are defined such that 
$\bar{\mathbf{U}}^{c*}_{\chi_{k+1}} = [\bar{\mathbf{U}}^{c_a*}_{\chi_{k+1}} \bar{\mathbf{U}}^{c_b*}_{\chi_{k+1}}]$.

Assume that $\boldsymbol{\chi}_{k+1}$ can be projected onto {its} leading $r$ singular vectors such that $\bar{\boldsymbol{\chi}}_{k+1}=\bar{\mathbf{U}}_{\chi_{k+1}} \bar{\mathbf{\Sigma}}_{\chi_{k+1}} \bar{\mathbf{V}}^{*}_{\chi_{k+1}}$ with $r \le p$. Then a reduced order model can be represented as:
\begin{eqnarray}
\tilde{\mathbf{x}}_{k+1}= \tilde{\mathbf{A}}^{w_c}_{k}\tilde{\mathbf{x}}_{k}+\tilde{\mathbf{B}}^{w_c}_{k} {\gamma}_k, \nonumber
\end{eqnarray}
where the lower-dimensional approximation of $\tilde{\mathbf{A}}^{w_c}_{k+1} \in \mathbb{R}^{r \times r}$ and $\tilde{\mathbf{B}}^{w_c}_{k+1} \in \mathbb{R}^{r \times l}$ can obtained by projecting $\bar{\mathbf{A}}^{w_c}_{k+1}$ and $\bar{\mathbf{B}}^{w_c}_{k+1}$ onto a subspace spanned by the columns of $\bar{\mathbf{U}}_{\chi_{k+1}}$ as follows:
\begin{align}
\tilde{\mathbf{A}}^{w_c}_{k+1}&=\bar{\mathbf{U}}_{\chi_{k+1}}^*\bar{\mathbf{A}}^{w_c}_{k}\bar{\mathbf{U}}_{\chi_{k+1}}+\bar{\mathbf{U}}_{\chi_{k+1}}^*\left(\mathbf{y}_{k+1}-\bar{\mathbf{A}}^{w_c}_{k}\mathbf{x}_{k+1}-\bar{\mathbf{B}}^{w_c}_{k}\boldsymbol{\gamma}_{k+1}\right)
\bar{\mathbf{v}}_{\hat{s}_{k,2}}^{c}(\bar{\mathbf{\Sigma}}^{c}_{\chi_{k+1}})^{-1} \bar{\mathbf{U}}^{c_a*}_{\chi_{k+1}} \bar{\mathbf{U}}_{\chi_{k+1}},  \nonumber \\                 
\tilde{\mathbf{B}}^{w_c}_{k+1}&=\bar{\mathbf{U}}_{\chi_{k+1}}^*\bar{\mathbf{B}}^{w_c}_{k}+\bar{\mathbf{U}}_{\chi_{k+1}}^*\left(\mathbf{y}_{k+1}-\bar{\mathbf{A}}^{w_c}_{k}\mathbf{x}_{k+1}-\bar{\mathbf{B}}^{w_c}_{k}\boldsymbol{\gamma}_{k+1}\right)
\bar{\mathbf{v}}_{\hat{s}_{k,2}}^{c}(\bar{\mathbf{\Sigma}}^{c}_{\chi_{k+1}})^{-1} \bar{\mathbf{U}}^{c_b*}_{\chi_{k+1}}. \nonumber   
\end{align}
For simplicity of exposition, in Section~\ref{Weighted Incremental DMDc} and~\ref{Windowed Incremental DMDc}, we assumed that the SVD of the data matrix $\mathbf{X}_{k}~(\boldsymbol{\chi}_{k})$ and the augmented data matrix $\mathbf{X}^{c}_{k}~(\boldsymbol{\chi}^{c}_{k})$ is known. However, if the dimension of the input $\boldsymbol \gamma_k$ is small, or if the large dimension of {the data matrix} leads to storage concerns for two SVDs, it may be beneficial to store only the SVD of $\mathbf{X}_{k}~(\boldsymbol{\chi}_{k})$ and subsequently, use incremental SVD updates to first compute SVD of $\mathbf{X}_{k+1}~(\boldsymbol{\chi}_{k+1})$ and then compute the SVD of $\mathbf{X}^{c}_{k+1}~(\boldsymbol{\chi}^{c}_{k+1})$. The former can be accomplished using the incremental SVD updates described in Section~\ref{Incremental SVD}, while the latter update is described in Appendix~\ref{app_Incremental DMD with input}.

\subsection{Numerical Illustration for the Incremental DMDc Algorithm}\label{Numerical Example}
\begin{figure}[ht!]
\centering
\subfloat[Weighted Incremental DMDc]
{
\includegraphics[width=0.5\linewidth]{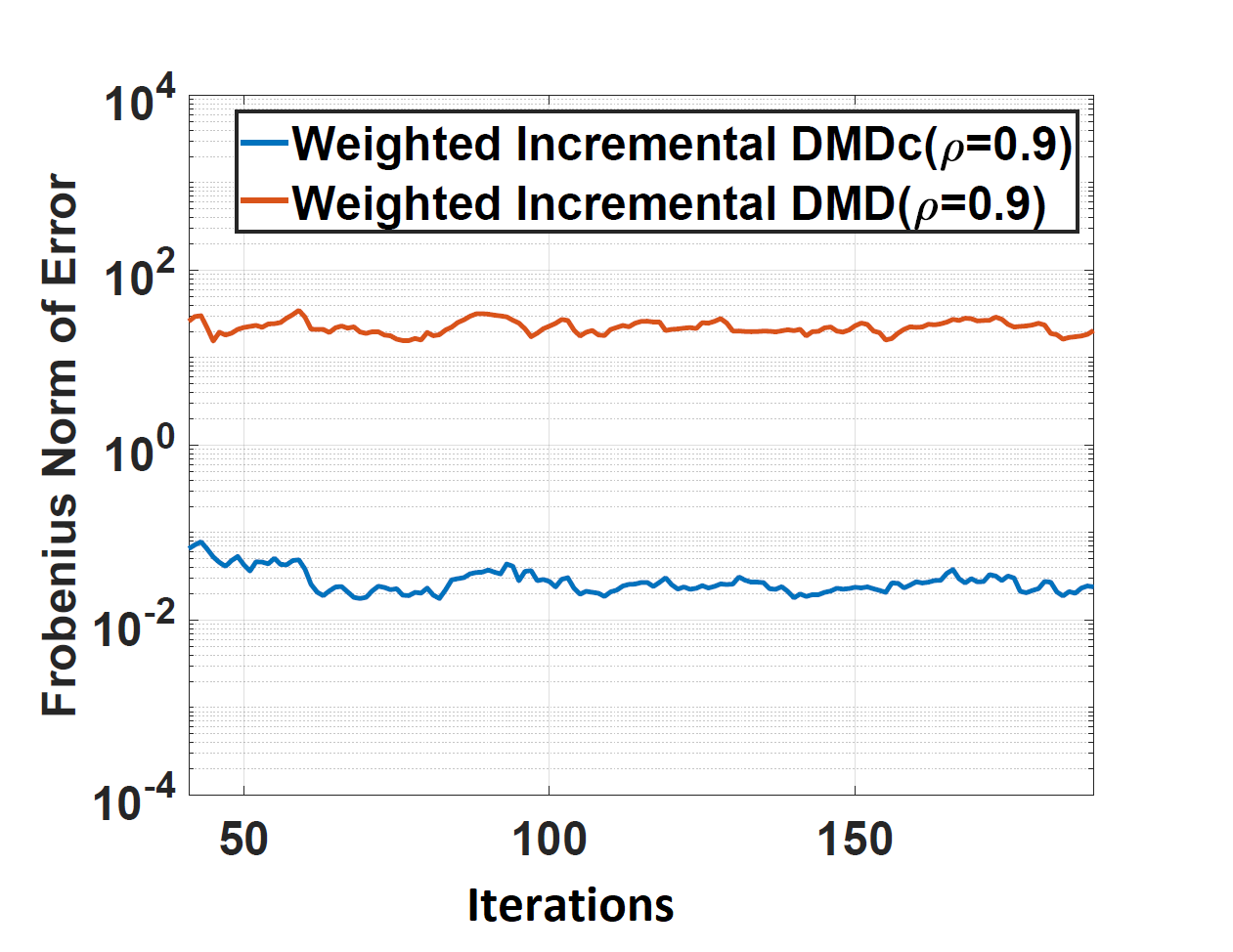}
} 
\subfloat[Windowed Incremental DMDc]
{
\includegraphics[width=0.5\linewidth]{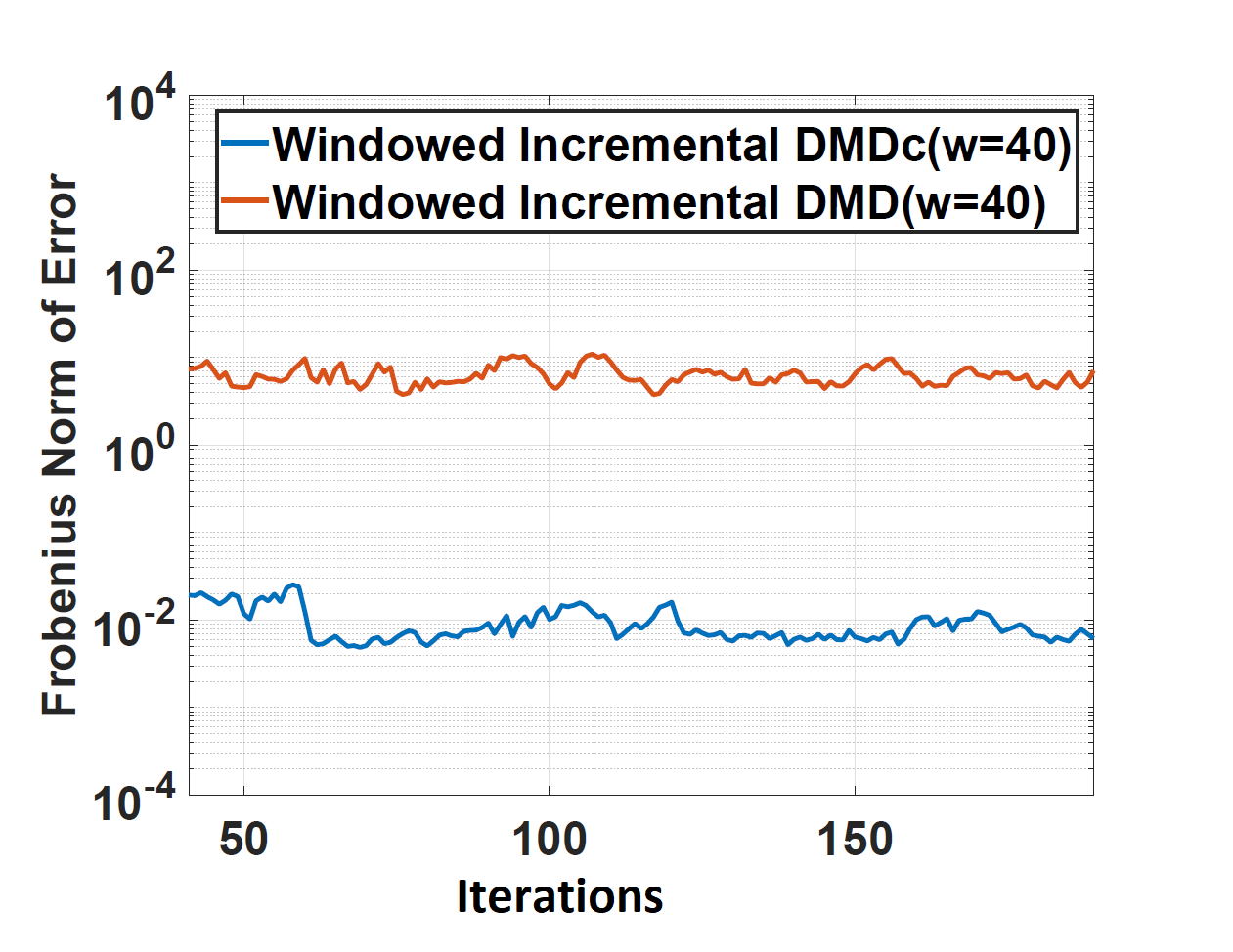}
} 
\caption{The Frobenius norm of prediction error for a future-window of 10 samples using (a) weighted incremental DMD (red line) and weighted incremental DMDc(blue line), and (b) windowed incremental DMD (red line) and windowed incremental DMDc (blue line).}
\label{future_frob_error}
\end{figure}

In this section, we illustrate the efficiency of incremental DMDc algorithms. To this end, we generate a time-varying linear dynamical system as follows. We generate random linear discrete time invariant system matrices $\mathbf{A} \in \mathbb{R}^{n \times n}$ and $\mathbf{B} \in \mathbb{R}^{n \times l}$ using MATLAB function \texttt{drss}. These system matrices were used to generate the following time-varying linear system:
% generated $\mathbf{A} \mathbf{B}$ and $\mathbf{\Gamma}$ matrices are used to construct the following system:
\begin{equation} \label{random dynamics}
    \mathbf{x}_{k+1}= \mathbf{A}_{k} \mathbf{x}_{k}+ \mathbf{B}_{k} \boldsymbol{\gamma}_{k} 
\end{equation}
where $\mathbf{A}_{k}= (1+\epsilon \sin{\omega k})\mathbf{A}$, and $\mathbf{B}_{k}= (1+\epsilon \sin{\omega k})\mathbf{B}$ and $\boldsymbol{\gamma}_k \in \real^l$ is selected as an i.i.d. sequence of standard Gaussian random vector. The system in \eqref{random dynamics} is used to run a simulation for $m\gg n$ time steps, to obtain two datasets $\mathbf{X}, \mathbf{Y} \in \mathbb{R}^{n \times (m-1)}$, with $\omega =1$, $\epsilon = 0.001$, $m=200$, $n=20$, and $l=2$.
Initial models are estimated using DMD and DMDc algorithms using  initial windows of $\mathbf{X} = \begin{bmatrix} \mathbf{x}_1, \cdots, \mathbf{x}_{40}\end{bmatrix}, \mathbf{Y} = \begin{bmatrix} \mathbf{y}_1, \cdots, \mathbf{y}_{40}\end{bmatrix}$, and $\mathbf{\Gamma} = \begin{bmatrix} \boldsymbol \gamma_1, \cdots, \boldsymbol \gamma_{40}\end{bmatrix}$.
The weighted ($\rho = 0.9$) and windowed ($w=40$ samples) incremental DMD and DMDc algorithms are applied at each iteration on the generated data $\mathbf{x}_k, \mathbf{y}_k, \boldsymbol \gamma_k$, for $k \in \{41, \cdots, 200 \}$, to {update the initial DMD model} and DMDc model. The two models are also used to predict $10$ future state vectors at each iteration.
The Frobenius norm of the prediction error for incremental DMDc (blue line) and the incremental DMD (red line) are shown in Figure \ref{future_frob_error}. The incremental DMDc models have  higher prediction accuracy because of its ability to characterize the relationship between the states and the control input which is vital for any predictive model. The windowed incremental DMDc algorithm appears to have smaller prediction error than the weighted incremental DMDc algorithm. 

\section{Conclusion and Future Direction} \label{sec:conclusions}
In this paper, we developed algorithms for incremental computation of time-varying dynamic mode decomposition for autonomous and non-autonomous systems. In contrast to existing algorithms, these algorithms rely on incremental singular value decomposition to update {singular values of the data matrix} and allow for {computation of a reduced order model} at each time step. These algorithms are particularly useful for the cases in which the data matrix is singular and incremental matrix inversion based algorithms cannot be applied. We applied the proposed algorithms to an EEG dataset associated with error-related potentials and showed the efficacy of the algorithms in terms of predicting the future EEG signal. We also illustrated that the principal DMD modes obtained were consistent with the EEG activity seen in the brain. 

There are several interesting future research directions for this work. First, the proposed approach can be adapted to apply the incremental updates on a nonlinear mapping of the measurements. This could be done in the spirit of extended dynamic mode decomposition proposed in~\cite{williams2015data}. Such extension may further improve the predictive power of the computed models. Second, it would be interesting to conduct human-in-the-loop experiments in which the EEG data is used in real-time to computed a DMD-based model of human performance and the control {is designed} to improve the performance.

\section*{Code} The MATLAB code used to generate the numerical results presented in this paper is available at \\
\url{https://github.com/MSU-dcypherlab/Incremental-DMD-for-EEG-Data.git}.   

% \clearpage
\appendix

\section{Proof of Proposition~\ref{weighted incremental SVD  proposition }}\label{proof of Proposition 1}
The incremented dataset $\mathbf{X}_{k+1}$  can be represented in term of $\mathbf{X}_{k}$  using the following additive updating formula
\begin{align} 
\mathbf{X}_{k+1} &=\rho\begin{bmatrix} \mathbf{X}_{k} & \mathbf{0} \end{bmatrix} + \mathbf{x}_{k+1} \mathbf{z}_{k+1}^{\text{T}}, \nonumber \\
&=
\rho \mathbf{U}_{x_k} \mathbf{\Sigma}_{x_k}\begin{bmatrix} \mathbf{V}^*_{x_k} & \mathbf{0} \end{bmatrix} + \mathbf{x}_{k+1} \mathbf{z}_{k+1}^{\text{T}}  \nonumber\\
&=
\begin{bmatrix}
\rho \mathbf{U}_{x_k} & \mathbf{x}_{k+1}
\end{bmatrix}
\begin{bmatrix}
\mathbf{ \Sigma}_{k} & \mathbf{0} \\
\mathbf{0} & 1
\end{bmatrix}
\begin{bmatrix}
\begin{bsmallmatrix} \mathbf{V}_{k} \\ \mathbf{0} \end{bsmallmatrix}& \mathbf{z}_{k+1}
\end{bmatrix}^{\ast} \nonumber \\
 &=
\rho \mathbf{U}_{x_k}
\begin{bmatrix}
\mathbf{I} & \rho^{-1} \mathbf{U}_{x_k}^* \mathbf{x}_{k+1}
\end{bmatrix}
\begin{bmatrix}
\mathbf{ \Sigma}_{k} & \mathbf{0} \\
\mathbf{0} & 1
\end{bmatrix}
\begin{bmatrix}
\mathbf{{V}}_{k}& \mathbf{0} \\
\mathbf{0}           & 1
\end{bmatrix}^{\ast} \nonumber \\
&=
\rho \mathbf{U}_{x_k}
\begin{bmatrix}
\mathbf{\Sigma}_{x_k} & \rho^{-1} \mathbf{U}_{x_k}^* \mathbf{x}_{k+1}
\end{bmatrix}
\begin{bmatrix}
\mathbf{{V}}_{x_k}& \mathbf{0} \\
\mathbf{0}           & 1
\end{bmatrix}^{\ast}.  \label{weighted1 X(k+1)}
\end{align}
where $\mathbf{z}_{k+1}= \begin{bmatrix} 0 & 0 &\cdots& 1 \end{bmatrix}^{\text{T}} \in \mathbb{R}^ {k+1}$.

Let the SVD of $\mathbf{S}_k \triangleq \begin{bmatrix}\mathbf{\Sigma}_{x_k} & \rho^{-1} \mathbf{U}_{x_k}^* \mathbf{x}_{k+1} \end{bmatrix}=\mathbf{U}_{s_k} \mathbf{\Sigma}_{s_k} \mathbf{V}^*_{s_k}$ with $\mathbf{U}_{s_k} \in \mathbb{C}^{n \times n}$, $\mathbf{\Sigma}_{s_k} \in\mathbb{C}^{n\times n}$, and $\mathbf{V}_{s_k} \in\mathbb{C}^{(n+1)\times n}$. 
Then, $\mathbf{V}_{s_k}$ can be partitioned as $\mathbf{V}_{s_k} =\begin{bmatrix}\mathbf{V}_{s_{k,1}} \\ \mathbf{v}_{s_{k,2}} \end{bmatrix}$ such that $\mathbf{v}_{s_{k,2}}  \in\mathbb{C}^{1 \times n}$ is orthogonal to $\mathbf{V}_{s_{k,1}} \in\mathbb{C}^{n\times n}$, and  $\mathbf{X}_{k+1}$ in \eqref{weighted X(k+1)} can be written as:
\begin{equation}
\mathbf{X}_{k+1}=\rho \mathbf{U}_{x_k} \mathbf{U}_{s_k} \mathbf{\Sigma}_{s_k} \begin{bmatrix}
\mathbf{V}_{x_k} \mathbf{V}_{s_{k,1}} \\
\mathbf{v}_{s_{k,2}}
\end{bmatrix}^\ast. \nonumber
\end{equation}
Thus, the SVD of $\mathbf{X}_{k+1}=\mathbf{U}_{x_{k+1}} \mathbf{\Sigma}_{x_{k+1}} \mathbf{V}^*_{x_{k+1}}$ is given by
\begin{eqnarray}
\mathbf{U}_{x_{k+1}} = \mathbf{U}_{x_k} \mathbf{U}_{s_k},~
\mathbf{\Sigma}_{k+1}=\rho \mathbf{\Sigma}_{s_k}
,~\text{and}~
\mathbf{V}_{x_{k+1}}=
\begin{bmatrix}
\mathbf{V}_{x_k} \mathbf{V}_{s_{k,1}} \\
\mathbf{v}_{s_{k,2}}
\end{bmatrix}.\nonumber
\end{eqnarray}

\section{Proof of Proposition~\ref{windowed incremental SVD  proposition}}\label{proof of Proposition 2}
%\label{windowed incremental SVD  proposition}
% In particular, in the decremental update, the dataset is updated to 
Consider the updated dataset
\begin{align}
\boldsymbol{\acute{\tilde{\chi}}}_{k} &=\boldsymbol{\chi}_{k}-\mathbf{x}_{k-w+1} \mathbf{z}_{1}^{\text{T}} \nonumber \\
& = \mathbf{U}_{\chi_k} \mathbf{U}_{\chi_k}^{*}(\boldsymbol{\chi}_{k}- \mathbf{x}_{k-w+1} \mathbf{z}_{1}^{\text{T}}) \mathbf{V}_{\chi_k} \mathbf{V}_{\chi_k}^{*} \nonumber\\
&= \mathbf{U}_{\chi_k}(\mathbf{\Sigma}_{\chi_k}- \mathbf{U}_{\chi_k}^{*}\mathbf{x}_{k-w+1} \mathbf{z}_{1}^{\text{T}} \mathbf{V}_{\chi_k})\mathbf{V}_{\chi_k}^{*}, \label{dataset decremental update11}
\end{align}
where $\boldsymbol{\acute{\tilde{\chi}}}_{k} \triangleq \begin{bmatrix} 0 & \mathbf{x}_{k-w+2} & \cdots & \mathbf{x}_{k} \end{bmatrix}$ is the $\boldsymbol{\acute{\chi}}_{k}$ dataset padded with $n$-dimensional zero column, and $\mathbf{z}_{1}= \begin{bmatrix} 1 & 0 &\cdots& 0 \end{bmatrix}^{\text{T}} \in \mathbb{R}^w$.

Let the SVD of $\mathbf{\acute{S}}_k \triangleq  \mathbf{\Sigma}_{\chi_k}- \mathbf{U}_{\chi_k}^{*}\mathbf{x}_{k-w+1} \mathbf{z}_{1}^{\text{T}} \mathbf{V}_{\chi_k}$ be
\begin{equation} \label{dec SVD}
\mathbf{\acute{S}}_k= \mathbf{U}_{\acute{s}_k} 
\begin{bmatrix}
{\mathbf{\Sigma}}_{\acute{s}_k} &  \mathbf{0}_{q \times (w-q)}
\end{bmatrix}
\begin{bmatrix} \mathbf{V}_{\acute{s}_k} & \tilde{\mathbf{V}}_{\acute{s}_k} \end{bmatrix} ^\ast,
\end{equation}
where $\mathbf{U}_{\acute{s}_k} \in \mathbb{C}^{q \times q}$, ${\mathbf{\Sigma}}_{\acute{s}_k} \in \mathbb{C}^{q \times q}$, $\mathbf{V}_{\acute{s}_k}\in \mathbb{C}^{w \times q}$ and $\tilde{\mathbf{V}}_{\acute{s}_k}\in \mathbb{C}^{w \times (w-q)}$. 
% are two subspaces which constitute the right singular vector space for $\mathbf{\acute{S}}$ such that the collection of column vectors of  $\tilde{\mathbf{V}}_{\acute{{s}}}$ is a basis for the null space $\text{Ker}\mathbf{\acute{S}}$, and $\mathbf{x}_{k-w+1} \in \text{Ran}\tilde{\mathbf{V}}_{\acute{{s}}}$.
Let $\mathbf{V}_{\chi_k}$ be partitioned into $\mathbf{V}_{\chi_k}=
\begin{bmatrix}
	\mathbf{v}_{\chi_{k,1}}\\
	\mathbf{V}_{\chi_{k,2}}
	\end{bmatrix}$, 
where $\mathbf{v}_{\chi_{k,1}} \in \mathbb{C}^{1 \times w}$ and $\mathbf{V}_{\chi_{k,2}} \in \mathbb{C}^{(w-1) \times w}$. 

Substituting the SVD from equation \eqref{dec SVD} into equation\eqref{dataset decremental update11}, the dataset $\mathbf{\acute{{\tilde{X}}}}_{k}$ can be expressed as:
\begin{align*}
   %\label{dataset decremental update22}
	\boldsymbol{\acute{\tilde{\chi}}}_{k} &= 
	\left(\mathbf{U}_{\chi_k} \mathbf{U}_{\acute{s}_k}\right) 
   \begin{bmatrix}
    {\mathbf{\Sigma}}_{\acute{s}_k} & \mathbf{0}_{q \times (w-q)}
   \end{bmatrix}
	\left(\mathbf{V}_{\chi_k} \begin{bmatrix}  \mathbf{V}_{\acute{s}_k} & \tilde{\mathbf{V}}_{\acute{s}_k} \end{bmatrix} \right)^{\ast} \\
	%%% 
	%\label{dataset decremental update3}
&=
\left(\mathbf{U}_{\chi_k} \mathbf{U}_{\acute{s}_k}\right)  
    \begin{bmatrix}
    {\mathbf{\Sigma}}_{\acute{s}_k} & \mathbf{0}_{q \times (w-q)}
   \end{bmatrix}
	\left( \begin{bmatrix}
	\mathbf{v}_{\chi_{k,1}}\\
	\mathbf{V}_{\chi_{k,2}}
	\end{bmatrix} 
    \begin{bmatrix} 
    \mathbf{V}_{\acute{s}_k} & \tilde{\mathbf{V}}_{\acute{s}_k}
    \end{bmatrix} \right)^\ast \\
	& =
\left(\mathbf{U}_{\chi_k} \mathbf{U}_{\acute{s}_k}\right)  
    \begin{bmatrix}
   {\mathbf{\Sigma}}_{\acute{s}_k} & \mathbf{0}_{q \times (w-q)}
    \end{bmatrix}
	 \begin{bmatrix}
	\mathbf{v}_{\chi_{k,1}} \mathbf{V}_{\acute{s}_k} & \mathbf{v}_{\chi_{k,1}} \tilde{\mathbf{V}}_{\acute{s}_k} \\
		\mathbf{V}_{\chi_{k,2}} \mathbf{V}_{\acute{s}_k} & 	\mathbf{V}_{\chi_{k,2}} \tilde{\mathbf{V}}_{\acute{s}_k}
	\end{bmatrix} ^\ast  \nonumber \\
	& =\begin{bmatrix}
	\mathbf{U}_{\chi_k} \mathbf{U}_{\acute{s}_k} {\mathbf{\Sigma}}_{\acute{s}_k} \mathbf{V}_{\acute{s}_k}^* \mathbf{v}_{\chi_{k,1}}^*   & \mathbf{U}_{\chi_k} \mathbf{U}_{\acute{s}_k} {\mathbf{\Sigma}}_{\acute{s}_k} \mathbf{V}_{\acute{s}_k}^* \mathbf{V}_{\chi_{k,2}}^* 
	\end{bmatrix}.\nonumber
\end{align*}
Since, $\boldsymbol{\acute{\tilde{\chi}}}_{k} = \begin{bmatrix} \mathbf{0} & 	\boldsymbol{\acute{{\chi}}}_{k} \end{bmatrix}$, it follows that $\mathbf{U}_{\chi_k} \mathbf{U}_{\acute{s}_k} {\mathbf{\Sigma}}_{\acute{s}_k} \mathbf{V}_{\acute{s}_k}^* \mathbf{v}_{\chi_{k,1}}^* = \mathbf{0}$ and $\boldsymbol{\acute{{\chi}}}_{k}= \mathbf{U}_{\chi_k} \mathbf{U}_{\acute{s}_k} {\mathbf{\Sigma}}_{\acute{s}_k} \mathbf{V}_{\acute{s}_k}^* \mathbf{V}_{\chi_{k,2}}^*  $. 
Thus, the SVD of $\boldsymbol{\acute{\chi}}_{k} = \mathbf{U}_{\acute{\chi}_{k}}  {\mathbf{\Sigma}}_{\acute{\chi}_{k}}  {\mathbf{V}}_{\acute{\chi}_{k}}^*$ is 
\begin{eqnarray}
 \mathbf{U}_{\acute{\chi}_{k}}= \mathbf{U}_{\chi_k} \mathbf{U}_{\acute{s}_k}, ~
 {\mathbf{\Sigma}}_{\acute{\chi}_{k}}={\mathbf{\Sigma}}_{\acute{s}_k},~ \text{and} ~
 {\mathbf{V}}_{\acute{\chi}_{k}}= 	\mathbf{V}_{\chi_{k,2}} \mathbf{V}_{\acute{s}_k}.
\end{eqnarray}

A noteworthy property of the above decomposition that we will use in the later developments is that $\mathbf{v}_{\chi_{k,1}} \mathbf{V}_{\acute{s}_k} = \mathbf{0}$. To establish this property, we note that since $\mathbf{U}_{\chi_k} \mathbf{\acute{S}}_k \mathbf{v}_{\chi_{k,1}}^* = \mathbf{0}$, $\mathbf{v}^*_{\chi_{k,1}} \in \text{Ker}(\mathbf{U}_{\chi_k} \mathbf{\acute{S}}_k)$. Furthermore, since  $\mathbf{U}_{\chi_k}$ is full rank, then $\mathbf{v}^*_{\chi_{k,1}} \in \text{Ker}(\mathbf{\acute{S}}_k) = \text{span}(\tilde{\mathbf{V}}_{\acute{s}_k})$, and consequently
\begin{align}\label{orthogonality}
   \mathbf{v}_{\chi_{k,1}}  \mathbf{V}_{\acute{s}_k}  = \mathbf{0}.
\end{align}

\section{Proof of Theorem~\ref{weighted DMD operator(k+1) theorem}}\label{proof of Theorem 1}
Let the SVD of $\mathbf{X}_{k+1}= \mathbf{U}_{x_{k+1}} {\mathbf{\Sigma}_{x_{k+1}}} \mathbf{V}^{*}_{x_{k+1}}$ be calculated by the incremental SVD update~\eqref{weighted2 updated SVD factors}. Then, the time-varying DMD operator~\eqref{unique lse Ak} at sampling time $t_{k+1}$ is:
\begin{eqnarray}\label{weighted DMD operator}
\mathbf{A}^{\rho}_{k+1} 
&=&\mathbf{Y}_{k+1}\mathbf{V}_{x_{k+1}} {\mathbf{\Sigma}_{x_{k+1}}^{-1}} \mathbf{U}_{x_{k+1}}^* \nonumber\\
&=&\begin{bmatrix}\rho \mathbf{Y}_k  & \mathbf{y}_{k+1}\end{bmatrix}  \mathbf{V}_{x_{k+1}} {\mathbf{\Sigma}_{x_{k+1}}^{-1}} \mathbf{U}_{x_{k+1}}^*.
\end{eqnarray}
Substituting the incrementally computed $\mathbf{V}_{k+1}$  from \eqref{weighted2 updated SVD factors} into the DMD operator~\eqref{weighted DMD operator} yields:
\begin{align} 
\mathbf{A}^{\rho}_{k+1} =& \begin{bmatrix}\rho \mathbf{Y}_k  & \mathbf{y}_{k+1}\end{bmatrix}  \begin{bmatrix}
\mathbf{V}_{x_k} \mathbf{V}_{s_{k,1}}\\
\mathbf{v}_{s_{k,2}}
\end{bmatrix}\mathbf{\Sigma}_{x_{k+1}}^{-1} \mathbf{U}_{x_{k+1}}^{*} \nonumber\\
=& \left(\rho \mathbf{Y}_k\mathbf{V}_{x_{k}} \mathbf{V}_{s_{k,1}}+ \mathbf{y}_{k+1}\mathbf{v}_{s_{k,2}}\right)\mathbf{\Sigma}_{x_{k+1}}^{-1} \mathbf{U} ^*_{x_{k+1}} \nonumber\\
=& \Big(\rho \mathbf{Y}_k\mathbf{V}_{x_k} 
\underset{= I_n}{\underbrace{(\mathbf{\Sigma}_{x_{k}}^{-1} \rho^{-1} \mathbf{U}^*_{x_{k}} \rho \mathbf{U}_{x_{k}} \mathbf{\Sigma}_{x_{k}})}}
\mathbf{V}_{s_{k,1}}+ \mathbf{y}_{k+1}\mathbf{v}_{s_{k,2}}\Big)\mathbf{\Sigma}_{x_{k+1}}^{-1} \mathbf{U}^*_{x_{k+1}}\nonumber\\ 
=& \left(\mathbf{A}^{\rho} _{k} \rho \mathbf{U}_{x_k} \mathbf{\Sigma}_{x_k} \mathbf{V}_{s_{k,1}}+ \mathbf{y}_{k+1}\mathbf{v}_{s_{k,2}}\right)\mathbf{\Sigma}_{x_{k+1}}^{-1} \mathbf{U}^*_{x_{k+1}} \nonumber\\  
=& \mathbf{A}^{\rho}_{k}-\mathbf{A}^{\rho}_{k}+\left(\mathbf{A}^{\rho}_{k} \rho \mathbf{U}_{x_k} \mathbf{\Sigma}_{x_k} \mathbf{V}_{s_{k,1}}+ \mathbf{y}_{k+1}\mathbf{v}_{s_{k,1}}\right)\mathbf{\Sigma}_{x_{k+1}}^{-1} \mathbf{U}^*_{x_{k+1}} \label{eq:dmd-op-3}\\  
=& \mathbf{A}^{\rho}_{k}-\mathbf{A}^{\rho}_{k} 
\underset{= I_n}{\underbrace{(\mathbf{U}_{x_{k+1}}\mathbf{\Sigma}_{x_{k+1}}\mathbf{\Sigma}_{x_{k+1}}^{-1} \mathbf{U}^*_{x_{k+1}})}}
+  \mathbf{A}^{\rho}_{k} \rho \mathbf{U}_{x_{k}} \mathbf{\Sigma}_{x_{k}} \mathbf{V}_{s_{k,1}}\mathbf{\Sigma}_{x_{k+1}}^{-1} \mathbf{U}^*_{x_{k+1}}+ \mathbf{y}_{k+1}\mathbf{v}_{s_{k,2}}\mathbf{\Sigma}_{x_{k+1}}^{-1} \mathbf{U}^*_{x_{k+1}} \nonumber \\  
=&\mathbf{A}^{\rho}_{k}+\mathbf{y}_{k+1}\mathbf{v}_{s_{k,2}}\mathbf{\Sigma}_{x_{k+1}}^{-1}  \mathbf{U}^*_{x_{k+1}}-\mathbf{A}^{\rho}_{k} \left(\mathbf{U}_{x_{k+1}}\mathbf{\Sigma}_{x_{k+1}}-\rho \mathbf{U}_{x_k} \mathbf{\Sigma}_{x_k}\mathbf{V}_{s_{k,1}}\right)\mathbf{\Sigma}_{x_{k+1}}^{-1} \mathbf{U}^*_{x_{k+1}}, \label{weighted2 DMD operator}      
\end{align} 
where $I_n$ is the identity matrix of order $n$, and $\mathbf{A}^{\rho}_{k}$ has been added and subtracted in~\eqref{eq:dmd-op-3}.
The term $\left(\mathbf{U}_{x_{k+1}}\mathbf{\Sigma}_{x_{k+1}}-\rho \mathbf{U}_{x_k} \mathbf{\Sigma}_{x_k}\mathbf{V}_{s_{k,1}}\right)$ can be rewritten as
\begin{align}
\mathbf{U}_{x_{k+1}}\mathbf{\Sigma}_{x_{k+1}}-\rho \mathbf{U}_{x_k} \mathbf{\Sigma}_{x_k}\mathbf{V}_{s_{k,1}} &=\left(\mathbf{U}_{x_{k+1}}\mathbf{\Sigma}_{x_{k+1}}-\rho \mathbf{U}_{x_k} \mathbf{\Sigma}_{x_k}\mathbf{V}_{s_{k,1}}\right)\mathbf{V}_{x_{k+1}}^*\mathbf{V}_{x_{k+1}}\nonumber\\
&=\left(\mathbf{U}_{x_{k+1}}\mathbf{\Sigma}_{x_{k+1}}\mathbf{V}_{x_{k+1}}^*- \rho \mathbf{U}_{x_k} \mathbf{\Sigma}_{x_k}\mathbf{V}_{s_{k,1}}\mathbf{V}_{x_{k+1}}^*\right)\mathbf{V}_{x_{k+1}}\nonumber\\
&=\left(\mathbf{X}_{k+1}-\rho \mathbf{U}_{x_k} \mathbf{\Sigma}_{x_k}\mathbf{V}_{s_{k,1}}\begin{bsmallmatrix}
\mathbf{V}_{s_{k,1}}^{*} \mathbf{V}_{x_k}^*  & \mathbf{v}_{s_{k,2}}^{*} 
\end{bsmallmatrix}\right)\mathbf{V}_{x_{k+1}}\label{eq:nwd-3}\\
&=\left(\mathbf{X}_{k+1}- \begin{bmatrix} \rho \mathbf{U}_{x_k} \mathbf{\Sigma}_{x_k}  \mathbf{V}_{x_k}^* & \mathbf{0} \end{bmatrix} \right)\mathbf{V}_{x_{k+1}}\label{eq:nwd-4}\\
&=\left(\mathbf{X}_{k+1}- \begin{bmatrix} \rho \mathbf{X}_k & \mathbf{0} \end{bmatrix} \right)\mathbf{V}_{x_{k+1}}\nonumber\\
&=\begin{bmatrix}\mathbf{0} & \dots & \mathbf{0} & \mathbf{x}_{k+1} \end{bmatrix}\begin{bmatrix}
\mathbf{V}_{x_k} \mathbf{V}_{s_{k,1}}\\
\mathbf{v}_{s_{k,2}} 
\end{bmatrix}=\mathbf{x}_{k+1}\mathbf{v}_{s_{k,2}}. \label{new weighted data}
\end{align}
where~\eqref{eq:nwd-3} is obtained by substituting incremental update for $\mathbf{V}_{x_{k+1}}$ from \eqref{weighted2 updated SVD factors}, and~\eqref{eq:nwd-4} is obtained by using the fact that the rows in $\mathbf V_{s_k} = \begin{bmatrix}
\mathbf V_{s_{k,1}} \\ \mathbf v_{s_{k,2}}
\end{bmatrix}$ are orthogonal to each other. Substituting equation \eqref{new weighted data} into the equation \eqref{weighted2 DMD operator}:
\begin{eqnarray}
\mathbf{A}^{\rho}_{k+1}&=&\mathbf{A}^{\rho}_{k}+\mathbf{y}_{k+1}\mathbf{v}_{s_{k,2}}\mathbf{\Sigma}_{x_{k+1}}^{-1} \mathbf{U}_{x_{k+1}}^*-\mathbf{A}^{\rho}_{k}\mathbf{x}_{k+1}\mathbf{v}_{s_{k,2}} \mathbf{\Sigma}_{x_{k+1}}^{-1} \mathbf{U}_{x_{k+1}}^*\nonumber\\  
&=&\mathbf{A}^{\rho}_{k}+\left(\mathbf{y}_{k+1}-\mathbf{A}^{\rho}_{k}\mathbf{x}_{k+1}\right)\mathbf{v}_{s_{k,2}}\mathbf{\Sigma}_{x_{k+1}}^{-1} \mathbf{U}_{x_{k+1}}^*. \nonumber
\end{eqnarray}

\section{Proof of Theorem~\ref{windowed DMD operator(k+1) theorem}}\label{proof of Theorem 2}
Let the SVD of $\boldsymbol{\chi}_{k+1}=\mathbf{U}_{\chi_{k+1}} \mathbf{\Sigma}_{\chi_{k+1}}\mathbf{V}^*_{\chi_{k+1}}$ be calculated by the windowed SVD update~\eqref{windowed updated SVD factors}. Then, the time-varying DMD operator \eqref{unique lse Ak} at sampling time $t_{k+1}$ is

\begin{eqnarray}\label{windowed DMD operator1}
\mathbf{A}^{w}_{k+1} &=& \boldsymbol{\Upsilon}_{k+1} \mathbf{V}_{\chi_{k+1}} \mathbf{\Sigma}_{\chi_{k+1}}^{-1} \mathbf{U}^{*}_{\chi_{k+1}}. 
\end{eqnarray}

Substituting the incrementally computed $\mathbf{V}_{k+1}$  from \eqref{windowed updated SVD factors} into the DMD operator~\eqref{windowed DMD operator1} yields:
\begin{eqnarray} \label{windowed DMD operator2}
\mathbf{A}^{w}_{k+1} &= &\begin{bmatrix} \mathbf{y}_{k-w+2}& \cdots & \mathbf{y}_{k+1}\end{bmatrix}  
\begin{bmatrix}
\mathbf{V}_{\chi_{k,2}} \mathbf{V}_{\acute{s}_k} \mathbf{V}_{\hat{s}_{k,1}} \\
\mathbf{v}_{\hat{s}_{k,2}} 
\end{bmatrix}\mathbf{\Sigma}_{\chi_{k+1}}^{-1} \mathbf{U}^{*}_{\chi_{k+1}}\nonumber\\
&= & \begin{bmatrix} \mathbf{y}_{k-w+1}& \mathbf{y}_{k-w+2}& \cdots & \mathbf{y}_{k+1}\end{bmatrix}  
\begin{bmatrix}
\mathbf{v}_{\chi_{k,1}} \mathbf{V}_{\acute{s}_k} \mathbf{V}_{\hat{s}_{k,1}}\\
\mathbf{V}_{\chi_{k,2}} \mathbf{V}_{\acute{s}_k} \mathbf{V}_{\hat{s}_{k,1}}\\
\mathbf{v}_{\hat{s}_{k,2}} 
\end{bmatrix}\mathbf{\Sigma}_{\chi_{k+1}}^{-1} \mathbf{U}^{*}_{\chi_{k+1}}\label{window0edeq:nwd-3}\\
&= & \begin{bmatrix} \boldsymbol{\Upsilon}_{k} & \mathbf{y}_{k+1}\end{bmatrix}  
\begin{bmatrix}
\mathbf{V}_{\chi_k} \mathbf{V}_{\acute{s}_k} \mathbf{V}_{\hat{s}_{k,1}}\\
\mathbf{v}_{\hat{s}_{k,2}} 
\end{bmatrix}\mathbf{\Sigma}_{\chi_{k+1}}^{-1} \mathbf{U}^{*}_{\chi_{k+1}}\nonumber\\
&= & \left(\boldsymbol{\Upsilon}_k\mathbf{V}_{\chi_k} \mathbf{V}_{\acute{s}_k} \mathbf{V}_{\hat{s}_{k,1}}+ \mathbf{y}_{k+1}\mathbf{v}_{\hat{s}_{k,2}}\right)\mathbf{\Sigma}_{\chi_{k+1}}^{-1} \mathbf{U} ^*_{\chi_{k+1}},\label{window0edeq:nwd-5}
\end{eqnarray} 
where equation \eqref{window0edeq:nwd-3} holds because $\mathbf{v}_{\chi_{k,1}}  \mathbf{V}_{\acute{s}_k}  = \mathbf{0}$ from equation \eqref{orthogonality}.

After following the same steps presented {in the proof} of Theorem \ref{weighted DMD operator(k+1) theorem}, equation \eqref{window0edeq:nwd-5} becomes:

\begin{align} \label{windowed DMD operator3}
\mathbf{A}^{w}_{k+1} = \mathbf{A}^{w}_{k}+\mathbf{y}_{k+1}\mathbf{v}_{\hat{s}_{k,2}} \mathbf{\Sigma}_{\chi_{k+1}}^{-1} \mathbf{U} ^*_{\chi_{k+1}}-\mathbf{A}_{k} \left(\mathbf{U}_{\chi_{k+1}}\mathbf{\Sigma}_{\chi_{k+1}}-\mathbf{U}_{\chi_k} \mathbf{\Sigma}_{\chi_k}\mathbf{V}_{\acute{s}_k}\mathbf{V}_{\hat{s}_{k,1}}\right)\mathbf{\Sigma}_{\chi_{k+1}}^{-1} \mathbf{U}_{\chi_{k+1}}^*.
\end{align} 

The term $\left(\mathbf{U}_{\chi_{k+1}}\mathbf{\Sigma}_{\chi_{k+1}}-\mathbf{U}_{\chi_k} \mathbf{\Sigma}_{\chi_k}\mathbf{V}_{\acute{s}_k} \mathbf{V}_{\hat{s}_{k,1}}\right)$ can be rewritten as
\begin{align}
\mathbf{U}_{\chi_{k+1}}\mathbf{\Sigma}_{\chi_{k+1}}-\mathbf{U}_{\chi_k} \mathbf{\Sigma}_{\chi_k}\mathbf{V}_{\acute{s}_k} \mathbf{V}_{\hat{s}_{k,1}}&=\left(\mathbf{U}_{\chi_{k+1}}\mathbf{\Sigma}_{\chi_{k+1}}-\mathbf{U}_{\chi_k} \mathbf{\Sigma}_{\chi_k}\mathbf{V}_{\acute{s}_k} \mathbf{V}_{\hat{s}_{k,1}}\right)\mathbf{V}_{\chi_{k+1}}^*\mathbf{V}_{\chi_{k+1}}\nonumber\\
&=\left(\mathbf{U}_{\chi_{k+1}}\mathbf{\Sigma}_{\chi_{k+1}}\mathbf{V}_{\chi_{k+1}}^*- \mathbf{U}_{\chi_k} \mathbf{\Sigma}_{\chi_k}\mathbf{V}_{\acute{s}_k}\mathbf{V}_{\hat{s}_{k,1}}\mathbf{V}_{\chi_{k+1}}^*\right)\mathbf{V}_{\chi_{k+1}}\nonumber\\
&=\left(\boldsymbol{\chi}_{k+1}-\mathbf{U}_{\chi_k} \mathbf{\Sigma}_{\chi_k}\mathbf{V}_{\acute{s}_k}\mathbf{V}_{\hat{s}_{k,1}}\begin{bsmallmatrix}
\mathbf{V}_{\hat{s}_{k,1}}^* \mathbf{V}_{\acute{s}_k}^* \mathbf{V}_{{\chi}_{k,2}}^*  & \mathbf{v}_{\hat{s}_{k,2}}^*
\end{bsmallmatrix}\right)\mathbf{V}_{\chi_{k+1}}\label{windowedeq:nwd-3}\\
&=\left(\boldsymbol{\chi}_{k+1}- \begin{bsmallmatrix} \mathbf{U}_{\chi_k} \mathbf{\Sigma}_{\chi_k}  \mathbf{V}_{{\chi}_{k,2}}^* & \mathbf{0} \end{bsmallmatrix} \right)\mathbf{V}_{\chi_{k+1}}\label{windowedeq:nwd-4}\\
&=\left(\boldsymbol{\chi}_{k+1}- \begin{bmatrix}  \mathbf{x}_{k-w+2} &\cdots& \mathbf{x}_{k} & \mathbf{0} \end{bmatrix} \right)\mathbf{V}_{\chi_{k+1}}\nonumber\\
&=\begin{bmatrix}\mathbf{0} & \dots & \mathbf{0} & \mathbf{x}_{k+1} \end{bmatrix}\begin{bmatrix}
\mathbf{V}_{\chi_{k,2}} \mathbf{V}_{\acute{s}_k} \mathbf{V}_{\hat{s}_{k,1}} \\
\mathbf{v}_{\hat{s}_{k,2}}  
\end{bmatrix}=\mathbf{x}_{k+1}\mathbf{v}_{\hat{s}_{k,2}}, \label{new windowed data}
\end{align}
where equation \eqref{windowedeq:nwd-3} is obtained {after substituting} the windowed incremental update formula from \eqref{windowed updated SVD factors}, and \eqref{windowedeq:nwd-4} is obtained after substituting $\mathbf{V}_{\acute{s}_k} \mathbf{V}_{\acute{s}_k}^*=\mathbf{V}_{\hat{s}_{k,1}} \mathbf{V}_{\hat{s}_{k,1}}^*= \mathbf{I}$ and $\mathbf{V}_{\hat{s}_{k,1}} \mathbf{v}_{\hat{s}_{k,2}}^* = \mathbf{0}$. The substitution of \eqref{new windowed data} in \eqref{windowed DMD operator3} yields:
\begin{equation}
\mathbf{A}^w_{k+1}=\mathbf{A}^w_{k}+\left(\mathbf{y}_{k+1}-\mathbf{A}^w_{k}\mathbf{x}_{k+1}\right)\mathbf{v}_{\hat{s}_{k,2}}\mathbf{\Sigma}_{\chi_{k+1}}^{-1} \mathbf{U}_{\chi_{k+1}}^*.\nonumber
\end{equation}

\section{EEG Prediction Error for Incremental DMD and Online DMD Algorithms}\label{EEG_Pred_nrmse}
The figures below show the normalized RMS for the predicted EEG signal at channel FCz during correct and erroneous events using incremental DMD with $\sigma_{thr}= \{ 0.01,0.001 \}$, $\rho=\{ 0.1,0.2,0.4,0.8\}$ and $w=512$, and online DMD with $\rho=\{ 0.1,0.2,0.4,0.8\}$ and $w=512$.    
\begin{figure}[ht!]
\centering
\subfloat[Correct event]
{
\label{EEG pred rms1}
\includegraphics[width=0.3\linewidth]{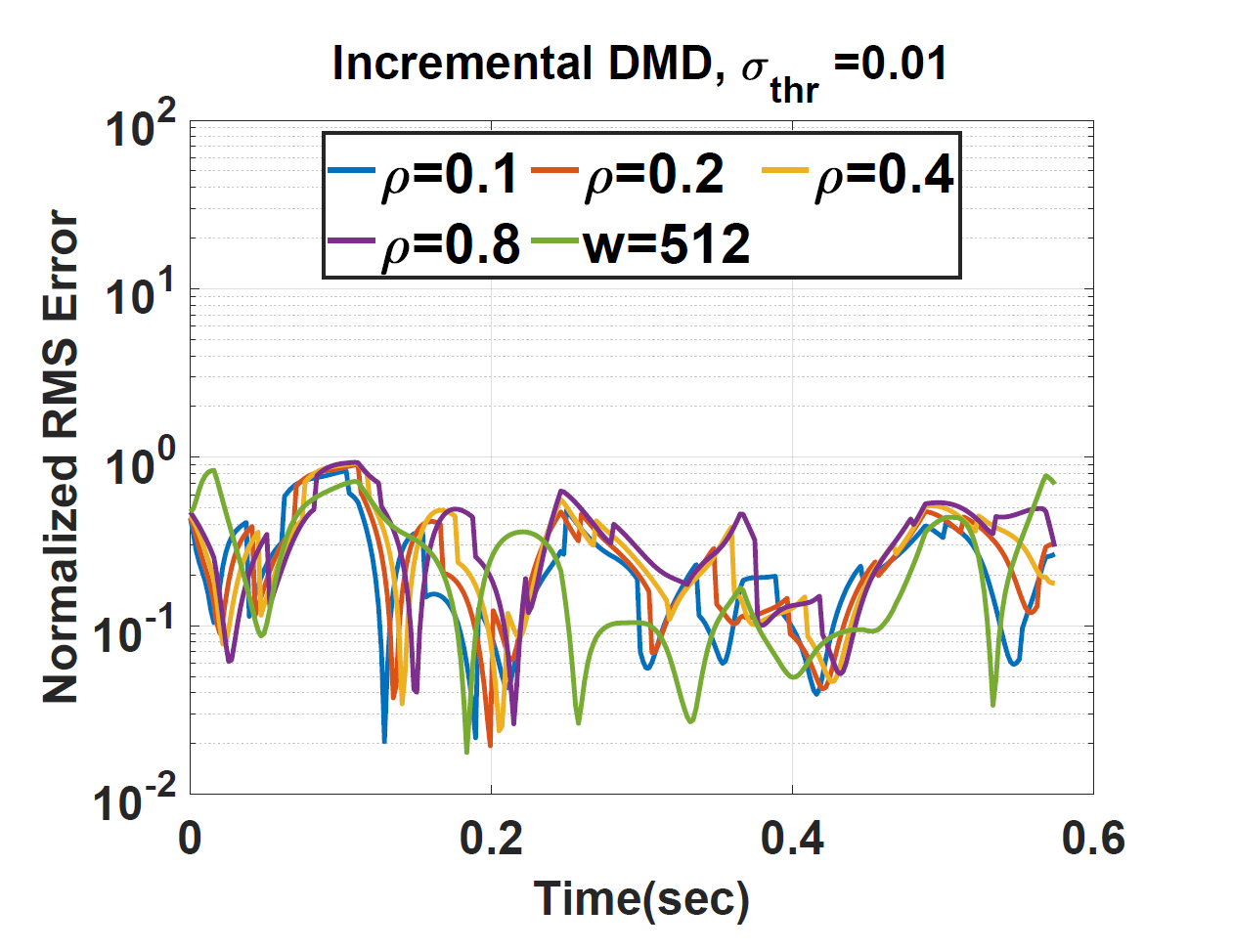}
\includegraphics[width=0.3\linewidth]{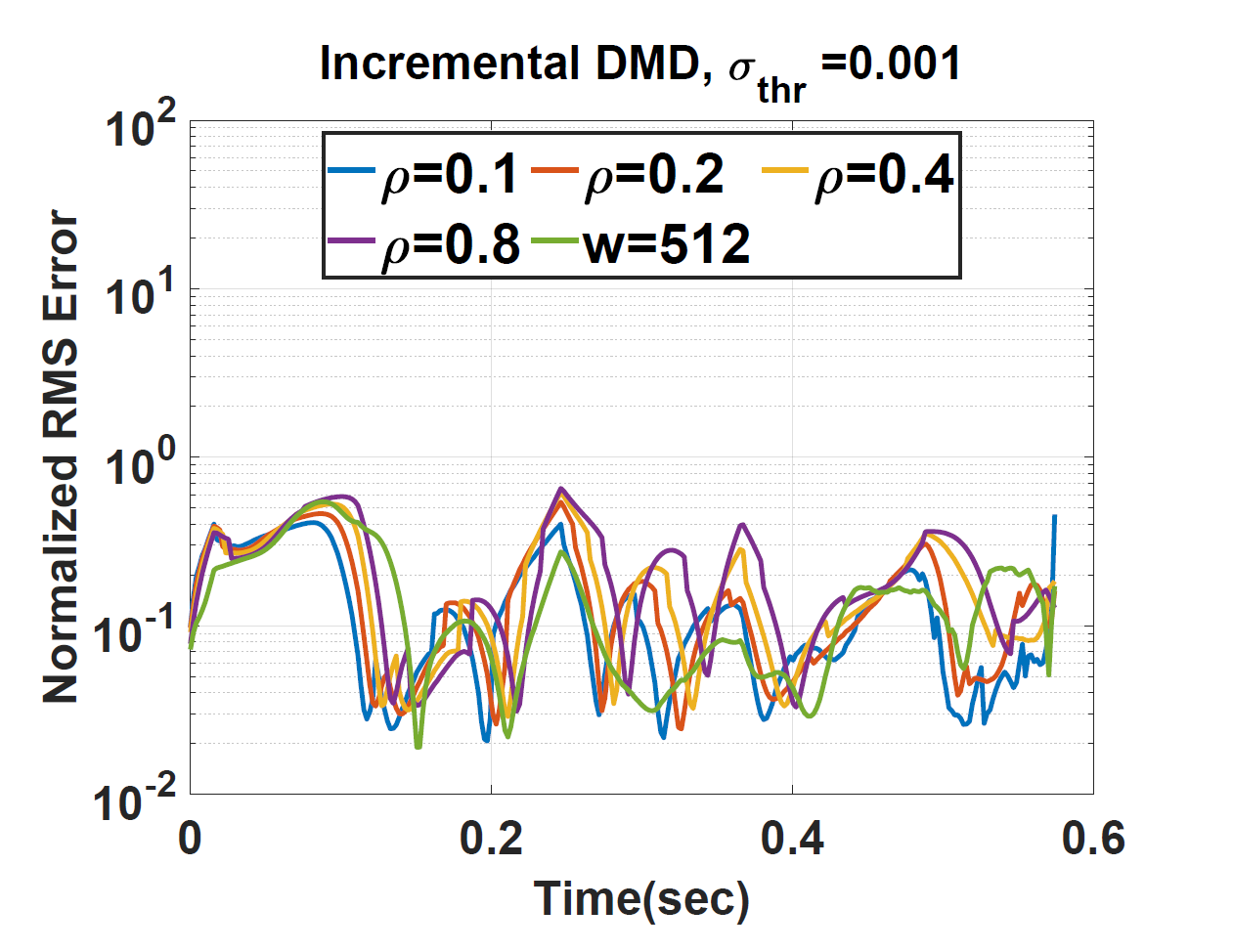}
\includegraphics[width=0.3\linewidth]{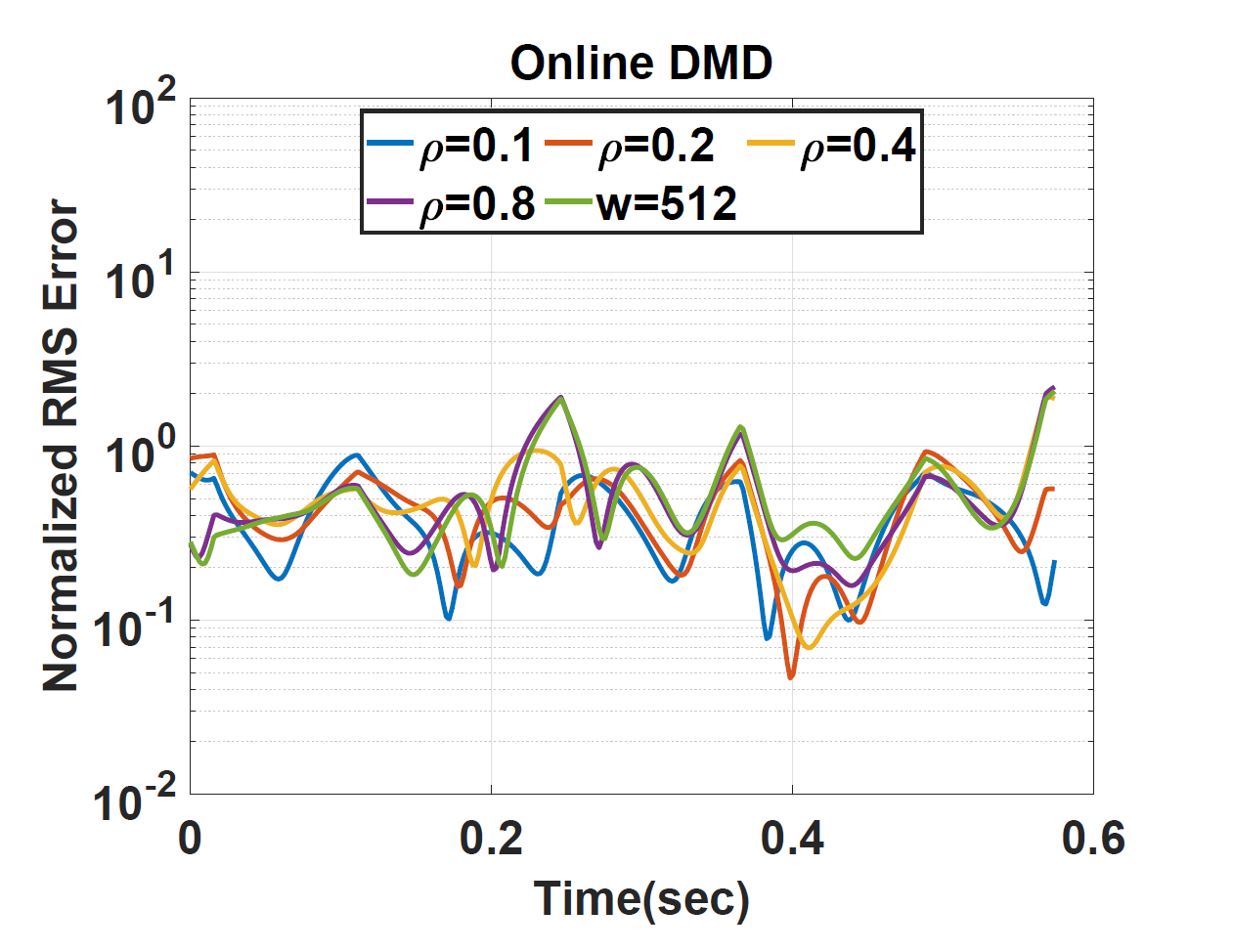}
} \\
\subfloat[Erroneous event]
{
\label{EEG pred rms2}
\includegraphics[width=0.3\linewidth]{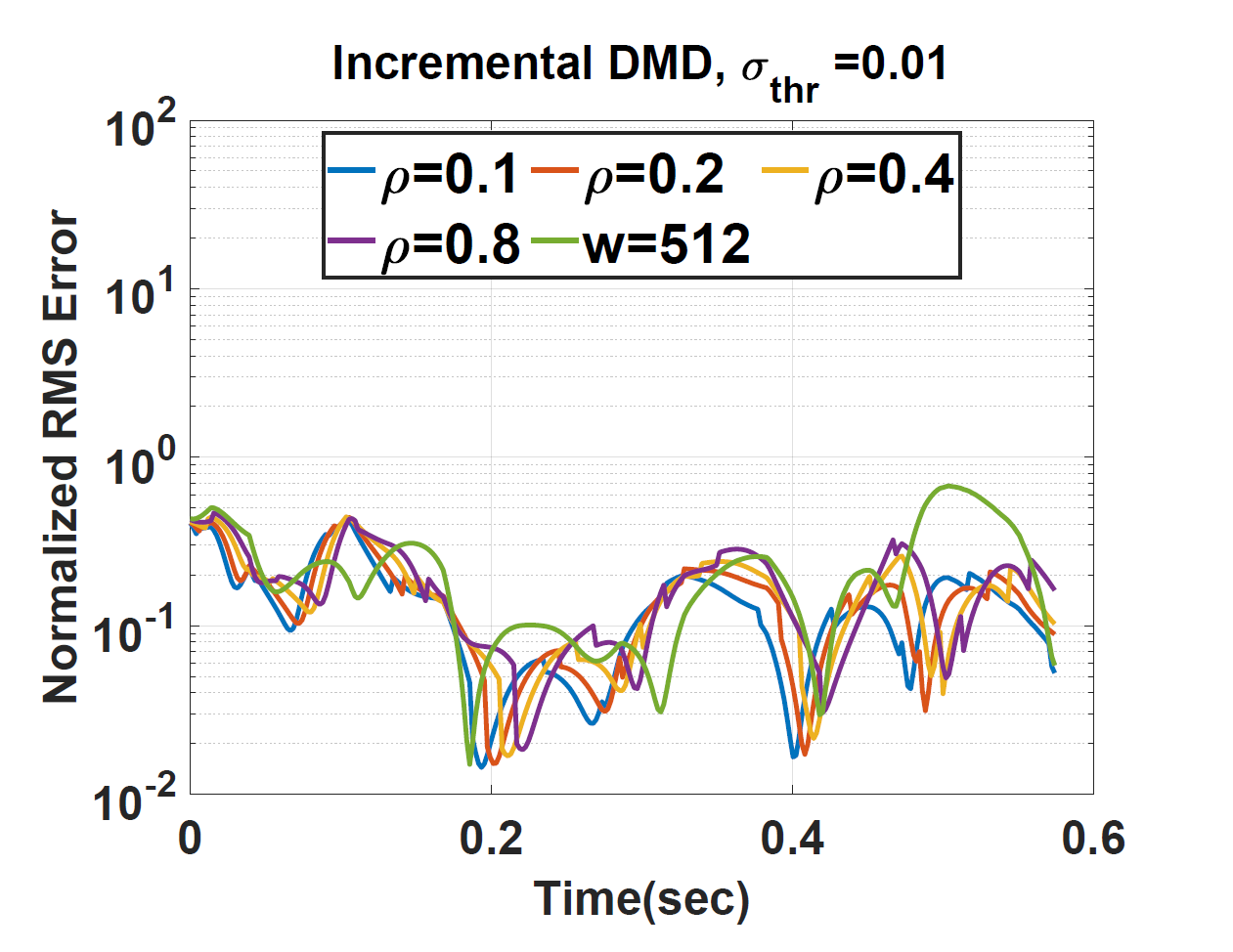}
\includegraphics[width=0.3\linewidth]{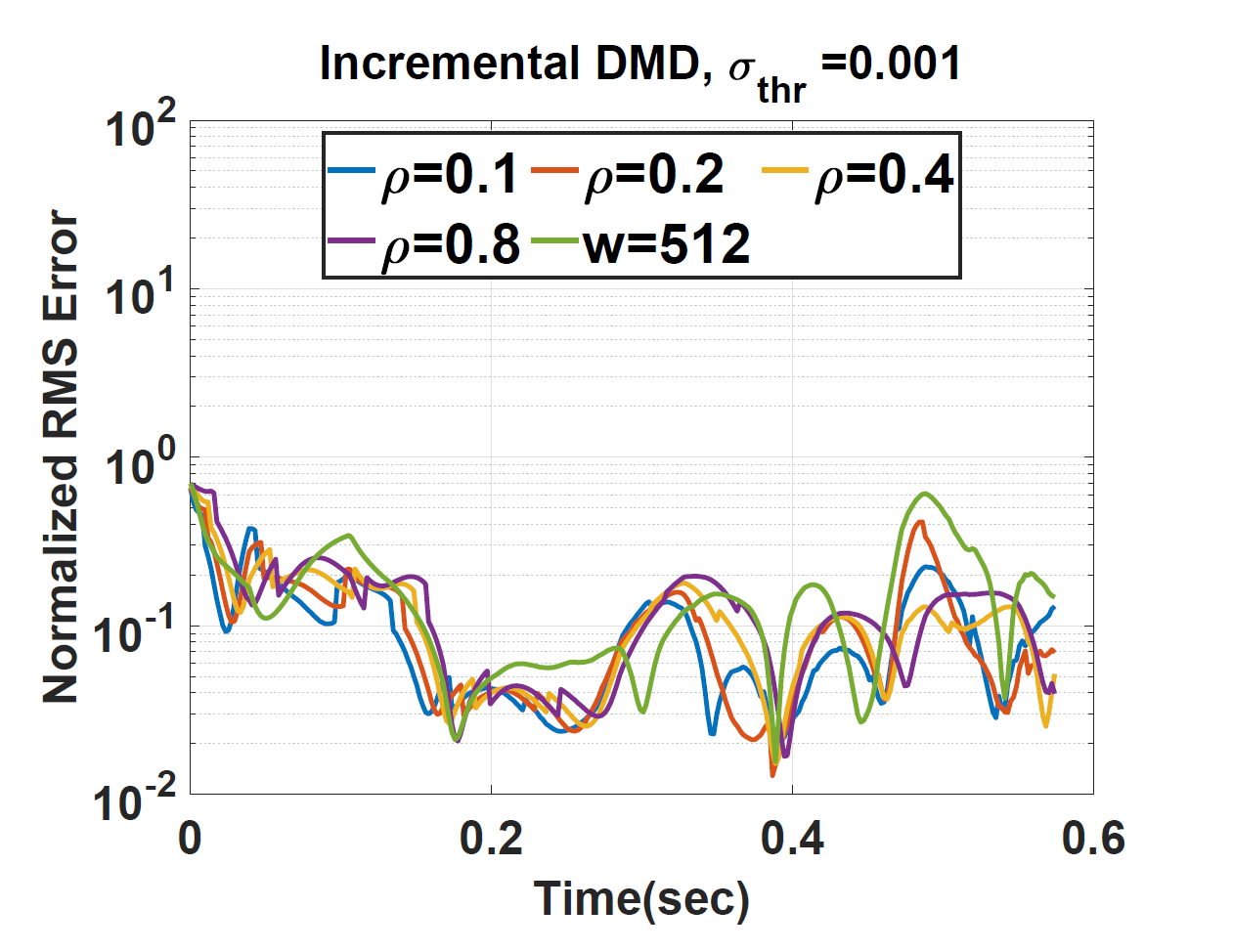}
\includegraphics[width=0.3\linewidth]{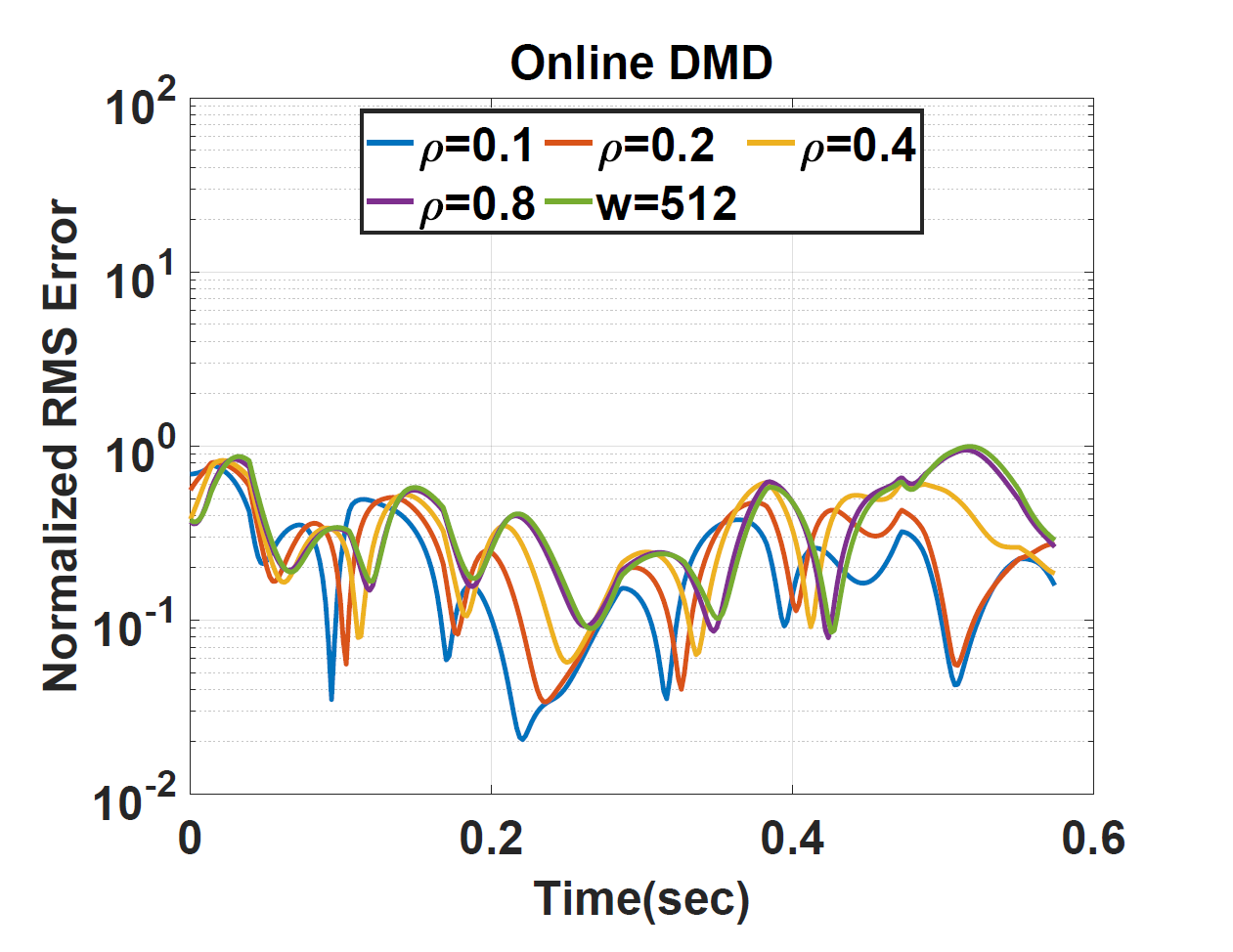}
} 
\caption{Normalized RMS error for a future-window of $64$ samples of EEG states at channel FCz using incremental DMD with $\sigma_{\text{thr}}=0.01$ (left panel), incremental DMD with $\sigma_{\text{thr}}=0.001$ (middle panel), and online DMD (right panel) for (a) correct event, and (b) erroneous event.}
\label{EEG pred rms}
\end{figure}

\section{ERP Prediction Error for Incremental DMD Algorithm with Different Initial Window Sizes}\label{app_wind DMD with different sizes}
\begin{figure}[ht!]
\centering
\subfloat[Correct Event]
{
\includegraphics[width=0.4\linewidth]{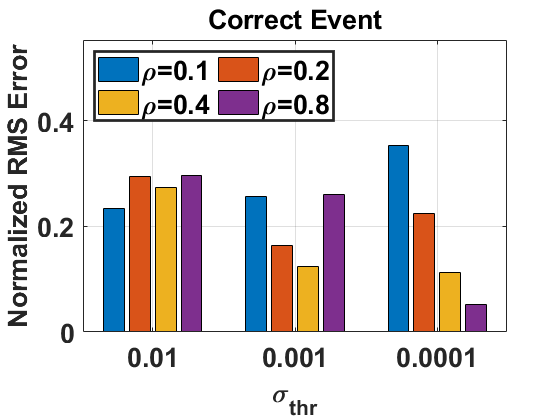}
} 
\subfloat[erroneous Event]
{
\includegraphics[width=0.4\linewidth]{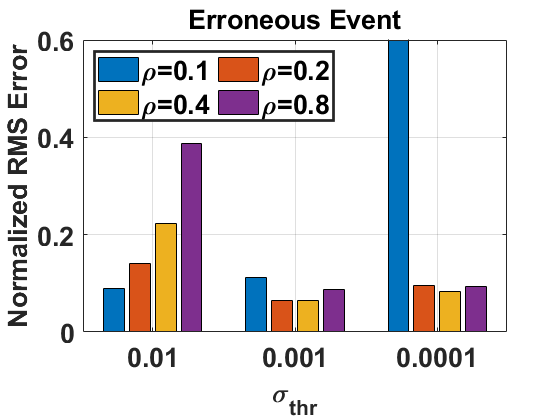}
} 
\caption{Normalized RMS error of ERP prediction using weighted incremental DMD with initial window of $128$ samples during (a) correct event, and (b) Erroneous event.}
\label{ERP_RMS_128_windows}
\end{figure}

\begin{figure}[ht!]
\centering
\subfloat[Correct Event]
{
\includegraphics[width=0.4\linewidth]{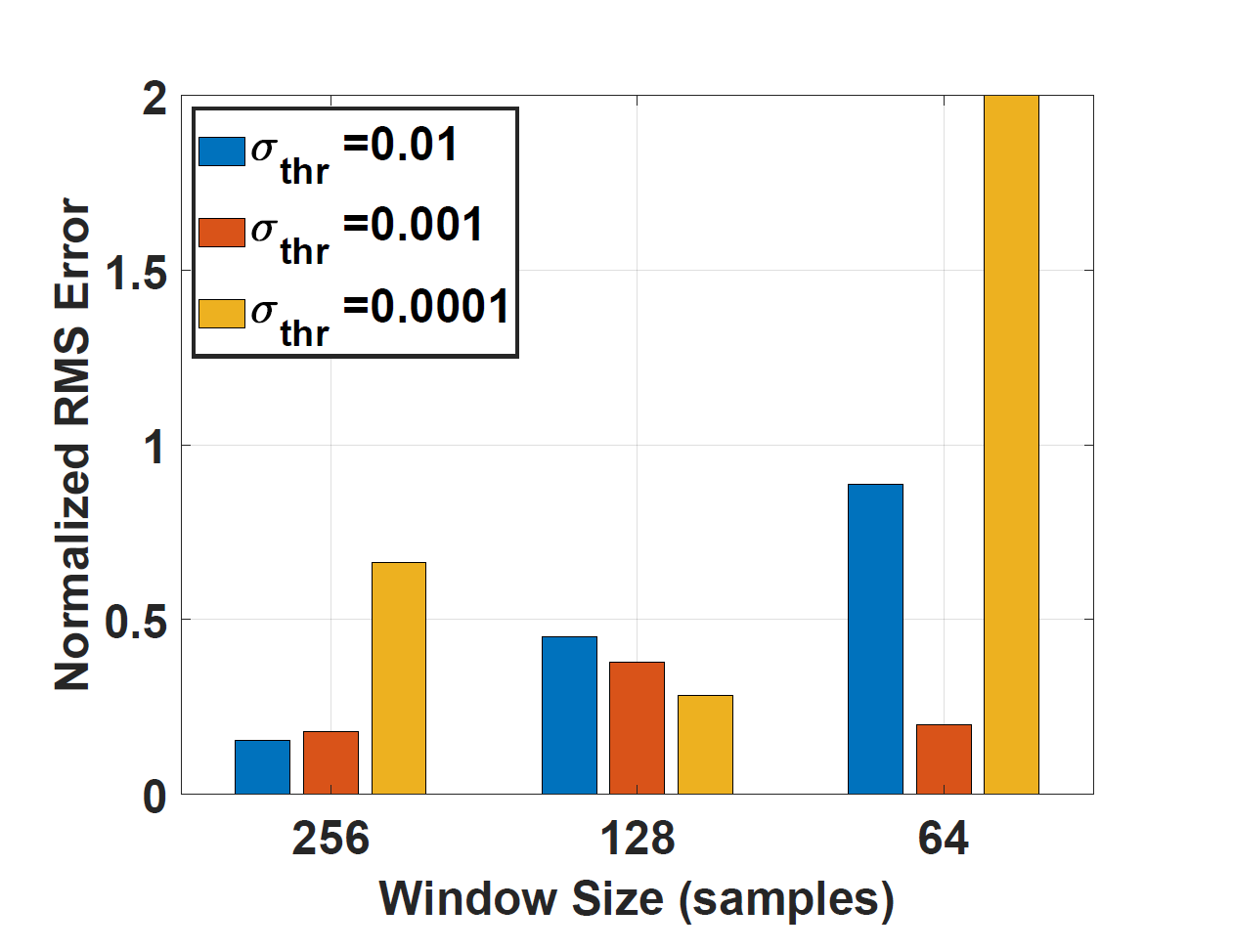}
} 
\subfloat[erroneous Event]
{
\includegraphics[width=0.4\linewidth]{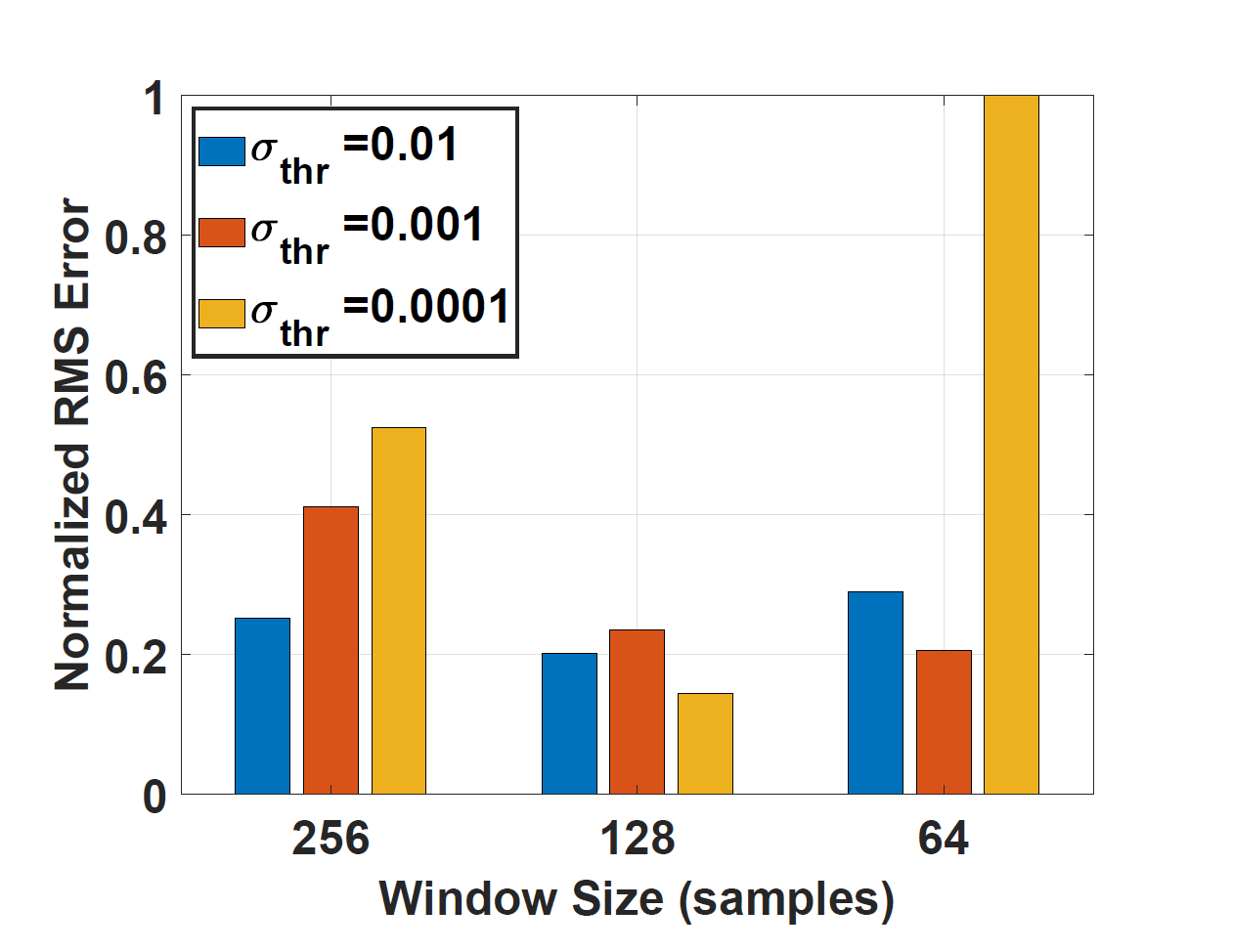}
} 
\caption{Normalized RMS error of ERP prediction using windowed incremental DMD with different window sizes during (a) correct event, and (b) Erroneous event.}
\label{ERP_RMS_windows}
\end{figure}

Figure \ref{ERP_RMS_128_windows} show the normalized RMS for the prediction error of the ERP signal during the correct and erroneous events using weighted incremental DMD with initial window size of $w_0=128$ samples, and Figure \ref{ERP_RMS_windows} shows the normalized RMS for the same error using windowed incremental DMD with $w=\{ 64,128,256\}$.   

% \clearpage

\section{Proof of Theorem~\ref{weighted incremental DMD with controlinput}}\label{proof of Theorem 3}

Consider the SVD of $\mathbf{X}^{c}_k= \mathbf{U}^{c}_{x_k} \mathbf{\Sigma}^{c}_{x_k} \mathbf{V}^{c*}_{x_k}$, where $\mathbf{U}_{x_k}^{c} \in \mathbb{C}^{(n+l) \times (n+l)}$, $\mathbf{\Sigma}_{x_k}^{c} \in \mathbb{C}^{(n+l) \times (n+l)}$, and $\mathbf{V}_{x_k}^{c} \in \mathbb{C}^{ k \times (n+l)}$, and the SVD of $\mathbf{S}^{c}_k \triangleq \begin{bmatrix}\mathbf{\Sigma}^c_{x_k} & \rho^{-1} \mathbf{U}_{x_k}^{c*} \mathbf{x}^c_{k+1} \end{bmatrix}=\mathbf{U}^c_{s_k} \mathbf{\Sigma}_{s_k}^c \begin{bmatrix}\mathbf{V}^{c*}_{s_{k,1}} & \mathbf{v}^{c*}_{s_{k,2}} \end{bmatrix}$. Then, the weighted incremental SVD presented in Proposition \ref{weighted incremental SVD  proposition } can be applied to obtain the SVD of $\mathbf{X}^{c}_{k+1}= \mathbf{U}^{c}_{x_{k+1}} \mathbf{\Sigma}^{c}_{x_{k+1}} \mathbf{V}^{c*}_{x_{k+1}}$ as follows
\begin{equation} \label{wighted SVD Xc(k+1)}
\mathbf{U}^{c}_{x_{k+1}} = \rho \mathbf{U}^{c}_{x_{k}} \mathbf{U}^{c}_{s_k},~
\mathbf{\Sigma}^{c}_{x_{k+1}}=\mathbf{\Sigma}_{s_k}^{c},~\text{and}~
\mathbf{V}^{c}_{x_{k+1}}=
\begin{bmatrix}
\mathbf{V}^{c}_{x_{k}} \mathbf{V}^{c}_{s_{k,1}} \\
\mathbf{v}^{c}_{s_{k,2}}
\end{bmatrix}. 
\end{equation}
Given the SVD of $\mathbf{X}_k= \mathbf{U}_{x_k} \mathbf{\Sigma}_{x_k} \mathbf{V}^{*}_{x_k}$, the updated SVD for $\mathbf{X}_{x_{k+1}}$ is given by  \eqref{weighted2 updated SVD factors}. Following the same steps presented in the proof of Theorem \ref{weighted DMD operator(k+1) theorem} with the SVD updates in \eqref{wighted SVD Xc(k+1)}, we can obtain the following weighted incremental DMDc update
\begin{eqnarray}\label{inc weighted DMDc(k+1)} 
\mathbf{G}^{\rho_c}_{k+1}&=&\mathbf{G}^{\rho_c}_{k}+\left(\mathbf{y}_{k+1}-\mathbf{G}^{\rho_c}_{k}\mathbf{x}^{c}_{k+1}\right)\mathbf{v}^{c}_{s_{k,2}}
({\mathbf{\Sigma}}^{c}_{x_{k+1}})^{-1} \mathbf{U}^{c*}_{x_{k+1}},
\end{eqnarray}
where $\mathbf{G}^{\rho_c}_{k} = \begin{bmatrix}
\mathbf{A}^{\rho_c}_k & \mathbf{B}^{\rho_c}_k
\end{bmatrix}$. Let ${\mathbf{U}}_{x_{k+1}}^{c*}$ be partitioned such that ${\mathbf{U}}_{x_{k+1}}^{c*} = [{\mathbf{U}}^{c_a*}_{x_{k+1}} {\mathbf{U}}^{c_b*}_{x_{k+1}}]$,  where ${\mathbf{U}}^{c_a}_{x_{k+1}} \in \mathbb{C}^{n \times n}$ and ${\mathbf{U}}^{c_b}_{x_{k+1}} \in \mathbb{C}^{l \times n}$. Then, the update in \eqref{inc weighted DMDc(k+1)} yields the following updates for $ {\mathbf{A}}^{\rho_c}_{k+1} \in \mathbb{R}^{n \times n}$ and ${\mathbf{B}}^{\rho_c}_{k+1} \in \mathbb{R}^{n \times l}$
\begin{align}
{\mathbf{A}}^{\rho_c}_{k+1}&={\mathbf{A}}^{\rho_c}_{k}+\left(\mathbf{y}_{k+1}-{\mathbf{A}}^{\rho_c}_{k}\mathbf{x}_{k+1}-{\mathbf{B}}^{\rho_c}_{k}\boldsymbol{\gamma}_{k+1}\right){\mathbf{v}}^{c}_{s_{k,2}} ({\mathbf{\Sigma}}^{c}_{x_{k+1}})^{-1} {\mathbf{U}}^{c_a*}_{x_{k+1}},  \nonumber \\                 
{\mathbf{B}}^{\rho_c}_{k+1}&={\mathbf{B}}^{\rho_c}_{k}+\left(\mathbf{y}_{k+1}-{\mathbf{A}}^{\rho_c}_{k}\mathbf{x}_{k+1}-{\mathbf{B}}^{\rho_c}_{k}\boldsymbol{\gamma}_{k+1}\right){\mathbf{v}}^{c}_{s_{k,2}} ({\mathbf{\Sigma}}^{c}_{x_{k+1}})^{-1} {\mathbf{U}}^{c_b*}_{x_{k+1}}. \nonumber                   
\end{align}

\section{Proof of Theorem~\ref{windowed incremental DMD with controlinput}}\label{proof of Theorem 4}
Let $q_c$ be defined by
\begin{equation}
  q_c \triangleq
    \begin{cases}
      n+l, & \text{if } n+l < w, \\
      w, & \text{otherwise}. \\
    \end{cases}\nonumber .     
\end{equation} 
Consider the SVD of $ \boldsymbol{\chi}_k^{c}= \mathbf{U}^{c}_{\chi_k} \mathbf{\Sigma}^{c}_{\chi_k} \mathbf{V}^{c*}_{\chi_k}$, where $\mathbf{U}_{\chi_k}^{c} \in \mathbb{C}^{(n+l) \times q_c}$, $\mathbf{\Sigma}_{\chi_k}^{c} \in \mathbb{C}^{q_c \times w}$, and $\mathbf{V}_{\chi_k}^{c} \in \mathbb{C}^{w \times w}$. Define the matrices $\mathbf{\acute{S}}^{c}_k$ and $\mathbf{\hat{S}}^{c}_k$ with their associated SVDs by
\begin{align}
\mathbf{\acute{S}}^{c}_k &\triangleq  \mathbf{\Sigma}^c_{\chi_k}- \mathbf{U}_{\chi_k}^{c*}\mathbf{x}^c_{k-w+1} \mathbf{z}_{1}^{\text{T}} \mathbf{V}^c_{\chi_k} =\mathbf{U}^c_{\acute{s}_k} 
\begin{bmatrix}
{\mathbf{\Sigma}}^c_{\acute{s}_k} &  \mathbf{0}_{q_c \times (w-q_c)}
\end{bmatrix}
\begin{bmatrix} \mathbf{V}^c_{\acute{s}_k} & \tilde{\mathbf{V}}^c_{\acute{s}_k} \end{bmatrix} ^*, \\
%%%
\mathbf{\hat{S}}^{c}_k &\triangleq \begin{bmatrix} {\mathbf{\Sigma}}^c_{\acute{\chi}_{k}} &  \mathbf{U}_{\acute{\chi}_{k}}^{c*} \mathbf{x}^c_{k+1} \end{bmatrix}=\mathbf{U}_{\hat{s}_k}^{c} \mathbf{\Sigma}_{\hat{s}_k}^c \begin{bmatrix}\mathbf{V}^{c*}_{\hat{s}_{k,1}} & \mathbf{v}^{c*}_{\hat{s}_{k,2}} \end{bmatrix}.
\end{align}
Then, the windowed incremental SVD presented in Proposition \ref{windowed incremental SVD  proposition} and equation~\eqref{windowed updated SVD factors} can be applied to obtain the SVD of $\boldsymbol{\chi}^{c}_{k+1}= \mathbf{U}_{\chi_{k+1}}^{c} \mathbf{\Sigma}_{\chi_{k+1}}^{c} \mathbf{V}^{c*}_{\chi_{k+1}}$ as follows
\begin{equation}\label{windowed updated SVDc factors Xc(k+1)}
\mathbf{U}_{\chi_{k+1}}^{c} = \mathbf{U}_{\chi_k}^{c} \mathbf{U}_{\acute{s}_k}^{c} \mathbf{U}_{\hat{s}_k}^{c}, ~
\mathbf{\Sigma}^{c}_{k+1}=\mathbf{\Sigma}^{c}_{\hat{s}_k}, ~ \text{and} ~
\mathbf{V}^{c}_{k+1}=
\begin{bmatrix}
\mathbf{V}_{\chi_{k,2}}^{c} \mathbf{V}_{\acute{s}_k}^{c} \mathbf{V}_{\hat{s}_{k,1}}^{c} \\
\mathbf{v}_{\hat{s}_{k,2}}^{c}
\end{bmatrix}. 
\end{equation}
Similarly, given the SVD of $\boldsymbol{\chi}_k= \mathbf{U}_{\chi_k} \mathbf{\Sigma}_{\chi_k} \mathbf{V}^{*}_{\chi_k}$, the SVD of $\boldsymbol{\chi}_{k+1}= \mathbf{U}_{\chi_{k+1}} \mathbf{\Sigma}_{\chi_{k+1}} \mathbf{V}^{*}_{\chi_{k+1}}$ can be obtained using the updates in \eqref{windowed updated SVD factors}. 

The same steps presented in the proof of Theorem \ref{windowed DMD operator(k+1) theorem} can be followed with the SVD update in \eqref{windowed updated SVDc factors Xc(k+1)} to obtain the following update
\begin{equation} \label{inc windowed DMDc(k+1)}  
\mathbf{G}^{w_c}_{k+1}=\mathbf{G}^{w_c}_{k}+\left(\mathbf{y}_{k+1}-\mathbf{G}^{w_c}_{k}\mathbf{x}^{c}_{k+1}\right)\mathbf{v}_{\hat{s}_{k,2}}^{c}(\mathbf{\Sigma}^{c}_{\chi_{k+1}})^{-1} \mathbf{U}^{c*}_{\chi_{k+1}}.  
\end{equation}
 Let ${\mathbf{U}}_{\chi_{k+1}}^{c*}$ be partitioned such that ${\mathbf{U}}_{\chi_{k+1}}^{c*} = [{\mathbf{U}}^{c_a*}_{\chi_{k+1}} {\mathbf{U}}^{c_b*}_{\chi_{k+1}}]$,  where ${\mathbf{U}}^{c_a}_{\chi_{k+1}} \in \mathbb{C}^{n \times n}$ and ${\mathbf{U}}^{c_b}_{\chi_{k+1}} \in \mathbb{C}^{l \times n}$. Then, the updates in \eqref{inc windowed DMDc(k+1)} yields the following updates for $ {\mathbf{A}}^{w_c}_{k+1} \in \mathbb{R}^{n \times n}$ and ${\mathbf{B}}^{w_c}_{k+1} \in \mathbb{R}^{n \times l}$
 \begin{align}
{\mathbf{A}}^{w_c}_{k+1}&={\mathbf{A}}^{w_c}_{k}+\left(\mathbf{y}_{k+1}-{\mathbf{A}}^{w_c}_{k}\mathbf{x}_{k+1}-{\mathbf{B}}^{w_c}_{k}\boldsymbol{\gamma}_{k+1}\right)
{\mathbf{v}}_{\hat{s}_{k,2}}^{c}({\mathbf{\Sigma}}^{c}_{\chi_{k+1}})^{-1} \mathbf{U}^{c_a*}_{\chi_{k+1}},  \nonumber \\                 
{\mathbf{B}}^{w_c}_{k+1}&={\mathbf{B}}^{w_c}_{k}+\left(\mathbf{y}_{k+1}-{\mathbf{A}}^{w_c}_{k}\mathbf{x}_{k+1}-{\mathbf{B}}^{w_c}_{k}\boldsymbol{\gamma}_{k+1}\right)
{\mathbf{v}}_{\hat{s}_{k,2}}^{c}({\mathbf{\Sigma}}^{c}_{\chi_{k+1}})^{-1} \mathbf{U}^{c_b*}_{\chi_{k+1}}. \nonumber 
\end{align}
\section{Incremental Singular Value Decomposition for Data Matrix Augmented with Exogenous Input}\label{app_Incremental DMD with input}
In this section, we present an incremental update to compute the SVD of the data matrix augmented with the exogeneous input. 
%A variant of incremental SVD with control input is presented here by which the SVD at time $t_k$ for the dataset of system state measurement $X_k$ is updated with the control input measurement $\mathbf{\Gamma}_k$ to obtain the SVD of $\mathbf{X}_k^{c}$. This variation 
%
This update can be beneficial in case of low-dimensional control input or limited storage capacity. Assume that at $t_k$, the state measurements are arranged in a matrix denoted by $\mathbf{X}_k \in \mathbb{R}^{n\times q_m}$, and the exogeneous input measurements are arranged in a matrix denoted by $\mathbf{\Gamma}_k \in \mathbb{R}^{l\times q_m}$, where $(q_m=k)$ for weighted incremental SVD and $(q_m=w)$ for windowed incremental SVD. Both $\mathbf{X}_k$ and $\mathbf{\Gamma}_k$ are arranged in an augmented matrix denoted by $\mathbf{X}^{c}_k\in \mathbb{R}^{(n+l)\times q_m}$ as follows:
\begin{equation}\label{XGamma(k)}
\mathbf{X}^{c}_k=
    \begin{bmatrix}
    \mathbf{X}_k\\
    \mathbf{\Gamma}_k
    \end{bmatrix},
\end{equation} 
where for weighted incremental SVD we have $\mathbf{\Gamma}_k= \begin{bmatrix} \rho ^{k-1} \boldsymbol{\gamma}_1 & \cdots & \boldsymbol{\gamma}_k\end{bmatrix}$ and $\mathbf{X}_k= \begin{bmatrix} \rho ^{k-1} \mathbf{x}_1 & \cdots & \mathbf{x}_k\end{bmatrix}$, and for windowed incremental SVD we have $\mathbf{\Gamma}_k= \begin{bmatrix} \boldsymbol{\gamma}_{k-w+1} & \cdots & \boldsymbol{\gamma}_k\end{bmatrix}$ and $\mathbf{X}_k= \begin{bmatrix} \mathbf{x}_{k-w+1} & \cdots & \mathbf{x}_k\end{bmatrix}$. {Let $q_n \triangleq n$ for weighted incremental SVD and $q_n \triangleq q$ for windowed incremental SVD, where $q$ is given in \eqref{generic dim}. Then, the SVD of $\mathbf{X}^{c}_k$ can be calculated by updating the SVD of $\mathbf{X}_k$ using the following proposition. 
\begin{proposition}
Let at sampling time $t_k$, the SVD of $\mathbf{X}_k=\mathbf{U}_k  \mathbf{\Sigma}_k \mathbf{V}_k^{*}$ be known, where $\mathbf{U}_k \in\mathbb{C}^{n\times q_n}$, $\mathbf{\Sigma}_k \in \mathbb{C}^{q_n \times q_m}$, and ${\mathbf{V}_k} \in\mathbb{C}^{q_m\times q_m}$. 
% Assume that at the same sampling time, a matrix of input measurements $\mathbf{\Gamma}_k$ is accessed. 
Then, the SVD of the dataset in \eqref{XGamma(k)}, defined by $\mathbf{X}^{c}_k=\mathbf{U}^{c}_k \mathbf{\Sigma}^{c}_k\mathbf{V}^{c*}_k$, is 
\begin{eqnarray}\label{SVD with control}
\mathbf{U}^{c}_k = 
\begin{bmatrix}
\mathbf{U}_{k} \mathbf{U}^{c}_{R_1}\\ 
\mathbf{U}^{c}_{R_2}
\end{bmatrix},~
\mathbf{\Sigma}^{c}_k = \mathbf{\Sigma}^{c}_{R},~\text{and} ~
\mathbf{V}^{c}_k= {\mathbf{V}_{k}} {\mathbf{V}}^{c}_{R},
\end{eqnarray}
where $\mathbf{U}^{c}_{R} = \begin{bmatrix}\mathbf{U}^{c}_{R_1}\\ \mathbf{U}^{c}_{R_2}\end{bmatrix}$, $\mathbf{\Sigma}^{c}_{R}$, and ${\mathbf{V}}^{c}_{R}$ are given by the following SVD:
\begin{equation}
 \begin{bmatrix}
\mathbf{\Sigma}_{k} \\
\left({\mathbf{V}^*_k} \mathbf{\Gamma}_k\right)^*
\end{bmatrix} = \mathbf{U}^{c}_{R} \mathbf{\Sigma}^{c}_{R} {\mathbf{V}}^{{c_l}*}_{{R}}.\nonumber
\end{equation} 
\begin{proof}

The new dataset matrix  $\mathbf{X}^{c}_k$ can be represented in term of $\mathbf{X}_k$ using the following additive update formula:
\begin{equation} \label{X_k with control}
 \mathbf{X}^{c}_k= \begin{bmatrix}\mathbf{X}_k \\ \mathbf{0}\end{bmatrix} + \mathbf{Z}_{n+1} \mathbf{\Gamma}_k^{\text{T}},
\end{equation}
where $\mathbf{Z}_{n+1}= \begin{bmatrix} \mathbf{0} & \mathbf{0} & \dots & \mathbf{I}_l\end{bmatrix}^{\text{T}} \in \mathbb{R}^{(n+l) \times l}$, where $\mathbf{I}_l$ is the identity matrix of order $l$. 
Then, the new dataset in \eqref{X_k with control} can be written as: 
\begin{eqnarray}\label{X_k with control1}
\mathbf{X}^{c}_k &=& \begin{bmatrix}\mathbf{U}_k \\ \mathbf{0} \end{bmatrix}
\mathbf{{\Sigma}}_k {\mathbf{{V}}_k} + \mathbf{Z}_{n+1}\mathbf{\Gamma}_k^{\text{T}}
= \begin{bmatrix}
\begin{bsmallmatrix}\mathbf{U}_k \\ \mathbf{0} \end{bsmallmatrix}  & \mathbf{Z}_{n+1}
\end{bmatrix}
\begin{bmatrix}
\mathbf{{\Sigma}}_k & \mathbf{0} \\
\mathbf{0} & \mathbf{I}_l
\end{bmatrix}
\begin{bmatrix}
\mathbf{{V}}_k & \mathbf{\Gamma}_k
\end{bmatrix}^{*}\nonumber \\ 
&=& 
\begin{bmatrix}
\mathbf{U}_{k} & \mathbf{0}\\
\mathbf{0} & \mathbf{I}_l
\end{bmatrix}
\begin{bmatrix}
\mathbf{\Sigma}_{k} & \mathbf{0}\\
 \mathbf{0} & \mathbf{I}_l
\end{bmatrix}
\left(\mathbf{V}_{k}\begin{bmatrix}
 \mathbf{I}_q &  \mathbf{V}^*_{k}\mathbf{\Gamma}_k
\end{bmatrix}\right)^*
=
\begin{bsmallmatrix}
\mathbf{U}_{k} & \mathbf{0}\\
\mathbf{0} & \mathbf{I}_l
\end{bsmallmatrix}
\begin{bsmallmatrix}
\mathbf{\Sigma}_{k}\\
\left(\mathbf{V}_{k}^* \mathbf{\Gamma}_k\right)^*
\end{bsmallmatrix}
\mathbf{V}_{k}^*.      
\end{eqnarray}
Define $ \mathbf{R}^{c} \triangleq \begin{bsmallmatrix}
\mathbf{\Sigma}_{k} \\
\left({\mathbf{V}^*_k} \mathbf{\Gamma}_k\right)^*
\end{bsmallmatrix}$, which has the SVD of $\mathbf{R}^{c}=\mathbf{U}^{c}_{R} \mathbf{\Sigma}^{c}_{R} {\mathbf{V}}^{{c_l}*}_{{R}}$, where $\mathbf{U}^{c}_{R} = \begin{bsmallmatrix}
\mathbf{U}^{c}_{R_1}\\ 
\mathbf{U}^{c}_{R_2}
\end{bsmallmatrix} \in \mathbb{C}^{(n+l) \times q_n}$, $\mathbf{U}^{c}_{R_1}\in \mathbb{C}^{n \times q_n}$, $\mathbf{U}^{c}_{R_2}\in \mathbb{C}^{l \times q_n}$, $\mathbf{\Sigma}^{c}_{R} \in \mathbb{C}^{q_n \times q_m}$, and $\mathbf{V}^{c}_{R} \in \mathbb{C}^{q_m \times q_m} $. Then, after substituting the SVD of matrix $\mathbf{R}^{c}$ in \eqref{X_k with control1}, the updated dataset becomes:
\begin{eqnarray}\label{update with control4}
\mathbf{X}^{c}_{k}=
\begin{bmatrix}
\mathbf{U}_{k} & \mathbf{0}\\
\mathbf{0} & \mathbf{I}_l
\end{bmatrix}
\begin{bmatrix}
\mathbf{U}^{c}_{R_1}\\ 
\mathbf{U}^{c}_{R_2}
\end{bmatrix}
\mathbf{\Sigma}^{c}_{R}
\left({\mathbf{V}_{k}} \mathbf{V}^{c}_{R}\right)^* 
=
\begin{bmatrix}
\mathbf{U}_{k} \mathbf{U}^{c}_{R_1}\\ 
\mathbf{U}^{c}_{R_2}
\end{bmatrix}
\mathbf{\Sigma}^{c}_{R}
\left({\mathbf{V}_{k}} {\mathbf{V}}^{c}_{R}\right) ^*, \nonumber
\end{eqnarray} 
Finally the SVD factors for $\mathbf{X}^{c}_{k}$ is defined by:
\begin{eqnarray}
\mathbf{U}^{c}_k = 
\begin{bmatrix}
\mathbf{U}_{k} \mathbf{U}^{c}_{R_1}\\ 
\mathbf{U}^{c}_{R_2}
\end{bmatrix},~
\mathbf{\Sigma}^{c}_k = \mathbf{\Sigma}^{c}_{R},~\text{and} ~
\mathbf{V}^{c}_k= {\mathbf{V}_{k}} {\mathbf{V}}^{c}_{R}.\nonumber
\end{eqnarray}
\end{proof}
\end{proposition}
}For $l=1$, the matrix $\mathbf{R}$ has a broken-arrow structure and its SVD can be computed efficiently~\cite{brand2006fast}. For $l>1$, the matrix is still sparse but does not carry the broken-arrow structure. In this case, either the above procedure can be applied recursively with one input added at a time or problem specific sparse SVD solvers can be employed.

\footnotesize 
%\bibliographystyle{AIMS}
%\bibliography{references.bib}

\providecommand{\href}[2]{#2}
\providecommand{\arxiv}[1]{\href{http://arxiv.org/abs/#1}{arXiv:#1}}
\providecommand{\url}[1]{\texttt{#1}}
\providecommand{\urlprefix}{URL }

\end{document}